\documentclass[rmp,aps,12pt]{revtex4-1}
\usepackage{graphicx}
\usepackage{amsfonts}
\usepackage[utf8]{inputenc}
\DeclareUnicodeCharacter{2212}{-}

\usepackage[normalem]{ulem}
\usepackage[colorlinks,citecolor=blue]{hyperref}

\usepackage[colorlinks,citecolor=blue]{hyperref}
\def\lsim{\mathrel{\raise.3ex\hbox{$<$\kern-.75em\lower1ex\hbox{$\sim$}}}}
\def\gsim{\mathrel{\raise.3ex\hbox{$>$\kern-.75em\lower1ex\hbox{$\sim$}}}}

\begin{document}


\bibliographystyle{apsrmp4-1}

\title{Using gravitational waves to see the first second of the Universe
}
\author{Rishav Roshan and Graham White}

\affiliation{School of Physics and Astronomy, University of Southampton,\\
	Southampton SO17 1BJ, United Kingdom}

\date{\today}

\begin{abstract}

Gravitational waves are a unique probe of the early Universe, as the Universe is transparent to gravitational radiation right back to the end of inflation. In this article, we summarise detection prospects and the wide scope of primordial events that could lead to a detectable stochastic gravitational wave background. Any such background would shed light on what lies beyond the Standard Model, sometimes at remarkably high scales. We overview the range of strategies for detecting a stochastic gravitational wave background before delving deep into three major primordial events that can source such a background. Finally, we summarize the landscape of other sources of primordial backgrounds.

\end{abstract}

\maketitle

\tableofcontents

\cleardoublepage






\section{Introduction}
\label{prehistorychapter}


Gravitational waves (GW) are one of the most striking predictions of Einstein's theory of general relativity \cite{einstein1916naherungsweise,einstein1918gravitationswellen}. Remarkably, the concept of gravitational waves predates Einstein, as physicists at the turn of the 19th century wondered if there was a gravitational analog of electromagnetic waves \cite{poincare1913dynamique}. Moreover, Einstein famously published an incorrect proof that gravitational waves do not exist \cite{kennefick2007traveling}.\footnote{This was in a paper that was originally rejected by physical review, with the referee later vindicated that Einstein made the mistake of making a pathological coordinate transformation \cite{blum2022einstein}. Eventually, Einstein published a corrected version that argued for the existence of cylindrical gravitational waves \cite{einstein1937gravitational}.} Even still, the recent discovery of gravitational waves by aLIGO \cite{LIGOScientific:2016aoc} is widely seen as one of his greatest triumphs.

Decades before it became possible to observe the Universe via gravitational waves, another consequence of Einstein's theory was being explored - that the Universe expanded from a hot dense initial state in a theory later coined the ``big bang theory'' \cite{lemaitre1931beginning}. The Big Bang theory predicts that the Universe should expand and cool until the point at which free protons and electrons combine into the first hydrogen atoms, after which the mean free path of photons rapidly grows to the point where its light is visible today in a black body spectrum. This cosmic microwave background radiation (CMB), was later seen in 1964, falsifying the steady state theory and verifying the Big Bang theory of cosmology \cite{penzias1979measurement}. Since then we have entered the era of precision measurements of nature's first light, which allows us to probe features of the primordial fireball \cite{Planck:2018vyg}. In particular, we can measure the expansion of the Universe and the asymmetry between matter and anti-matter\cite{Riotto:1998bt,Riotto:1999yt,Dine:2003ax,Cline:2006ts,Canetti:2012zc,Morrissey:2012db,balazs2014baryogenesis,White:2016nbo,Bodeker:2020ghk} at the time when the CMB forms. We have an alternative measurement of these two observables. About a second after the moment of creation, the first nucleons were synthesized and we can cross-check the primordial abundances of hydrogen, helium, and deuterium with our predictions, the latter of which are sensitive to the precise value of the baryon asymmetry and the expansion rate of the Universe \cite{cyburt2016big}. In other words, we can calculate the expansion rate and baryon asymmetry during Big Bang nucleosynthesis (BBN) by measuring the primordial abundances. Remarkably, in a triumph of modern cosmology, we find concordance between the cosmic microwave background and Big Bang nucleosynthesis \cite{ParticleDataGroup:2004fcd,Planck:2018vyg}.


Despite the success of our current model of cosmology (up to some recent tensions with data \cite{di2021realm}), physicists face a huge ``gap problem'' where we know nothing about the period between a early epoch of rapid expansion, known as inflation and BBN. Measured in SI units, the problem does not seem alarming - we are in the dark by a mere second of our history. Measured in terms of temperature demotes cosmology to being a field in its infancy. The history of the Universe potentially spans twenty-two orders of magnitude in temperature between the end of inflation and the onset of BBN \cite{deSalas:2015glj,DelleRose:2015bpo}. Being ignorant of this period renders us incapable of understanding why there is more matter than antimatter \cite{Riotto:1998bt,Riotto:1999yt,Dine:2003ax,Cline:2006ts,Canetti:2012zc,Morrissey:2012db,balazs2014baryogenesis,White:2016nbo,Bodeker:2020ghk}, what dark matter is and how it came to be \cite{Bergstrom:2000pn,Bertone:2004pz,Feng:2010gw,garrett2011dark,peter2012dark,Bertone:2016nfn,arun2017dark,bertone2018new,arbey2021dark}, how hot the Universe was and how did it become so hot after inflation \cite{bassett2006inflation,Allahverdi:2010xz}, was this period always dominated by radiation or were there early periods of matter domination \cite{allahverdi2020first}, was it always in thermal equilibrium or did the Universe come to boil \cite{linde1979phase,grojean2007gravitational,boyanovsky2006phase,weir2018gravitational,mazumdar2019review,caprini2020detecting}?


Gravitational wave cosmology is the unrivaled method to make progress on the gap problem - the Universe is transparent to gravitational waves right up to the instant of its birth. Any violent event in that first second will leave a trace in the stochastic gravitational wave background (SGWB) which we can hope to detect today. The next generation promises to birth a new dawn of gravitational wave detection, with ground and space-based interferometers pledging to search for primordial spectra in the $\mu$Hz to kHz range \cite{LIGOScientific:2014pky,LIGOScientific:2016emj,aLIGO:2020wna,Tse:2019wcy,VIRGO:2014yos,Virgo:2019juy,KAGRA:2018plz,Aso:2013eba,Reitze:2019iox,LISA:2017pwj,Sesana:2019vho,Kawamura:2019jqt} whereas astrometry \cite{Kaiser:1996wk,Gaia:2018ydn,Theia:2017xtk,Book:2010pf,Moore:2017ity,Mihaylov:2018uqm,Mihaylov:2019lft,Garcia-Bellido:2021zgu,Fedderke:2021kuy,Caliskan:2023cqm,Moore:2017ity,Klioner:2017asb,Wang:2020pmf,Wang:2022sxn} and pulsar timing arrays \cite{EPTA:2023sfo,NANOGrav:2023hde,Zic:2023gta,Weltman:2018zrl,Zhu:2015ara,Burke-Spolaor:2018bvk,Dahal:2020dvb,sazhin1978opportunities} promise to be sensitive to low-frequency gravitational waves down to the nanoHz range. Moreover, there are even a rich set of ideas percolating to probe high-frequency gravitational wave sources which, if successful, would allow us to be even more ambitious with how close we can get to probing the instant after inflation \cite{Aggarwal:2020olq}.

In this review, we review how the scientific community pulled off the immense task of detecting gravitational waves and what efforts exist on the horizon to detect the SGWB in section \ref{sec:detector}. We then review three types of events that can lead to a SGWB - cosmic first-order phase transitions, topological defects, and scalar-induced gravitational waves, whether from a period of ultra-slow roll or a sudden change in the equation of the state of the Universe. In all three sections \ref{sec:phase}, \ref{sec:top} and \ref{sec:SIGW} respectively, we review applications of such sources to some of the deeper, outstanding questions in the field. Finally, we give a brief overview of other sources of gravitational waves in section \ref{sec:HFsources } before concluding.

\section{Detection of GW backgrounds}\label{sec:detector}

\begin{figure}
    \centering
    \includegraphics[width=0.65\textwidth]{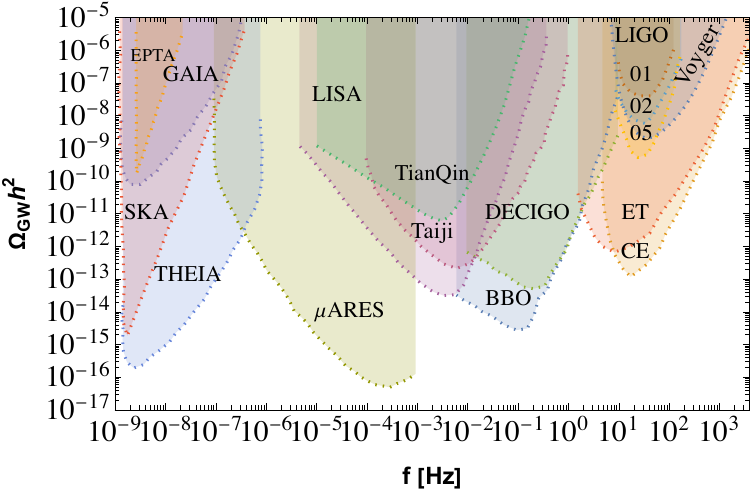}
    \caption{Power law integrated sensitivity curve of current and future gravitational wave detectors. These include future pulsar timing arrays including the NANOGrav \cite{NANOGrav:2023hde} (16 years), the EPTA \cite{EPTA:2023sfo} (5 years), IPTA \cite{Hobbs:2009yy} and SKA \cite{Weltman:2018zrl} (10 years), astrometry proposal like Gaia \cite{Moore:2017ity,Gaia:2018ydn} (5 years) and Theia \cite{Theia:2017xtk,Garcia-Bellido:2021zgu} (20 years), space-based interferometers including $\mu$-Ares \cite{Sesana:2019vho} (10 years), LISA \cite{LISA:2017pwj} (4 years) and DECIGO \cite{Kawamura:2020pcg} (3 years) and ground-based interferometers including LIGO \cite{LIGOScientific:2014pky} (20 years), TianQin \cite{TianQin:2015yph}, Taiji \cite{Hu:2017mde}, Voyger \cite{LIGO:2020xsf}, cosmic explorer \cite{Reitze:2019iox} (20 to 50 years) and the Einstein telescope \cite{Maggiore:2019uih} (3 years). In the parenthesis, we mention the tentative mission lifetime of each detector.}
    \label{fig: GW_bounds}
\end{figure}

It was many decades after gravitational waves were first proposed that they were finally discovered, which speaks to the difficulty of the task. All experimental designs experience noise which dwarfs the amplitude of gravitational waves of any known signal.  The ambitious effort to bring about an era of gravitational wave cosmology has therefore required remarkable ingenuity on both the theoretical and experimental fronts. In this section, we explain how to model gravitational waves and the main strategies the community has come up with to detect them.\footnote{For a very informative review, we highly recommend the seminal textbook by Maggiore \cite{maggiore2007gravitational} together with some review articles \cite{Ricci:1997cx,Aufmuth:2005sv,Romano:2016dpx,Aggarwal:2020olq}}

Let us begin by showing the large payoff to accurately modeling predicted gravitational wave spectra. We will begin by explaining how to extract transient sources from the noisy background to introduce key concepts of filtering before moving on to the more relevant stochastic gravitational wave backgrounds. \par A gravitational wave detector can be thought of as a \emph{linear system} whose output $s(t)$ is considered as a combination of a Gravitational wave signal ($h(t)$) and noise ($n(t)$),
    \begin{equation}
        s(t)=h(t)+n(t).
        \label{output}
    \end{equation}
    Assuming the noise to be stationary\footnote{In a more realistic situation, the noise is not stationary for example each detector has a period where it is quieter with less environmental disturbance and a period where environmental disturbance is more.} (does not change over time), one can define a noise spectral density or noise power spectrum ($S_n(f)$) as 
   \begin{equation}
      \langle\tilde{n}^\star(f)\tilde{n}(f^\prime)\rangle=\delta(f-f^\prime)\frac{1}{2}S_n(f),
    \label{nsr_1}
    \end{equation}
    where $\langle..\rangle$ represents the ensemble average obtained by repeating the process of measuring the noise over a given time interval and $\tilde{n}(f)$ denotes the Fourier noise component. Factor 1/2 is inserted so that the ensemble average is obtained integrating $S_n(f)$ over the physical range if frequency $0\leq f<\infty$. Next, one can define the \emph{spectral strain sensitivity}, $\sqrt{S_n(f)}$, that characterizes the noise of the detector. 
    As mentioned above, with the present sensitivity of all gravitational wave detectors on the horizon, it is always true that $|h(t)|<<|n(t)|$. In such a situation, digging out the GW signal from a much larger noise becomes important. It is possible to detect a GW signal whose amplitude is much smaller than the noise if we have some knowledge of its form. Assuming we know the form of the signal, we can define a filter function, $K(T)$,
    \begin{equation}
       \hat{s}=\int_{-\infty}^{\infty}dt~ s(t)K(t). 
       \label{filter-function}
    \end{equation}
    that maximizes the signal-to-noise ratio for such a signal. \emph{Match filtering} is a technique where a filter function is chosen to match the signal that we are looking for. 
    We are also aided in our search for gravitational waves by the fact that the average noise vanishes. That is,
$\langle n(t)\rangle=0$. Under such a condition, the signal is
    \begin{equation}
        S=\int_{-\infty}^\infty dt \langle s(t)\rangle K(t)\\
        =\int_{-\infty}^\infty dt ~h(t) K(t)\\
         =\int_{-\infty}^\infty df ~\tilde{h}(f) \tilde{K}^\star(f)\ .
    \end{equation}
whereas the noise is by fiat defined in the absence of a signal, $\langle \hat{s}(t)\rangle=0$, so
we have
    \begin{equation}
      N^2=[\langle \hat{s}^2(t)\rangle]_{h=0}=\int_{-\infty}^\infty dt dt^\prime \langle n(t)n(t^\prime)\rangle K(t)K(t^\prime).
        \label{N_2}   
    \end{equation}
    Following Eq.~(\ref{nsr_1}), one can rewrite the above equation as,
    \begin{equation}
        N^2=\int_{-\infty}^\infty df \frac{1}{2}S_n(f) |\tilde{K}(f)|^2,
    \end{equation}
    and so the signal-to-noise ratio is then
    \begin{equation}
        \frac{S}{N}=\frac{\int_{-\infty}^\infty dt ~\tilde{h}(f) \tilde{K}^\star(f)}{[\int_{-\infty}^\infty df \frac{1}{2}S_n(f) |\tilde{K}(f)|^2]^{1/2}}.
        \label{SNR_1}
    \end{equation}
If we know the signal we are looking for, we can choose a filter function that optimizes the signal-to-noise ratio
\begin{equation}
    \tilde{K}(f)=\text{const.}\frac{\tilde{h}(f)}{S_n(f)}.
    \label{filter_function}   
\end{equation}
Using this, one can write the optimal value of $S/N$ \cite{maggiore2007gravitational}, 
\begin{equation}
    \bigg(\frac{S}{N}\bigg)^2\leq 4\int_{0}^\infty df~\frac{|\tilde{h}(f)|^2}{S_n(f)}.
    \label{SNR}
\end{equation}
However, one can only achieve this bound if we can accurately predict the signal. Primordial sources are difficult to model - the formidable challenges in doing so we discuss in later sections. For now, we just make note of the big payoff in developing the theoretical technology to predict gravitational waves.


Primordial gravitational wave sources will be stochastic backgrounds which will have additional challenges compared to transient sources. To see why, let us first use plane wave expansion and write,
\begin{eqnarray}
    h_{ij}(t,\textbf{x})=\sum_{A=+,x}\int_{-\infty}^{\infty}df\int d^2\hat{\textbf{n}}~\tilde{h}_A(f,\hat{\textbf{n}})e^{A}_{i,j}(\hat{\textbf{n}})e^{-2\pi if(t-\hat{\textbf{n}}.\textbf{x}/c)},
    \label{plane_wave}
\end{eqnarray}
where $e^{A}_{i,j}$ is the polarization tensor with $A=+,x$ denoting the polarization. In TT gauge, $h^i_i=0$ and $\partial^jh_{i,j}=0$. Next we assume that these stochastic gravitational wave backgrounds (SGWB)s are stationary, Gaussian, isotropic, and unpolarized and hence can be characterized by a spectral density of stochastic background $S_h(f)$ (defined analogously to the noise spectral density) as in Eq. \ref{nsr_1}
\begin{equation}
\langle\tilde{h}_A^\star(f,\hat{\textbf{n}})\tilde{h}_{A^\prime}^\star(f^\prime,\hat{\textbf{n}}^\prime)\rangle=\delta(f-f^\prime)\frac{\delta^2(\hat{\textbf{n}},\hat{\textbf{n}}^\prime)}{4\pi}\delta_{AA^\prime}\frac{1}{2}S_h(f),
    \label{sdGW_1}
    \end{equation}
where $\tilde{h}_A(f,\hat{\textbf{n}})$ are the  amplitudes of the stochastic GW with polarization $A$, coming from all possible propagation directions $\hat{\textbf{n}}$. The dependence of $\langle\tilde{h}_A^\star(f,\hat{\textbf{n}})\tilde{h}_{A^\prime}^\star(f^\prime,\hat{\textbf{n}}^\prime)\rangle$ on $\delta^2(\hat{\textbf{n}},\hat{\textbf{n}}^\prime)$ and $\delta_{AA^\prime}$ is a result of uncorrelated and unpolarized nature of these waves. Looking at Eqns.~(\ref{nsr_1}) and~(\ref{sdGW_1}), it is clear that a detector sees a SGWB as an additional source of the noise, so the distinction of a SGWB signal from noise becomes a big challenge! This suggests that one needs to set a relatively higher threshold on SNR while detecting a SGWB signal.

The next goal is to determine the minimum value of the energy density of the GW that can be
measured at a given S/N. The energy density of the GW ($\rho_{\rm GW}$) is related to $h_{ij}$ as,
\begin{eqnarray}
    \rho_{\rm GW}=\frac{c^2}{32\pi G}\langle\dot{h}_{ij}\dot{h}^{ij}\rangle.
    \label{rho_gw}
\end{eqnarray}
The intensity of the stochastic GW can be expressed using a dimension less quantity,
\begin{eqnarray}
    \Omega_{\rm GW}(f)=\frac{1}{\rho_c}\frac{d \rho_{\rm GW}}{d\log{f}},
\end{eqnarray}
with $\rho_c$ being the critical energy density of the Universe. Substituting Eq. \ref{plane_wave} in Eq. \ref{rho_gw} and using Eq. \ref{sdGW_1} one can calculate the ensemble average and that results in,
\begin{eqnarray}
    \rho_{\rm GW}=\frac{c^2}{8\pi G}\int_{f=0}^{f=\infty}d(\log{f})f(2\pi f)^2 S_h(f).
\end{eqnarray}
using this equation we can calculate, 
\begin{equation}
\Omega_{\text{GW}}(f)=\frac{4\pi^2}{3 H_0}f^3S_h(f) .   
\label{omega_2}
\end{equation}

 The minimum detectable value of $\Omega_{\text{GW}}$ for a single detector is given by
\begin{equation}
  \Omega_{\text{GW}}(f)|_{\text{min}}=  \frac{4\pi^2}{3 H_0}f^3[S_h(f)]_{\text{min}},
  \label{omega_min}
\end{equation}
with
\begin{equation}
    [S_h(f)]_{\text{min}}=S_n(f)\frac{(S/N)^2}{F}.
    \label{sh_min}
\end{equation}
Here $F$ denotes the angular efficiency factor~\cite{maggiore2007gravitational} whose value varies depending on the choice of detector. The presence of $f^3$ in the above expression plays a nontrivial role. High-frequency signals from a primordial source will have a very similar peak value of $\Omega _{\rm GW}$ as low-frequency signals. However, detectors are sensitive to $S_h$, so it becomes progressively more difficult to resolve a gravitational wave source as the frequency gets higher.

Due to the unpredictable nature of the SGWB signal, the match-filtering technique is not easy for a single detector. The advantage of using two or more detectors over a single detector is that one can use a modified form of match-filtering where the output of one detector can be matched with the output of another. Analogous to Eq.~(\ref{SNR_1}), for a set of two detectors, one can follow the procedure for a single source and assume we know the signal shape we are looking for, and therefore the optimal filter function, to derive the optimal signal to noise ratio
\begin{equation}
        \bigg(\frac{S}{N}\bigg)^2=2T\int_{0}^\infty df \Gamma^2(f) \frac{S_h^2(f)}{S_n^2(f)}.
        \label{SNR_SGWB_2}
    \end{equation}
   Here $\Gamma(f)$~\cite{maggiore2007gravitational} is the overlap reduction function that takes into account the fact that two detectors can see different gravitational wave signals, either because they are at a different location or because they have different angular sensitivity.


Finally, one can also think of a situation where the number of identical detectors is more than two, $i.e.~N>2$. In such a case, with these $N$ detectors we can form $N(N-1)/2$ independent two-point correlators. It is interesting to point out that, for a stationary background, a scenario with $N$ detectors running for time $T$ is identical to a situation in which a pair of detectors run for a total time: $T\times N(N-1)/2$. Hence, one can simply obtain the $S/N$ with $N$ identical detectors just by replacing,
\begin{equation}
    T\to\frac{N(N-1)}{2}T
\end{equation}
in Eq.~(\ref{SNR_SGWB_2}). This boost in the signal-to-noise ratio from correlating more than two detectors gets used in, for example, the $\mu$Ares proposal \cite{Sesana:2019vho} which we will discuss in section \ref{sec:interfer}.

\subsection{GW Detectors}
\subsubsection{Pulsar Timing arrays}
Pulsar timing arrays (PTA) \cite{Zhu:2015ara,Burke-Spolaor:2018bvk,Dahal:2020dvb} are a promising method of detecting an ultra-low frequency ($10^{-9}\rm{Hz} -10^{-6}Hz $) GW. They can measure a tiny variation induced by a GW, passing in between the Earth and a pulsar, in the time of arrival of the pulses emitted by the milliseconds pulsar by exploiting the telescope used for radio astronomy \cite{sazhin1978opportunities}. 

To understand how PTAs can detect SGWB we need to have information about the time interval ($\Delta t$) of two successive pulses reaching the Earth, the distance ($d_a$, label $a$ is useful for generalizing to several pulsars ) between the Earth and the pulsar that emits the photons, the direction of propagation ($\hat{\bf{n}}$) of the GW and the timing residual denoted by $R_a$ that describes the observed variation in the time of arrival of the pulses as a result of passing by GW. Now, working in a TT gauge and choosing a reference frame such that the Earth is at the origin of the coordinate system, one can model how the time interval, ($\Delta t$), between two successive pulses reaching the Earth from a pulsar depends upon passing through the gravitational wave background. To do that one can write,
\begin{equation}
    \Delta t = T_a+\Delta T_a
\end{equation}
where $T_a$ denotes the rotational period of the pulsar which is typically of the order of a few milliseconds and $\Delta T_a$ represents the delay introduced as a result of a passing GW, 


\begin{eqnarray}
    \frac{\Delta T_a}{T_a} &=& \frac{n_a^in_a^j}{2}\int_{t_{em}}^{t_{em}+d_a}dt^\prime \bigg[\frac{\partial}{\partial t^\prime}h_{ij}^{\text{TT}}(t^\prime,\textbf{x})\bigg]_{\textbf{x}=\textbf{x}_0(t^\prime)},\label{delay}\\
    \textbf{x}_0(t^\prime)&=&(t_{em}+d_a-t^\prime)\hat{\bf{n}}_a,
\end{eqnarray}
where $t_{em}$ is the time of emission of the photons emitted from the pulsar. Assuming a monochromatic GW propagating along the $\hat{\textbf{n}}$ direction, we can write
\begin{equation}
    h_{ij}^{\text{TT}}(t,\textbf{x})=\mathcal{A}_{ij}(\hat{\textbf{n}})\cos[\omega_{\text{gw}}(t-\hat{\textbf{n}}.\textbf{x})],
\end{equation}
where $n^i\mathcal{A}_{ij}(\hat{\textbf{n}})=0$. 
Let us also define a quantity
\begin{equation}
    z_a(t)\equiv \bigg(\frac{\nu_0-\nu(t)}{\nu_0}\bigg)_a, 
\end{equation}
such that $ z_a(t)=-(\Delta \nu_a/\nu_a)(t)=(\Delta T_a/T_a)$ where $\nu_a=1/T_a$ and substitute our form of the gravitational wave into Eq. \ref{delay} as,
\begin{eqnarray}
    z_a(t) &=& \frac{n_a^in_a^j}{2(1+\hat{\textbf{n}}.\hat{\textbf{n}}_a)}[h_{ij}^{\text{TT}}(t,\textbf{x}=0)-h_{ij}^{\text{TT}}(t-\tau_a,\textbf{x}_a)].
    \label{delay3}
\end{eqnarray}
Here, $t_{ob}$ is simply replaced by $t$, and $\textbf{x}=0$ represents the observer's position. Finally, one can define the timing residuals $R_a$ of the $a^{th}$ pulsar as
\begin{equation}
    R_a(t)=\int_{0}^{t}dt^\prime z_a(t^\prime).
\end{equation}

With this, we can model the detection of SGWB using PTAs. Following Eq.~\ref{sdGW_1} and Eq.~\ref{delay3}, we can calculate ensemble average $\langle z_a(t)z_b(t)\rangle$ of the pulse redshift of a pair of millisecond pulsars,
\begin{equation}
 \langle z_a(t)z_b(t)\rangle=\frac{1}{2}\int_{-\infty}^{\infty}df S_h(f)\int\frac{d^2\hat{\textbf{n}}}{4\pi} \sum_{A=+,\rm x}F^A_a(\hat{\textbf{n}}) F^A_b(\hat{\textbf{n}}), 
 \label{SGWB_PTA}
\end{equation}
where $F^A_a(\hat{\textbf{n}})=\frac{n_a^in_a^j e^{A}_{ij}}{2(1+\hat{\textbf{n}}.\hat{\textbf{n}}_a)}$ represents the pattern function with $e^{A}_{ij}$ begin the polarization tensor (see \cite{maggiore2007gravitational} for detail). We can now write the angular part of the integral as
\begin{equation}
    C(\theta_{ab})\equiv \int\frac{d^2\hat{\textbf{n}}}{4\pi} \sum_{A=+, \rm x}F^A_a(\hat{\textbf{n}}) F^A_b(\hat{\textbf{n}}).
    \label{HD_func}
\end{equation}
Here $C(\theta_{ab})$ is known as the Hellings and Downs function~\cite{hellings1983upper}. Substituting Eq.~\ref{HD_func} in Eq.~\ref{SGWB_PTA} we get
\begin{equation}
 \langle z_a(t)z_b(t)\rangle=C(\theta_{ab})\int_{0}^{\infty}df S_h(f). 
 \label{SGWB_PTA2}
\end{equation}
Finally, it is useful to write the results in terms of the timing residual as they are directly related to the time of arrival of the pulses. The ensemble average between the timing residuals of a pair of millisecond pulsars can be determined as
\begin{equation}
 \langle R_a(t)R_b(t)\rangle=2C(\theta_{ab})\int_{0}^{\infty}df \frac{S_h(f)}{(2\pi f)^2}[1-\cos{(2\pi ft)}]. 
 \label{SGWB_PTA3}
\end{equation}




Next, we would also like to comment briefly on the different kinds of noises that can affect the detection of GW signals by PTA and how they can be reduced. PTAs are subjected to the noises that can be generated due to the intrinsic timing irregularities in the spindown of the different pulsars or can be instrumental in origin like radiometer noise \textit{etc.} Monitoring a large number of pulsar pairs provides us with the possibility of enhancing the GW signal concerning the noise. 

The main international collaborations working on the detection of GW using PTAs are the European Pulsar Timing Array (EPTA) \cite{EPTA:2023sfo}, the North American Nanohertz Observatory for Gravitational Waves (NANOGrav) \cite{NANOGrav:2023hde}, the Indian Pulsar Timing Array (InPTA) \cite{Tarafdar:2022toa} and the Parkes Pulsar Timing Array (PPTA) \cite{Zic:2023gta}. All these collaborations work together as the International Pulsar Timing Array (IPTA). The EPTA uses five 100 m class telescopes across Europe and monitors 41 pulsars at several frequencies. Using these five telescopes as a single telescope gives the highest overall sampling rate of pulsars among the existing PTAs. NANOGrav is a combined effort of US and Canada which uses single-dish telescopes that are among the world's largest single-dish telescopes. NANOGrav currently monitors 49 pulsars. InPTA is an Indo-Japanese collaboration that makes use of the unique capabilities of the upgraded Giant Meterwave Radio Telescope (uGMRT) for monitoring a sample of nearby millisecond pulsars. PPTA uses the Parkes Observatory in Australia and can access pulsars in the Southern Hemisphere. It currently monitors 25 pulsars. Finally, the square kilometer array (SKA) \cite{Weltman:2018zrl} is an international radio telescope project that is being built in Australia (low-frequency) and South Africa (mid-frequency). It is expected to begin in 2028–29. Once ready, the SKA telescopes will be, by far, the most powerful instrument in the field of radio astronomy. Sensitivities for current and projected experiments we show in Fig. \ref{fig: GW_bounds}

\subsubsection{Astrometry}

 The behaviour of light passing through a tensor perturbation is different to light passing through a scalar metric perturbation. The tensor perturbation is moving and the waves are transverse. The correct proceedure to recover the behaviour of light through a stochastic gravitational wave background is to make use of a geodesic equation. When one does so, as we will briefly show, one does not find that light undergoes a random walk such that the deviation grows with distance. Instead, the deviation is approximately independent of distance and depends only upon the strain sensitivity \cite{Kaiser:1996wk}

This remarkable fact allows us to search for correlated shifts in the angular velocity of many stars in order to reconstruct a gravitational wave background. The sensitivity of this approach grows linearly with time rather than the usual $1/\sqrt{T}$ convergence as the precision in measuring angular velocity kicks grows with time.

In this section, we will briefly explain these two facts before mentioning the reach of current and future astrometry experiments. For a more in-depth look, the reader is directed to refs.~\cite{Kaiser:1996wk,Caliskan:2023cqm,Moore:2017ity,Klioner:2017asb,Wang:2020pmf,Wang:2022sxn}. Let us begin with demonstrating the independence of the kicks with the distance from Earth. We begin with the geodesic equation in a gravitational wave background,
\begin{equation}
    \ddot{d}(z) = \dot{H}(z) 
\end{equation}
where under the assumption of isotropy, we have
\begin{equation}
    H(z) \sim \int k^2 dk h_a (k) T_a (k) e^{i(1-\mu) kz} + c.c.
\end{equation}
and
\begin{equation}
    T_+ = - \frac{1}{2} (1- \mu) \sqrt{1-\mu ^2)} (\cos \phi , \sin \phi ) , \ T_x = -\frac{1}{2} \sqrt{1- \mu ^2} (\sin \phi , \cos \phi ) \ .
\end{equation}
We can then write a power spectrum of velocity kicks stars will have due to the gravitational wave background, as perceived by an observer on Earth, 
\begin{eqnarray}
P_{\dot{d}}(k) & \sim & \langle H(0) \cdot H(z) \rangle \\
&\sim & \int dz e^{i k z} \int q^2 dq P(q) (T_+^2 + T_x^2)e^{i (1-\mu) k z} \\
&\sim & k \int dp P(p) (1- \cdots) \ ,
\end{eqnarray}
where we have expanded to leading order in $k$ in order to eventually obtain the $k$ dependence of the power spectrum.
In the above, $P(p)$ is the power spectrum for tensor perturbations
\begin{eqnarray}
    \langle h_a (k) h_b^\ast (p) \rangle = (2 \pi )^3 \delta (k-p) \delta_{ab} P(k) . 
\end{eqnarray}
The power spectrum of displacements will be related to the spectrum for kicks by $P_d \sim P_{\dot{d}}/k^2$ and we get the scaling
\begin{equation}
    P_d \sim \frac{1}{k} \frac{\langle h^2\rangle }{k_h^2 } \ .
\end{equation}
The mean squared deviation averaged over a distance $D$ scales as
\begin{eqnarray}
    \langle d^2 \rangle \sim \left. (k P_d) \right |_{k = 1/D} \sim D^0 \ .
\end{eqnarray}
In other words, the size of the kick is independent of the distance between the source and the observer. We should therefore see a kick of the same size from a stochastic gravitational wave background which grows with the size of the strain. Looking at a large number of stars and searching for such correlated kicks therefore becomes an efficient method for finding a SGWB.

As for the impressive scaling of astrometry, to reconstruct the background, we need to track $N$ sources, for a time $T$, resolving angles with a precision of $\Delta \Theta $ and therefore the angular velocity to a precision $\Delta \Theta / (T \sqrt{N}) $ \cite{Book:2010pf}. Our precision with the gravitational wave background then scales (remembering that $\Omega \sim h^2$)
\begin{equation}
    \Omega_{\rm GW} \lesssim \frac{\Delta \Theta ^2}{N T^2 H_0^2} \ .
\end{equation}
Thus we see the powerful potential of using astrometry for gravitational wave detection - the sensitivity to a background scales much more efficiently with the lifetime of the experiment than other detectors.

Launched in 2013 by the European Space Agency, Gaia is a global space astrometry mission that maps more than a billion objects in the Milky Way. Detectors like Gaia~\cite{Gaia:2018ydn} use astrometric measurement to look for the low-frequency gravitational wave background \cite{Book:2010pf,Moore:2017ity,Mihaylov:2018uqm,Mihaylov:2019lft,Garcia-Bellido:2021zgu}. Recently, it was also shown that Gaia can be an important tool for probing GW with frequencies in range $\sim [10^{-9}-10^{-7}]$ Hz where the upper-frequency cut is set by the cadence and the lower frequency by the inverse lifetime of the experiment \cite{Moore:2017ity,Mihaylov:2018uqm,Mihaylov:2019lft,Garcia-Bellido:2021zgu}. Gaia is expected to observe a billion stars at an angular velocity resolution of about $100 \mu$as resulting in a strain sensitivity of 
\begin{equation}
h_{\rm GW} \sim 10^{-14} \left( \frac{5 {\rm years}}{ T_M }  \right)    \ ,
\end{equation}
with $T_M$ the mission lifetime. This makes it currently competitive with pulsar timing array measurements.
There is a proposed upgrade to Gaia called Theia \cite{Theia:2017xtk} which is expected to observe significantly more stars at a much higher angular resolution, which results in a strain sensitivity of
\begin{equation}
h_{\rm GW} \sim 10^{-14} \left( \frac{5 {\rm years}}{ T_M }   \right) \left( \frac{\Delta \Theta _{\rm Theia}}{\Delta \Theta _{\rm Gaia}} \right) \sqrt{ \left( \frac{N_{\rm stars, \ Gaia}}{N_{\rm stars, \ Theia}} \right) }   \ ,
\end{equation}
where $(\Delta \Theta _x,N_x)$ are the angular resolution and number of stars observed in experiment x. Exactly how much Theia will improve from Gaia is not yet clear, and recent estimates range from a factor of $100-600$ improvement in the strain sensitivity \cite{Garcia-Bellido:2021zgu,Caliskan:2023cqm}. A more exotic proposal aims to look for correlated kicks in asteroids in asteroid belt \cite{Fedderke:2021kuy} which is projected to achieve an impressive strain sensitivity of $10^{-19}$ at frequencies of $10 \mu$Hz. The sensitivity of Gaia as well as an approximate projection for Theia is given in Fig. \ref{fig: GW_bounds}. 

We end this section by briefly mentioning another exotic idea to indirectly measure a stochastic gravitational wave background indirectly by observing their effect on astrophysical objects. Absent perturbations, binary systems follow elliptical orbits obeying Kepler's laws. A stochastic gravitational wave background can perturb the system, giving and searches for deviations in the orbits of binary systems can be a window into gravitational wave backgrounds at the $\mu$Hz range \cite{Blas:2021mqw,Blas:2021mpc}.
\subsubsection{Interferometers}\label{sec:interfer}

\begin{figure}[tbh!]
    \centering
    \includegraphics[scale=0.5]{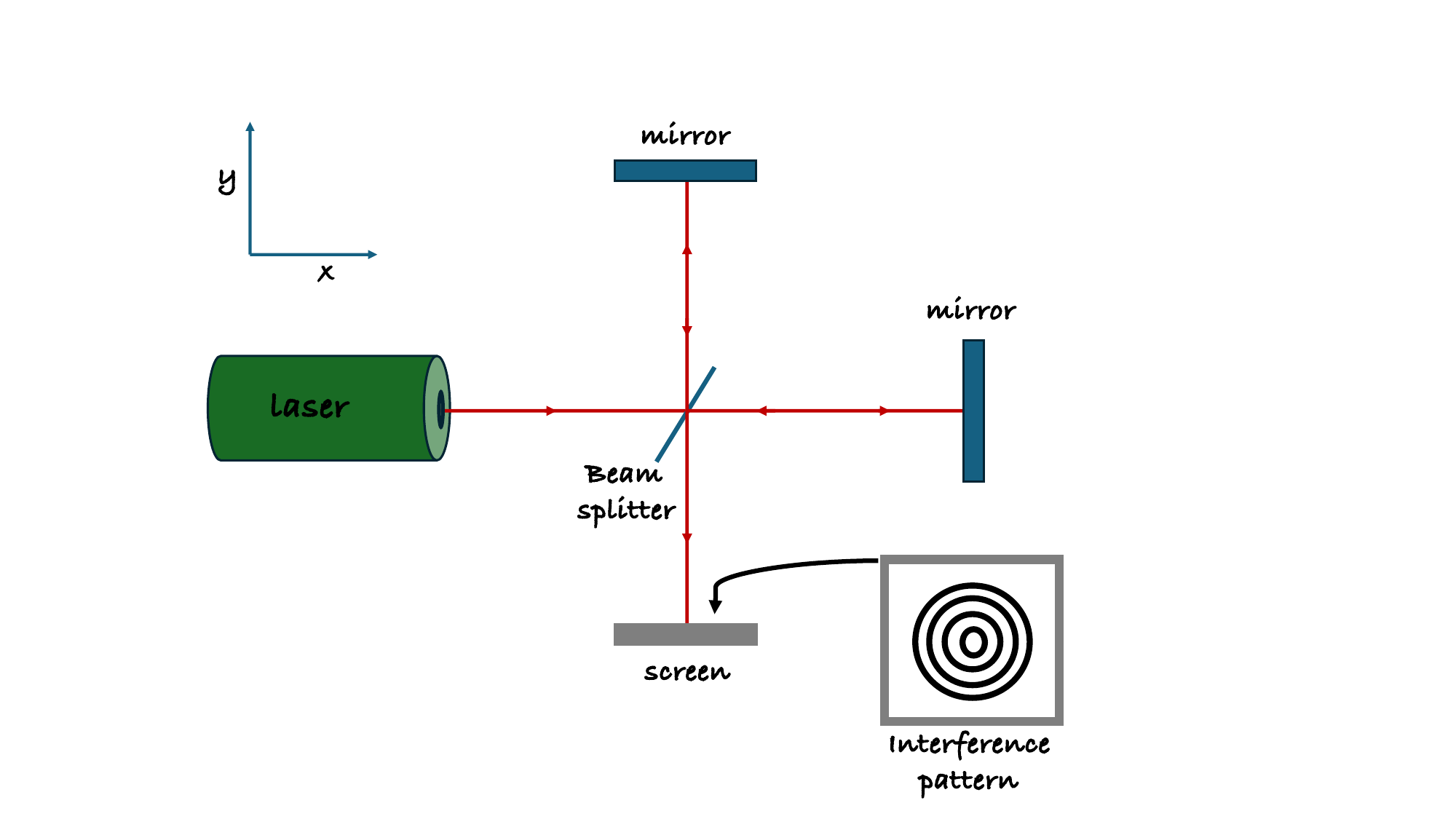}
    \caption{Schematic representation of a Michelson-type interferometer.}
    \label{schematic_1}
\end{figure}

In 1887, Michelson-Morley for the first time used an interferometer to demonstrate the non-existence of ether in the Universe \cite{michelson1881art}. As interferometers are very sensitive to even tiny changes in a path difference, they are very useful in measuring small fluctuations in spacetime. In Fig.~\ref{schematic_1}, we show a simple schematic of a Michelson-type interferometer.  

In order to understand how the presence of GW affects the propagation of light in the interferometer \cite{Ricci:1997cx,Aufmuth:2005sv,Maggiore:2019uih}, we first start with an assumption that GW has only a plus polarization and it arrives from the $z$ direction, so we have,
\begin{equation}
    h_+(t) = h_0 \cos \omega _{\rm GW} t.
\end{equation}
Since the photons travel along a null geodesic $i.e.~ds=0$ this allows us to solve the change in the path length along a single direction in TT frame \footnote{A similar calculation can also be performed in the proper frame. The computation
of higher-order corrections becomes much more involved in the proper frame of the detector and hence for simplicity, we stick to the TT frame.}
\begin{equation}
    dx = \pm c dt (1- \frac{1}{2} h_+(t))
 \end{equation}
 reflection and transmissions at the beam splitter give an overall phase shift of a $1/2$. Let us consider a Michelson interferometer with arm length L where the beam splitter is at the origin at time $t_0$ with a phase $e^{-i \omega _L t_0}$. The phase shift from the gravitational wave due to the field going through the x and y arms are 
 \begin{equation}
     \Delta \phi _x = -\Delta \phi _y = h_0 \frac{\omega _L L}{c} {\rm sinc} (\omega _{\rm GW} L/c) \cos [\omega _{\rm GW} (t- L/c) ] \ . 
     \label{phase_shift_MI}
 \end{equation}
The total power at the detector is related to a phase, $\phi _0$, set by the experimenter 
\begin{equation}
    P = \frac{P_0}{2} \left\{ 1- \cos[2 \phi_0 + \Delta \phi _{\rm Mich}(t)] \right\} \ .
\end{equation} 
For a given signal that peaks at a a frequency $f_p=\omega _{\rm GW}/2\pi $, the boost in power is maximal when
\begin{equation}
    L= 750 {\rm km} \left( \frac{100 {\rm Hz}}{f_p} \right)
\end{equation}
Ground-based detectors don't achieve an arm length 100s of kilometers long, Ligo and Virgo has arms of 4 and 3km respectively \cite{LIGOScientific:2014pky,LIGOScientific:2016emj,aLIGO:2020wna,Tse:2019wcy,VIRGO:2014yos,Virgo:2019juy}. However, to achieve a larger {\it effective} path length they use a Fabry-Perot cavity \cite{LIGOScientific:2014pky,LIGOScientific:2016emj}. Such a cavity traps light within it for a long time. If a Fabry-Perot cavity is a few kilometers long and has light trapped within it for around 100 trips between the edges of the cavity, then the gravitational wave detector can be sensitive to frequencies as low as $1-100$ Hz. The phase shift in the Fabry-Perot cavity in the presence of the GW along the $x$ and $y$ axis is given by
\begin{equation}
    \Delta\phi_x=-\Delta\phi_y\simeq h_0 2\frac{\omega_L   L}{c}\frac{\mathcal{F}}{\pi}\frac{1}{[1+(4\pi f_p\tau_s)^2]^{1/2}},
    \label{phase_shift_FP}
\end{equation}
where $\mathcal{F}$ is known as the finesse of the cavity which is defined as the ratio of free spectral range ($\Delta\omega_L$) to the full-width half maximum and $\tau_s$ is the storage time of the cavity, $i.e.$ the average time spent by a photon in the cavity.
\begin{figure}
    \centering
    \includegraphics[scale=0.5]{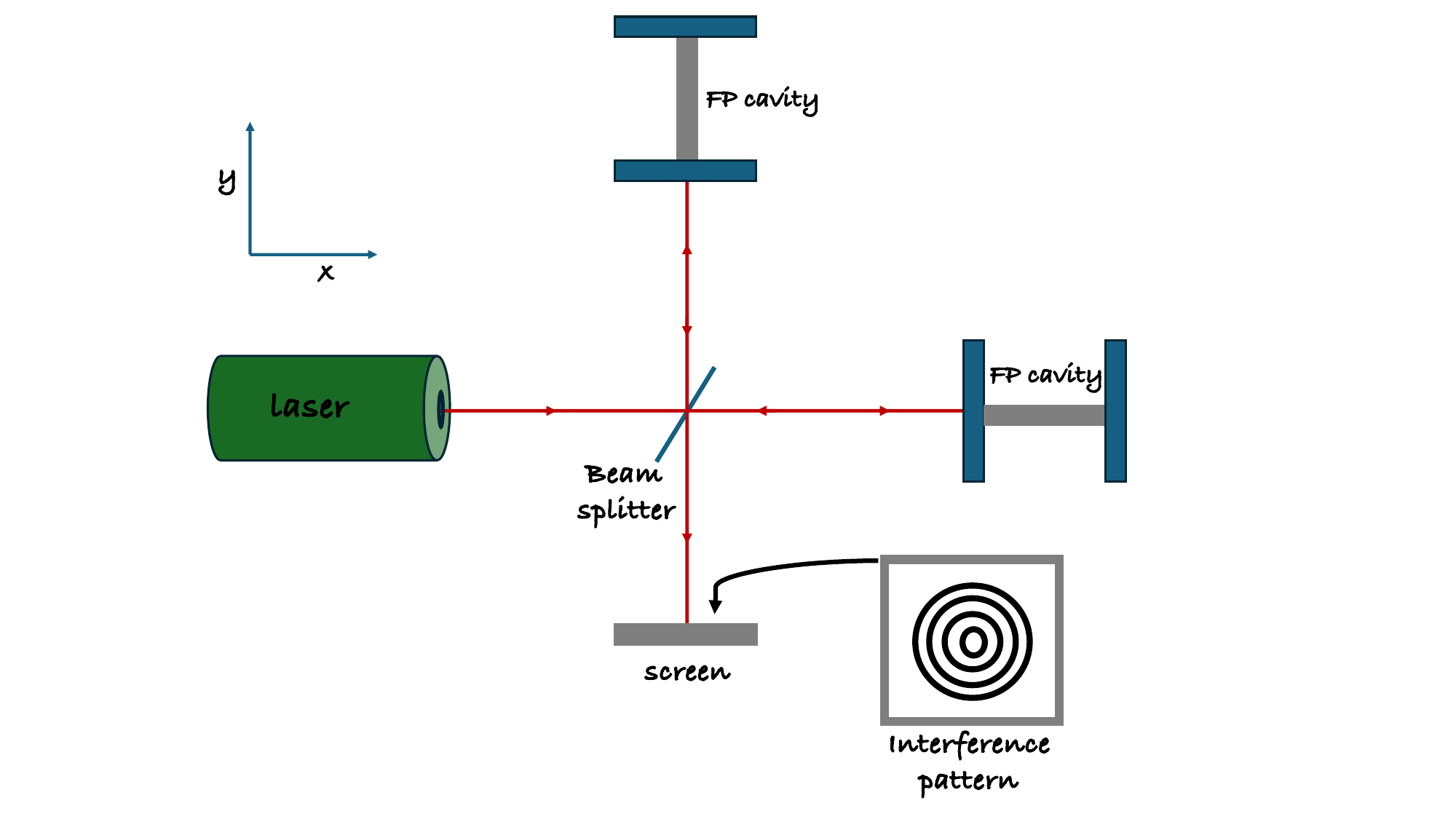}
    \caption{Schematic representation of an interferometer with Fabry-Perot cavities .}
    \label{schematic_2}
\end{figure}
Comparing Eq.~(\ref{phase_shift_FP}) with Eq.~(\ref{phase_shift_MI}) we find that in a Fabry-Perot, the sensitivity of a phase shift is enhanced by a factor of $(2/\pi)\mathcal{F}$ in comparison to what is obtained in a Michelson interferometer.

These interferometers are also subjected to various types of noises that can affect the detection of GW signals, for example, optical read-out noises which are a combination of the shot noises of a laser (that results from the discretization of photons) and radiation pressure (generated from pressure exerted by the laser beam on the mirrors), displacement noises which are generated by the movement of a test mass and are not induced by the GWs. Other noises that can influence the detection of GW signals are Seismic and Newtonian noise that results from local movements like traffic, trains, or surface waves that shake the suspension mechanism of the interferometer, thermal noises that are induced by the vibrations in mirrors and suspensions or the fluctuations in the test masses, etc. Apart from these sources, there exist several other sources of noise that can affect the detection of GW by an interferometer. Irrespective of the kind of noises, all of them can be characterized using their NSR which is further used to calculate the signal-to-noise ratios. For a detailed discussion, we refer the readers to~\cite{maggiore2007gravitational}.

With the first observation of the GW signals in 2015 by LIGO collaborations, the role and significance of interferometers in GW detection have increased manyfold. At present, there exist several earth-based interferometers that are currently running and are aiming to detect more signals. Among them, the most popular ones are at Hanford and Livington both located in the US with 4 km arms which are run by LIGO collaborations \cite{LIGOScientific:2014pky,LIGOScientific:2016emj,aLIGO:2020wna,Tse:2019wcy}, and the VIRGO interferometer ~\cite{VIRGO:2014yos,Virgo:2019juy} with arms of 3 km located near Pisa, Italy. KAGRA \cite{KAGRA:2018plz,Aso:2013eba} is another recently built underground laser interferometer with 3 km arm lengths located in Japan. Other smaller detectors like GEO600 (arms with 600 m) and TAMA (arms with 600 m) are situated in Hannover, Germany, and Tokyo, Japan respectively. Due to the longer arm lengths, better sensitivity can be achieved by detectors like LIGO, VIRGO, and KAGRA.  

Apart from these existing detectors, there have been proposals to build next-generation detectors with better sensitivity. Cosmic Explorer~\cite{Reitze:2019iox} is one of them which features two \emph{L-shaped} detectors one with arm lengths of 40 km and another with 20 km arm lengths. The Einstein Telescope~\cite{Maggiore:2019uih} is another proposed underground interferometer with arms of 10 km in length and it is expected to start its observations in 2035. Other than these ground-based detectors, there are several proposals for space-based detectors like LISA (Laser Interferometer Space Antenna)~\cite{LISA:2017pwj}, $\mu-$ARES~\cite{Sesana:2019vho}, TianQin \cite{TianQin:2015yph}, Taiji \cite{Hu:2017mde}, Voyger \cite{LIGO:2020xsf}, DECIGO ( DECi-hertz Interferometer Gravitational wave Observatory)~\cite{Kawamura:2019jqt}. These space-based interferometers have a few advantages over ground-based detectors. For starters, larger arm lengths can be achieved in space which in turn can help in increasing the sensitivity, and various noises that provide a hindrance for the earth-based detectors can easily be avoided. Space-based detectors are the future of GW detection as they will help us further in unfolding the mysteries of the early Universe. For the sensitivity of a selection of these detectors see Fig. \ref{fig: GW_bounds}
\subsection{Cosmological detectors}

In the era of precision cosmology, state-of-the-art analysis of the cosmic microwave background (CMB) and the abundance of light elements shed light on the gravitational wave background. First, both the production of light elements and the CMB are highly sensitive to the expansion rate of the Universe during each epoch (around 1 second and 1 million years respectively). If there is a gravitational wave background, the total radiation density is slightly higher than it would be otherwise. Precision fits to data for both the CMB and the abundances of light elements put a limit on the extra radiation allowed that is usually written in terms of the effective number of extra relativistic freedom,

\begin{equation}
    N_{\rm eff} \leq  \left\{  \begin{array}{cc}2.88 \pm 0.54 & \text{ BBN~\cite{Pitrou:2018cgg}}   \\ 3.00 \pm 0.34 & \text{ CMB ~(lensing+BAO) ~\cite{Planck:2018vyg}}   \\ 3.01 \pm 0.15 & \text{ (BBN+CMB) ~\cite{Pitrou:2018cgg}}   \end{array} \right. 
 \end{equation}
To make use of this bound, we need to convert the additional gravitational wave radiation from its stochastic background into an effective amount of relativistic degrees of freedom. After electron-positron annihilation at 0.5 MeV, the ratio of gravitational radiation to electromagnetic radiation becomes constant. This means for both BBN and CMB observables we can, to a good approximation, use
\begin{equation}
    \Delta N_{\rm eff} = \frac{8}{7} \left( \frac{11}{4} \right)^{\frac{4}{3}} \frac{\Omega _{\rm GW} ^0}{\Omega _\gamma ^0} 
\end{equation}
where the 0's indicate the values of their abundances today. The total gravitational radiation is of course
\begin{eqnarray}
    \Omega ^0 _{\rm GW} = \int \frac{df}{f} \Omega _{\rm GW}^0 (f) \ .
\end{eqnarray}

The second cosmological observable is temperature-polarization B-modes in the CMB, which are induced by gravitational waves (for a review see \cite{Kamionkowski:2015yta}). The CMB is sensitive to very low frequencies, around $10^{-17}$ Hz \cite{Lasky:2015lej}. The only known candidate to produce a source at such frequencies is inflation itself. We will cover higher frequency signals from gravitational waves due to a blue tilted spectrum in section \ref{sec:HFsources }

\subsection{High frequency proposals}


Many cosmological gravitational wave sources are strongly peaked at a frequency that approximately corresponds to the Hubble scale at the time the source was produced. We will get into specifics in later sections. For now, we wish to draw attention to the fact that the higher the frequency, the higher the energy scale a gravitational wave detector can probe. The maximum temperature the Universe can reach is at around the GUT scale, depending upon the details of preheating.\footnote{It depends upon the details of the model. If there is an instability scale, then finite temperature corrections can render the Universe unstable \cite{DelleRose:2015bpo}, alternatively, gravitational freezeout could clash with $\Delta N_{\rm eff}$ bounds for a modestly higher temperature \cite{Ringwald:2020ist}} This means that gravitational wave cosmology has an incredible scope to probe physics at scales we could only dream of on Earth. On the other hand, the challenge is enormous. 

As mentioned in the previous subsection, gravitational waves act as a source of dark radiation in the early Universe, mildly changing the expansion rate during Big Bang nucleosynthesis and recombination. The degree to which this expansion rate can change is restricted by observation. To shed light on any cosmological events before Big Bang nucleosynthesis, a gravitational wave detector must be more sensitive than limits on extra dark radiation. Since the abundance of gravitational radiation is related to the strain by $\Omega _{\rm GW} \sim h^2 f^2$, the target strain sensitivity needed to beat limits on dark radiation grows quadratically with the frequency. 
As such, no proposal currently exists that can uncontroversially probe gravitational waves at higher than around 1 kHz and beat the bounds on dark radiation. Even still, the opportunities presented by high-frequency gravitational wave physics have led to some ingenious proposals which are summarized in a living review in ref. \cite{Aggarwal:2020olq}. As yet, unfortunately there are no concrete proposal that can beat the bounds from $\Delta N_{\rm eff}$ on high frequency gravitational waves, so we do not cover any proposal in detail here. However, this is an active field to keep an eye on and we refer the reader to the above mentioned review for the current state of affairs.

\section{Gravitational waves from cosmic phase transitions}\label{sec:phase}
Gravitational waves from first-order cosmic phase transitions is a scenario  that has been the focus of intense attention for the last decade \cite{mazumdar2019review}. The concepts involved are in some ways quite familiar, everyone encounters phase transitions in their daily lives, most notably when water boils or freezes. A phase transition can occur when the ground state of a system is dependent on the temperature. In the case of boiling water, the high temperature phase is vapour which nucleates in a medium of the low temperature phase which is liquid. As the expanding Universe describes a system that is cooling, we will instead be considering cases where the Universe boils as it cools, and the bubbles contain the low temperature phase. As familiar as the concepts are, theoretically handling such an out-of-equilibrium quantum field theory has challenged the field for decades. In this section, we first make use of classical phase transitions to build an intuition for the subject before outlining the narrative of a cosmic phase transition, and the range of motivations for considering them and we review our current best understanding of how to model them.


\subsection{Nucleation of bubbles}

It is instructive to start with classical nucleation theory. Consider a bubble of a low-temperature phase nucleating in the background of the high-temperature phase. The Energy of the bubble of radius $R$ can be expressed in terms of a pressure term, $\Delta p$, that prompts expansion and a surface tension term, $\sigma$, that attempts to collapse the bubble \cite{abraham1974homogeneous}
\begin{equation}
    E= -\frac{4 \pi }{3} R^3 \Delta p + \sigma 4 \pi R^2 \ .
\end{equation}
 Such a bubble has a maximum in energy when the radius is at a critical value where the pressure and surface tension cancel each other out at $R=2\sigma/\Delta p$. Above this critical radius, the pressure begins to overwhelm the surface tension and the bubble expands. The nucleation rate is the exponential of the energy of a bubble with a critical radius divided by the temperature
 \begin{equation}
     \Gamma \sim e^{- \frac{16 \pi\sigma ^3}{3\Delta p ^2 T}} 
 \end{equation}
so we see that a large surface tension suppresses nucleation and is the main proxy for the strength of a transition. The pressure difference between the two phases will grow as the system cools, so the nucleation rate will increase at least as long as $\Delta p$ grows faster than the square root of temperature. In practice, $\Delta p$ can reach a maximum and it is possible for a phase transition to never complete.
\begin{figure}
    \centering
    \includegraphics[width=0.7\textwidth]{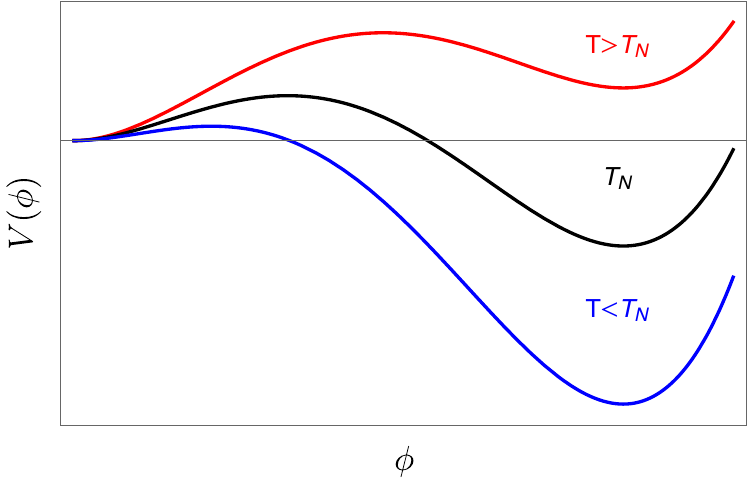}
    \caption{Schematic of an effective potential for some scalar field, $\phi$, evolving with temperature with a nucleation temperature denoted by $T$}
    \label{fig:potsketch}
\end{figure}
A cosmic phase transition has three additional ingredients to classical nucleation theory - a field-theoretic treatment, an expanding background, and quantum corrections. Let us begin by considering how field theory modifies the classical nucleation prescription. It is most common to consider phase transitions driven by the thermal evolution of the effective potential, sketched in Fig. \ref{fig:potsketch}, for a scalar field which we will denote as $\phi$. In such a case, we can derive the existence and behaviour of a bubble as a classical solution to the equations of motion. We simplify our lives by considering only spherical bubbles,  Wick rotating and assuming the thermal field theory is periodic in time with a period $1/T$. This allows us to make the approximation $\int dt \to \frac{1}{T}$, in which case the equations of motion are
\begin{equation}
    \frac{\partial ^2 \phi}{\partial r^2} + \frac{2}{r} \frac{\partial \phi}{\partial r} =\frac{\partial V}{\partial \phi } . 
\end{equation}
A solution to this is of course that the field occupies an extremum where the gradient of the potential vanishes. However, imagine the above equation with the $r$ coordinate replaced with time - it now describes the equations of motion of a ball on a hill of shape $-V$ with a strange time-dependent friction term. If we place the ball close to its maximum value, it will roll down towards the other local maximum. An initial condition that is too high will result in the ball rolling past the local maximum and continuing forever. If the initial condition is too low, the ball will roll around forever in the minimum connecting the two maxima. However, there is a Goldilocks solution where the ball starts in just the right place to land on the top of the local maximum. \par Returning our imagination to the field theory case with radial rather than time derivatives, the solution we have found is of a spherically symmetric object that is in the true vacuum at the center and becomes more like the false vacuum as we venture further out in radius. Or, in other words, it is a bubble of true vacuum nucleating in the medium of false vacuum \cite{coleman1977fate}. \par 
So let us connect with what we learnt in nucleation theory. First, we can estimate the nucleation probability with the path integral formalism. We can take a saddle point approximation to then estimate the path integral as the exponential of the action \cite{coleman1977fate,callan1977fate,linde1981fate} \footnote{It is interesting to point out that a phase transition may also proceed via bubbles seeded by impurities. This idea has been recently discussed in \cite{Blasi:2022woz,Agrawal:2023cgp} the context of the electroweak phase transition and and the corresponding implications for gravitational waves from sound waves have been investigated in \cite{Blasi:2023rqi}},
\begin{equation}
    \Gamma \sim e^{-S} \sim e^{-S_E/T}
\end{equation}
where in the second line we have again used the approximation that $\int _0 ^\beta dt \to \beta$, leaving a Euclidean action that involves only a spatial integral. If we approximate the bubble profile as having a very thin wall, it is possible to make a direct analogue of classical nucleation theory,
\begin{equation}
    S_E(\phi _{\rm bubble}) = \frac{16 \pi \sigma ^3}{\Delta p^2 }
\end{equation}
where $\Delta p = \Delta V$ and the surface tension can be calculated by integrating the potential from the false vacuum $v_1$ to the true one $v_2$,
\begin{equation}
    \sigma = {\rm Re}[\int _{v_1} ^{v_2}d\phi \sqrt{2 V}] \ .
\end{equation}
The thin wall approximation is only accurate very close to the critical temperature, however, we can nonetheless learn physical insight from the above. Recall that the strength of the transition is controlled by the surface tension. The surface tension grows with the height and width of the barrier separating the two phases. In the case of a scalar field theory, the surface tension will grow with the number of bosonic degrees of freedom that are made heavy by the change in phase as the change in mass of such Bosons. It is also possible in scalar field theories for there to be a large contribution to the surface tension from the tree-level shape of the potential if cubic terms are permitted by the symmetries of a theory. Specifically, one can have a field direction, $\phi$, where the tree level potential has the form $a \phi ^2 + b \phi ^3 + c \phi ^4$ and there is a tree level barrier between the minima. We thus expect these two paths to be the main avenues for producing a strong first-order phase transition - a large number of bosonic degrees of freedom gaining a large mass in the new phase and the symmetries of the theory allowing the potential to have a tree-level barrier. \par Confining transitions are much more difficult to theoretically model than transitions involving the vacuum expectation value of a scalar field. However, the methods we do have seem to find similar wisdom where large changes in mass \cite{Croon:2019iuh} of large degrees of freedom can lead to stronger transitions \cite{Lucini:2005vg,Datta:2010sq,Croon:2019iuh,Garcia-Bellido:2021zgu}. \par 
The last pieces of the picture needed to mimic the situation of the early Universe are the quantum effects and the expanding background. A modest quantum effect arises from the fluctuations around the bubble solution to the classical equations of motion. This results in a logarithmic correction to the effective action, and therefore a modification of the prefactor. A more dramatic effect is in the evolution of the effective potential - all thermal corrections are due to loops.\footnote{We acknowledge that the precise demarcation between what is truly quantum and what isn't can be subtle. For this review, we use the term quantum to refer to things that would vanish if $\hbar \to 0$ which, perhaps counterintuitively, is the case for all finite temperature corrections to a quantum field as all finite temperature terms are loop-induced} The fact that these loop corrections need to be large enough to qualitatively change the features of the tree level effective potential signals that we should worry about the efficacy of perturbation theory! This issue we will return to shortly. Finally, the expansion of the Universe means that the phase transition does not begin the moment it becomes energetically favourable for there to be a change in the ground state. Instead, we should compare the nucleation rate to the expansion of the Universe. When the nucleation rate is large enough that there is at least one critical size bubble per Hubble volume per Hubble time, the phase transition can be thought to begin.  However, if the nucleation rate does not further increase from this minimum rate, the phase transition could never be complete as the bubbles cannot catch up to the expanding volume around them. Because of this, the time scale of the phase transition, $\beta ^{-1}$\footnote{Unfortunately it is a confusing convention to have both the inverse time scale and the inverse temperature denoted by $\beta$. We will follow this convention in this review.} is controlled not by the nucleation rate, but its first (and occasionally second \cite{Athron:2023aqe,Athron:2023rfq}) derivative,
\begin{equation}
   \beta \sim \frac{1}{\Gamma} \frac{d\Gamma}{dt}\to \frac{\beta}{H} \sim T \frac{d(S_E/T)}{dT} \ .
\end{equation}

\subsection{Effective potential and action at finite temperature}
\begin{figure}
    \centering
    \includegraphics[width=0.7\textwidth]{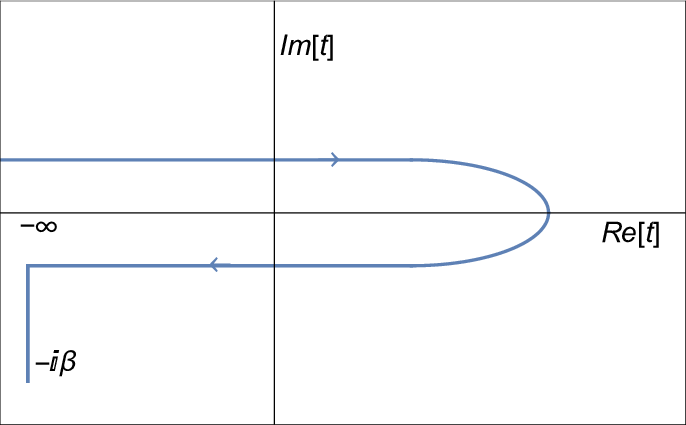}
    \caption{The contour used in the real-time formalism is known as the closed-time path.}
    \label{fig:CTP}
\end{figure}
There are in general two ways people tend to include temperature corrections to quantum field theory - the imaginary time formalism and the real time or closed time path formalism. The closed time path formalism assumes that the system was at once in equilibrium in the past. Starting with the finite temperature density matrix, one can then derive using the interaction picture of quantum mechanics that the expectation value of an operator is evaluated using the time contour shown in Fig. \ref{fig:CTP}. A consequence of this unusual time contour is that there are no longer just time-ordered and time-anti-ordered propagators, there are also mixed two-point correlators that involve on field moving in opposite directions in time. The propagators in equilibrium are just a zero temperature contribution (which vanishes in the case of the mixed correlators) summed with a finite temperature piece. This means that when one writes the corrections to the effective potential by calculating the 1-particle-irreducible loop diagrams, the result is a sum of the loop corrections at zero temperature and the finite temperature corrections. The closed time path formalism has the potential to handle departures from equilibrium, though this review is not aware of any attempt to take advantage of this in gravitational wave calculations. \par 
In equilibrium, an alternative formalism considers the propagators in imaginary time and in equilibrium. In this case, after Wick rotating, the thermal field theory becomes periodic in time. As a result, the $k_0$ modes become discrete, and are known as Matsubara modes and the $dk_0$ integrals are replaced by a sum over these modes. The result is not obviously equivalent to the sum of zero and finite temperature pieces that one derives in the closed time path formalism. However, as long as one approximates the background plasma as being in equilibrium\footnote{calculating the effective potential in a non-equilibrium background was recently done at zero temperature \cite{Garbrecht:2017idb}, though we are not aware of an attempt to do the equivalent at finite temperature}, then after some massaging, the result of both approaches are the same
\begin{equation}
    V_{T} = \sum _{i \in {\rm bosons}} n_i \frac{T^4}{2\pi ^2} J_B \left( \frac{m_i^2}{T^2} \right) - \sum _{i \in {\rm fermions}} n_i \frac{T^4}{2\pi ^2} J_F \left( \frac{m_i^2}{T^2} \right) \ ,
\end{equation}
where the thermal functions have the form
\begin{eqnarray}
    J_B(\bar{z}^2) &=&\int _0 ^\infty dx x^2 \log \left[ 1+ e^{-\sqrt{x^2 + \bar{z}^2}} \right] \sim \frac{\bar{z}^2}{24}-\frac{\bar{z}^3}{12 \pi} + \cdots \\
    J_F&=& \int _0 ^\infty dx x^2 \log \left[ 1- e^{-\sqrt{x^2 + \bar{z}^2}} \right] \sim -\frac{\bar{z}^2}{48} \  .
\end{eqnarray}
In the above, the right-hand approximations are the high-temperature expansions when $\bar{z}=m/T$ is small. \par 
We can now sketch how a first-order phase transition is generated. Consider a Bosonic field whose mass scales as $m_g=g \phi$ where $\phi$ is the scalar field undergoing a change in its vacuum expectation value during the phase transition. In such a case, the thermal corrections to the potential due to this field in the high-temperature limit is
\begin{equation}
    \Delta V_T = \frac{g^2 \phi^2T^2 }{24} - \frac{g^3 \phi ^3 T}{12 \pi } + \cdots
\end{equation}
The second term in the above equation produces a barrier as the full potential can have three consecutive terms with alternating signs. Let the tree-level potential be of the form
\begin{equation}
    V = \mu ^2 \phi ^2 + \lambda \phi ^4 \ .
\end{equation}
After rescaling $V\to \Lambda ^4 \tilde{V}$ and $\phi \to v \varphi$ where $v$ is the vacuum expectation value, then absorbing a redundant parameter into the definition of $\Lambda$, one can rewrite the potential and mass as\footnote{The specific redefinitions are $\mu^2 = \frac{1}{2v^2} \Lambda^4$ and $\lambda = \frac{\Lambda ^4}{4 v^4}$}
\begin{eqnarray}
    \tilde{V} &=& \Lambda ^4 \left(-\frac{1}{2}\left[ \frac{\varphi}{v} \right]^2 + \frac{1}{4}\left[ \frac{\varphi}{v} \right]^4  \right) \\  m_\phi &=& \sqrt{2}\frac{\Lambda ^2}{v}
    \\ m_g &=& g v  \ ,
\end{eqnarray}
where $m_\phi$ is the mass of the field $\phi$.
If we consider the temperature at which the potential is double-welled, that is the critical temperature, the surface tension scales as
\begin{equation}
    \sigma \approx \frac{ g m _g^5}{m_\phi ^2 }
\end{equation}
as long as $3 \pi ^2 m_\phi^2 >> g^2 m_g^2$. This limit is chosen to simplify our expression but is similar to the regime at which the high-temperature regime is valid. Regardless, we are only interested in a qualitative picture and indeed we see from the above that the strength of the phase transition grows with the ratio of the mass of the masses as we expected from our discussion in the previous section. Repeating the above for $n$ copies of $m_g$ finds that the strength of the transition grows with $n$. So we find a rule of thumb - {\it the strength of the transition grows with the number of bosonic degrees of freedom becoming nonrelativistic during the transition and the ratio of their masses to the mass of the dynamical scalar}. Remarkably this seems to be true of chiral transitions as well \cite{Croon:2019iuh}. From this rule of thumb, we can see why the Standard model is expected to have a crossover transition. The Higgs boson is in fact heavier than all gauge bosons that acquire a mass during electroweak symmetry breaking and the size of the gauge group, $SU(2)\times U(1)$, doesn't have enough gauge bosons to compensate for such a small ratio.\footnote{Strictly speaking, there is no surface tension for a crossover transition. However, analysing the surface tension does give the correct wisdom about how to make the transition stronger} \par 

\subsection{Possible sources}

If the reheating temperature reached at least the electroweak scale, then it is likely that our Universe experienced at least two changes in its ground state - an epoch of electroweak symmetry breaking at temperatures of around 100 GeV and an epoch where the quark-gluon plasma hadronizes at around 100 MeV. As mentioned above, both transitions are crossovers in the Standard Model at the low densities expected in the early Universe \cite{Kajantie:1995kf,Kajantie:1996mn,Kajantie:1996qd,Csikor:1998eu,DOnofrio:2015gop,Aoki:2006we,Bhattacharya:2014ara}. However, this picture can change if the physics beyond the Standard Model permits it, and these transitions could occur via the nucleation and percolation of bubbles of the low-temperature phase. Such a spectacular process of the early Universe boiling for an epoch naturally leads to a gravitational wave signal that could be visible today.

In the case of electroweak symmetry breaking, the most common ways to modify it to catalyze a strong first-order phase transition are
\begin{itemize}
    \item[1] Introduce more bosonic degrees of freedom that couple to the Higgs. The canonical examples are the two Higgs doublet model \cite{Dorsch:2013wja,Basler:2016obg,Dorsch:2016tab,Dorsch:2016nrg,Bernon:2017jgv,Dorsch:2017nza,Andersen:2017ika,Kainulainen:2019kyp,Wang:2019pet,Su:2020pjw,Davoudiasl:2021syn,Biekotter:2021ysx,Zhang:2021alu,Aoki:2021oez,Goncalves:2021egx,Phong:2022xpo,Biekotter:2022kgf,Anisha:2022hgv,Atkinson:2022pcn,Biekotter:2023eil,Goncalves:2023svb,Graf:2021xku,Arcadi:2022lpp} (including the inert doublet model \cite{Blinov:2015vma,Huang:2017rzf,Huang:2017rzf,Paul:2019pgt,Paul:2019pgt,Fabian:2020hny,Benincasa:2022elt,Paul:2022nzx,Jiang:2022btc,Astros:2023gda,Benincasa:2023vyp}) and the light stop mechanism \cite{Carena:1996wj,Espinosa:1996qw,Delepine:1996vn,Cline:1996cr,Losada:1996ju,Laine:1996ms,Bodeker:1996pc,deCarlos:1997tma,Cline:1998hy,Losada:1998at,Laine:2000rm,Carena:2008vj,Delgado:2012rk,Carena:2012np,Chung:2012vg,Huang:2012wn,Laine:2012jy,Liebler:2015ddv}. However, it is also possible with a scalar electroweak triplet \cite{Patel:2012pi,Huang:2017rzf,Chala:2018opy,Baum:2020vfl,Kazemi:2021bzj,Baldes:2021vyz,Azatov:2021irb}, and complex \cite{Jiang:2015cwa,Chiang:2017nmu,Ahriche:2018rao,Kannike:2019mzk,Chen:2019ebq,Cho:2021itv,Schicho:2022wty,DiBari:2023upq}, a real scalar singlet(s) \cite{Espinosa:1993bs,Profumo:2007wc,Cline:2009sn,Cline:2012hg,Cline:2013gha,Katz:2014bha,Profumo:2014opa,Vaskonen:2016yiu,Hashino:2016xoj,Chao:2017vrq,Matsui:2017ggm,Kang:2017mkl,Alves:2018jsw,Shajiee:2018jdq,Beniwal:2018hyi,Matsui:2018tpp,Kannike:2019mzk,Demidov:2021lyo,Cline:2021iff,Cao:2022ocg,Athron:2023xlk,Alanne:2020jwx} or a composite Higgs model \cite{Bian:2019kmg,Xie:2020bkl}. 
    \item[2] Introduce an effective tree-level barrier between the true and false vacuum. This means that the surface tension between the two phases may not vanish even at zero temperature. The two most common ways of achieving this  are through an effective cubic coupling from a gauge singlet, $s H^\dagger H$, or having a heavy degree of freedom integrated out such that the effective potential of the Higgs has alternating signs of the form \cite{Grojean:2004xa,Bodeker:2004ws,Delaunay:2007wb,Balazs:2016yvi,deVries:2017ncy,DeVries:2018aul,Chala:2018ari,Ellis:2019flb}
    \begin{equation}
        V=\mu^2 H^\dagger H - \lambda (H^\dagger H)^2 +\frac{1}{\Lambda^2} (H^\dagger H)^3 \ .
    \end{equation}
    Unfortunately, the Standard Model requires such a dramatic change to its effective potential to catalyze a first-order transition, that there is only one UV complete model that can be accurately characterized by the above potential - the real scalar singlet extension of the Standard Model \cite{Postma:2020toi}. This renders an effective field theory approach to be of questionable utility.
\end{itemize}
 A more exotic approach is to have a multi-step transition \cite{Weinberg:1974hy,Land:1992sm,Hammerschmitt:1994fn,Patel:2012pi,Patel:2013zla,Blinov:2015sna,Inoue:2015pza,Blinov:2015sna,Ramsey-Musolf:2017tgh,Croon:2018new,Angelescu:2018dkk,Niemi:2018asa,Morais:2018uou,Morais:2019fnm,Niemi:2020hto,Cao:2022ocg} where the direction the field goes during the phase transition is no longer from the origin.
Turning to the case of the QCD transition, it was first argued by Pisarski and
Wilczeck that one needs a minimum number of light flavours in order for chiral symmetry breaking to be strongly first order \cite{Pisarski:1983ms}, this has been demonstrated on the Lattice for $N_F\geq 3$ \cite{Iwasaki:1995ij}.  All models of the QCD transition seem to suggest that at large baryon chemical potential, the QCD transition becomes strongly first order \cite{Costa:2008gr,Marquez:2017uys,Vovchenko:2018eod,chen2019criticality,CamaraPereira:2020rtu,Gao:2020qsj,Gao:2020fbl,Gao:2021nwz,Gao:2023djs}. However, the lattice is yet to confirm the existence of a critical endpoint at which the baryon density is large enough to change the nature of the transition \cite{Bazavov:2017dus,Ding:2020rtq}. Finally, the pure Yang Mills theory is easier to model due to the absence of fermions. We know that pure Yang Mills always yields a first-order phase transition. We therefore have three possible methods for inducing a strong QCD phase transition
\begin{itemize}
    \item[1] Render the mass of the strange quark dynamical such that it is much lighter in the early Universe. This can be achieved through a flavon field which renders the effective coupling between the Higgs and strange quark to be proportional to the vacuum expectation value of the flavon field \cite{Davoudiasl:2019ugw}. Another option is to have Higgs field supercool until the QCD era, at which the QCD transition catalyzes both electroweak and chiral symmetry breaking \cite{Witten:1980ez,vonHarling:2017yew,Wong:2023qon,Li:2023yzq,Zhao:2022cnn}. 
    \item[2] In principle one could use flavon fields to render all quarks heavy in the early Universe. In such a case one would have a strong first order glueball transition. This option has not been analyzed in the literature, though glueball transitions have \cite{Halverson:2020xpg,Bigazzi:2020phm,Bigazzi:2020avc,Huang:2020crf,Wang:2020zlf,Kang:2021epo,Garcia-Bellido:2021zgu,Morgante:2022zvc}. We include it for completeness.
    \item[3] A third option is to have a large baryon chemical potential. The only known way to do this is to have a large lepton asymmetry \cite{Schwarz:2009ii,Caprini:2010xv,Middeldorf-Wygas:2020glx}.
\end{itemize}
 
\begin{figure}
    \centering
    \includegraphics[width=\textwidth]{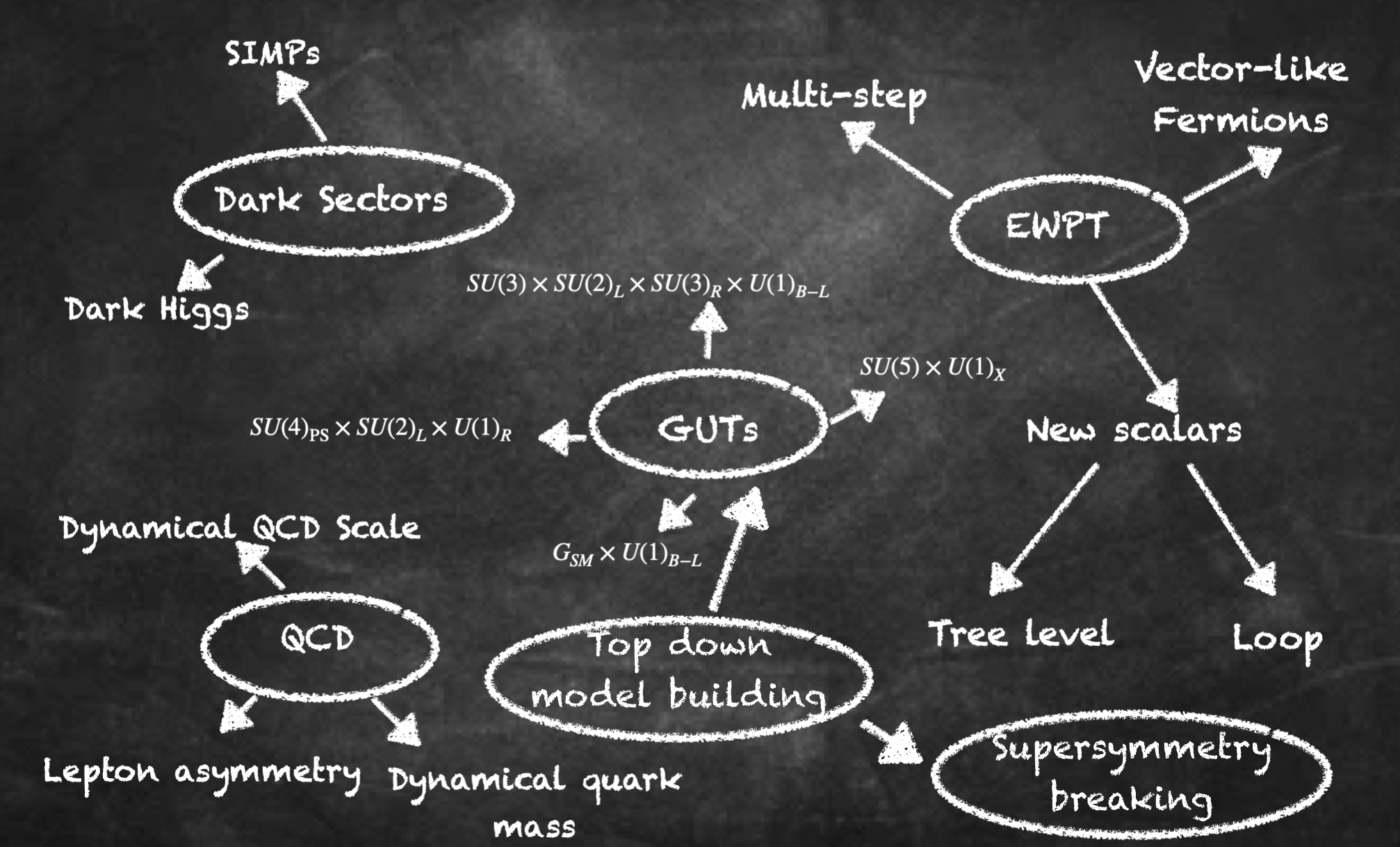}
    \caption{Current landscape of possible sources of gravitational waves from a cosmic first order phase transition}
    \label{fig:landscape}
\end{figure}

While the QCD and electroweak eras are arguably the most well-motivated epochs to search for (as we know that at least a change in ground state likely occurred) there are many more motivations for a strong first-order phase transition in the early Universe.  A popular model to realize a scale-invariant potential as described in section \ref{sec:col} is B-L breaking \cite{Jinno:2016knw,Chao:2017ilw,Brdar:2018num,Okada:2018xdh,Marzo:2018nov,Bian:2019szo,Hasegawa:2019amx,Ellis:2019oqb,Okada:2020vvb} (or B/L breaking \cite{Fornal:2020esl,Bosch:2023spa}). Additionally people have considered phase transitions in  neutrino mass models \cite{Li:2020eun,DiBari:2021dri,Zhou:2022mlz,Costa:2022lpy}, GUT symmetry breaking chains \cite{Hashino:2018zsi,Huang:2017laj,Croon:2018kqn,Brdar:2019fur,Huang:2020bbe,Graf:2021xku,Fornal:2023hri}, flavour physics \cite{Greljo:2019xan,Fornal:2020ngq}, supersymmetry breaking \cite{Fornal:2021ovz,Craig:2020jfv,Apreda:2001us,Bian:2017wfv}, hidden or dark sectors \cite{Schwaller:2015tja,Baldes:2018emh,Breitbach:2018ddu,Croon:2018erz,Hall:2019ank,Baldes:2017rcu,Croon:2019rqu,Hall:2019rld,Hall:2019ank,Chao:2020adk,Dent:2022bcd,Helmboldt:2019pan,Aoki:2019mlt,Helmboldt:2019pan,Croon:2019ugf,Croon:2019iuh,Alanne:2020jwx,Garcia-Bellido:2021zgu,Huang:2020crf,Halverson:2020xpg,Kang:2021epo,Fornal:2022qim,Costa:2022oaa,Benincasa:2023vyp} and axions \cite{Dev:2019njv,VonHarling:2019rgb,DelleRose:2019pgi}. The full landscape of ideas we show in Fig. \ref{fig:landscape}.
 
Most of the above involve phase transitions involving scalar fields. However, it takes different technology to model a confining transition. The typical approach in the literature is to model some condensate as a scalar field and consider its effective potential \cite{Croon:2019iuh,Helmboldt:2019pan}. In the case of the linear Sigma model, one writes an effective potential for the quark condensate where the parameters of the potential are free and the one-loop corrections are derived in exact analogue with a scalar field theory \cite{Croon:2019iuh}
\begin{equation}
    V(\Sigma ) = - m_\Sigma ^2 {\rm Tr} \left( \Sigma \Sigma ^\dagger \right) - (\mu _\Sigma {\rm det}\Sigma + h.c.) + \frac{\lambda}{2} \left[ {\rm Tr}(\Sigma \Sigma ^\dagger ) \right]^2 + \frac{\kappa}{2} {\rm Tr } \left( \Sigma \Sigma^\dagger \Sigma \Sigma ^\dagger \right)
\end{equation}
where $\Sigma _{ij} \sim \langle \bar{\psi} _{Rj}\psi _{Li} \rangle$ is the quark condensate that is decomposed as
\begin{equation}
    \Sigma _{ij} = \frac{\phi + i \eta ^\prime }{\sqrt{2 N_F}} \delta _{ij} + X^a T^a _{ij} + i \pi ^a T_{ij}^a
\end{equation}
and $\phi$ is what acquires a vacuum expectation value. The determinant term in the effective potential above arises from instanton effects.
In this model, it seems that the strength of the gravitational wave signal grows with the number of light flavours (at least going from 3 flavours to 4) and the hierarchy between the meson and the axion mass.

For the Nambu-Jona-Lasino model (NJL), again an effective potential is written for mesons but the mesons are non-propagating and the loop corrections are derived from considering quarks running in the loop \cite{Nambu:1961tp,Nambu:1961fr,Klevansky:1992qe,Holthausen:2013ota}
\begin{eqnarray}
    V_{0}^{\rm NJL} (\sigma) &=& \frac{3}{8G} \bar{\sigma ^2} - \frac{G_D}{16 G^3} \sigma^3 \\ 
    V_{\rm CW}^{\rm NJL} (\sigma) &=& - \frac{3n_c}{16 \pi ^2} \left[ \Lambda ^4 \log \left( 1+ \frac{M^2}{\Lambda^2} \right) - M^4 \log \left( 1+ \frac{\Lambda^2}{M^2}+ \Lambda^2 M^2 \right) \right] \\ 
    V_{\rm T}^{\rm NJL} (\sigma,T) &=& 6 n_c \frac{T^4}{\pi ^2} J_F(M^2/T^2) \\ 
    M&=& \bar{\sigma} - \frac{G_D}{8 G^2} \bar{\sigma ^2}
\end{eqnarray}
where $G$ and $G_D$ are free parameters. Unlike the linear sigma model.
The latter model can be enhanced by adding the Polyakov loop resulting in a gluon potential \cite{Fukushima:2017csk}. If the model in question has three colours, lattice data can be used to model the gluon contribution to the potential,
\begin{equation}
    {\cal L}_{\rm PNJL}^{\rm MFA} =     {\cal L}_{\rm NJL}^{\rm MFA} - V_{\rm Glue}(L,T) 
\end{equation}
\begin{equation}
    T^{-4} V_{\rm glue}(L,T) = - \frac{1}{2}a(T) L \bar{L} + b(T) \log \left[ 1-6 L \bar{L} - 3 (L \bar{L})^2 + 4 (L^3 + \bar{L}^3 \right]
\end{equation}
\begin{eqnarray}
    a(T) &=& 3.51 - 2.47 \frac{T_{\rm glue}}{T} + 15.2 \left( \frac{T_{\rm glue}}{T} \right)^2 \\
    b(T) &=& -1.75 \left( \frac{T_{\rm glue}}{T} \right)^3
\end{eqnarray}
In this case, the strength of the phase transition grows as $G$ goes near a critical point that depends on $G_D$ (or vice versa) where the nucleation rate becomes too slow compared to the expansion rate of the Universe.
In the case of Yang-Mills the surface tension and latent heat have been calculated on the lattice for $N_C\lesssim 8$ \cite{Lucini:2005vg,Datta:2010sq},
\begin{eqnarray}
    \sigma = (0.013 N_C^2 - 0.104)T_C^3, \quad L = \left( 0.549 + \frac{0.458}{N_C^2} \right) T_C^4
\end{eqnarray}
which in principle allows one to use critical nucleation theory to estimate the properties of the phase transition using an extrapolation for the pressure difference using the above result for the latent heat as was done in ref. \cite{Garcia-Bellido:2021zgu}. Alternatively, there exist attempts to model glueball transitions using lattice data to inform an ansatz for a potential for the Polyakov loop as was done in refs \cite{Huang:2020crf,Kang:2021epo,Halverson:2020xpg}.  Most methods tend to find that the more colours, the stronger the transition and that one needs more than 3 colours to be seen by any next-generation gravitational wave detector.

\subsection{Narrative of a phase transition and modeling of the spectrum} 

There are in principle three sources that lead to a gravitational wave signal that roughly correspond to the three sources of energy released when boiling a kettle. There is a source from the bubble nucleation itself \cite{Kosowsky:1991ua,Kosowsky:1992vn}, the scalar shells expand and crash into each other. A second source is an acoustic contribution, where sound shells collide \cite{Hindmarsh:2016lnk}. The third source is from an afterparty of turbulence \cite{Pen:2015qta}, where large eddy currents break down into smaller eddy currents. It is customary for each stage of the narrative to be described separately, though simulations may suggest that there may be no demarcation between the acoustic and turbulence sources \cite{Auclair:2022jod}. Usually, the acoustic source dominates, therefore so much of the theoretical modeling has been dedicated to understanding it. However, if the bubble walls become ultra-relativistic the collision term can dominate \cite{Ellis:2019oqb}. This has motivated a flurry of recent work focused on understanding this term as well as the kind of phase transition that can achieve such enormous Lorentz factors for the bubble walls. We will briefly cover the basic modeling techniques for each source before venturing into some of the current challenges in reproducing simulations.


\subsubsection{Acoustic source}

The acoustic source is widely believed to be the dominant source of gravitational waves and in the usual narrative is from the collision of sound shells before the onset of turbulence. In reality, it might be difficult to separate the turbulence and acoustic eras. The contribution to the stress-energy tensor from the acoustic source is controlled by the fluid velocity \cite{Hindmarsh:2016lnk,Hindmarsh:2017gnf,Guo:2020grp},
\begin{equation}
    T^{\mu \nu }\supset (e+p) U^\mu U^\nu + g^{\mu \nu} p
\end{equation}
where the 4-velocity of the fluid $U^\mu=\gamma (1, \vec{v})$ and the energy and momentum densities are given by
\begin{eqnarray}
    e &=& a _B T^4 + V - T \frac{\partial V}{\partial T} \\
    p &=& \frac{1}{3} a_B T^4 - V
\end{eqnarray}
\begin{equation}
    h^2 \Omega _{\rm GW } (f) = 1.2 \times 10^{-6} \left( \frac{100}{g_\ast (T_p)} \right) ^{1/3} \Gamma ^2 U_f^4 \left[ \frac{H_p}{\beta} \right] v_w S_{\rm sw} \Upsilon (y)
\end{equation}
where the spectral shape has the form
\begin{equation}
    S_{\rm SW} = \left( \frac{f}{f_{\rm sw}} \right)^{3} \left[ \frac{7}{4+3 (f/f_{\rm sw})^2} \right]^{7/2} \ .
\end{equation}
 Here, $f_{\rm sw}$ denotes the peak frequency, $U_f$ the RMS fluid velocity as described above, $v_w$ the wall velocity, $T_p$ the percolation temperature, $\Gamma$ the adiabatic index, $\Upsilon (y)$ a suppression factor that takes into account that the transition does not last a Hubble time, $g_\ast$ and $H_p$ the effective number of relativistic degrees of freedom and the value of Hubble at the percolation temperature.
\begin{figure}
    \centering
    \includegraphics[width=0.7\textwidth]{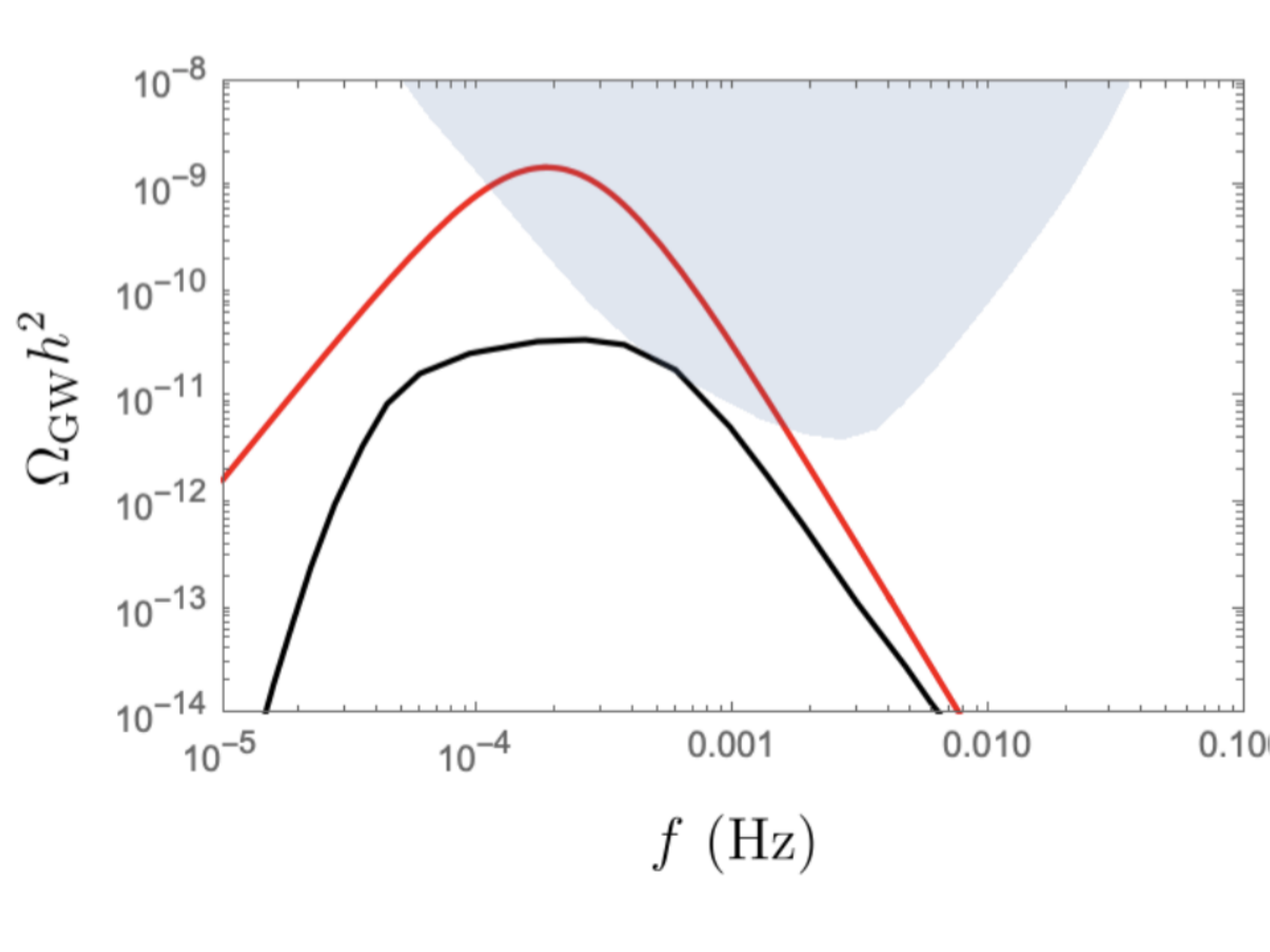}
    \caption{An example of how poor the simplest calculations of the sound wave source can be. In the above, we compare the sound shell model using the RMS fluid velocity (red/upper) with a numerical treatment that uses the whole velocity profile (black/lower). Note that the latter curve also the suppression factor from the energy lost to vorticity, likely due to the fact that some sound shells do not have enough time to reach a self-similar solution \cite{Cutting:2019zws}. Figure taken from \cite{White:2022ufa} and data taken from \cite{Gowling:2021gcy}}
    \label{fig:compsw}
\end{figure}
In reality, the finite width of the velocity profile results in there being two peaks to the spectrum corresponding to the two scales - the mean bubble separation and the width of the sound shell. So calculating the RMS of the fluid velocity rather than using the full velocity profile turns out to be quite a crude approximation\cite{Gowling:2021gcy} (see fig. \ref{fig:compsw} for a comparison). In this case, the gravitational wave spectrum has the form \cite{Hindmarsh:2019phv}
\begin{equation}
\Omega _{\rm GW} = F_{\rm GW,0} \Omega _p M(s,r_b,b)    
\end{equation}
 where $\Omega _p$ is the peak amplitude and $F_{\rm GW,0}$ is the dilution factor relating the original source to the source today.
The spectral shape now has two peaks
\begin{eqnarray}
    M(s,r_b,b  )= s^9 \left( \frac{1+r_b^4}{r_b^4+s^4} \right)^{(9-b)/4} \left( \frac{b+4}{b+4-m+m s^2} \right)^{(b+4)/2}
\end{eqnarray}
where $s=f/f_p$, $r_b$ is the ratio between the peaks, $b$ controls the slope and $m$ is chosen so that the peak is located at $s=1,M=1$ for $r_b<1$,
\begin{equation}
    m= (9 r_b^4 + b)/(r_b^4+1) \ .
\end{equation}
The parameters $(r_b,b)$ need to be obtained numerically from the numerical calculation of the gravitational wave spectrum.
Numerical simulations demonstrate that the sound shell model overestimates the gravitational wave spectrum as some sound shells do not have enough time to reach a self-similar solution before colliding \cite{Cutting:2019zws}. This suppression becomes large when the trace anomaly is large and the wall velocity is small. Other recent improvements come from allowing the speed of sound to vary from its equilibrium value \cite{Giese:2020rtr,Tenkanen:2022tly} and considering how the nucleation rate is affected by the reheating of the plasma \cite{Jinno:2021ury}. 
\subsubsection{Collision source}\label{sec:col}

This source captures the collision of the scalar shells and was the original focus of early research into gravitational waves from phase transitions \cite{Kosowsky:1991ua,Kosowsky:1992vn}. It arises from the field component of the stress-energy tensor,
\begin{equation}
    T_{\mu \nu}= \left( \frac{\partial {\cal L}}{ \partial ^\mu \phi} \partial _\nu \phi + c.c. \right)+\cdots \ .
\end{equation}
The fraction of energy that becomes dumped into the collision term is severely suppressed due to friction terms that are related to the Lorentz boost factor \cite{Bodeker:2017cim}. The exact dependence is a matter of ongoing debate \cite{Hoche:2020ysm,Ai:2021kak}, which we will return to shortly. For a linear dependence on the boost factor, the pressure difference between the two phases has to be an incredible trillion times larger than the radiation energy density for this source to dominate \cite{Ellis:2019oqb}. Surprisingly, this does not necessarily require fine-tuning. One option is to have a classical conformal symmetry such that the mass terms in the tree-level potential vanishes. In the case of scalar electrodynamics charged under a U(1) symmetry, the one-loop corrections have the approximate form under the high-temperature expansion \cite{Espinosa:2007qk,Espinosa:2008kw,Konstandin:2011dr,Konstandin:2011ds,Servant:2014bla,Fuyuto:2015jha,Sannino:2015wka,Lewicki:2020azd,Lewicki:2022pdb}
\begin{equation}
    V \sim \frac{ g^2}{2} T^2\phi ^2 + \frac{3 g^4 \phi ^4}{4 \pi ^2}\left(\log \left[ \frac{\phi ^2}{v^2} \right] - \frac{1}{2} \right) \ . 
\end{equation}
In reality, such use of a high-temperature expansion is a little scandalous if we are considering supercooling so severe that the collision term dominates. In fact, it is not clear how to accurately calculate the dynamics of the phase transition. We will later discuss methods of resummation to tame finite temperature field theory. The most successful method, dimensional reduction \cite{Ginsparg:1980ef,Appelquist:1981vg,Nadkarni:1982kb,Landsman:1989be,Farakos:1994kx,Braaten:1995cm,Braaten:1995jr,Kajantie:1995dw}, requires a hierarchy between the Matsubara modes and other masses of the theory \cite{Curtin:2022ovx}. Another method using gap equations is expected to be most useful precisely when the high-temperature expansion breaks down \cite{Dine:1992wr,Espinosa:1992gq,Boyd:1992xn,Espinosa:1992kf,Boyd:1993tz,Curtin:2016urg,Curtin:2022ovx}, however, the high-temperature expansion is breaking down so badly that the momentum dependence would need to be taken into account. This is not insurmountable, but whether gap equation methods are reliable in such an extreme regime is yet to be tried let alone tested. Regardless, taking the above potential at face value one finds that for modestly small values of the gauge coupling, $g\lesssim0.4$, one can achieve a large enough amount of supercooling that the collision term dominates \cite{Lewicki:2020azd}.

\subsubsection{Velocity of ultrarelativistic walls}
Let us briefly digress to discuss the current status of an ongoing debate on how one achieves an enormous Lorentz factor needed for the collision term to dominate. It was originally argued that a heuristic check of whether a bubble wall runs away was all that was needed \cite{Bodeker:2009qy}. Bodecker and Moore considered the $1\to 1$ interactions with the bubble wall, that is, interactions where a particle interacts with a bubble wall only through a change in its mass. For $\gamma >>1$, if the pressure on the bubble wall from $1\to 1$ processes is $V_{0-T}+V_{T}(\rm mean \ field)$. If this pressure is less than zero, the bubble wall will run away and the Lorentz factor will diverge. In a follow-up work \cite{Bodeker:2017cim}, they found that $1\to 2$ interactions with the bubble wall scale linearly with $\gamma$. If the scalar field undergoing a change in vacuum expectation value has any gauge charges, such a process will exist and the bubble will reach a terminal velocity \cite{Bodeker:2017cim}.
Recent work painted a more pessimistic picture, where interactions with the bubble wall involving multiple gauge bosons were shown to yield a pressure contribution quadratic in the Lorentz factor, $P\gamma ^2 T^4$ \cite{Hoche:2020ysm}.  By contrast, refs. \cite{Baldes:2020kam,Azatov:2020ufh,Gouttenoire:2021kjv,Azatov:2023xem} argued that the pressure scaled linearly with the Lorentz factor. This debate appears to still be in flux and a resolution is needed. 

A recently debated issue that {\it does} appear to be settled is the behaviour of the friction before one enters the relativistic regime \cite{Ai:2021kak,Ai:2023suz,Ai:2023see,Ai:2024shx}. Originally, it was claimed in Ref. \cite{BarrosoMancha:2020fay} that there is a friction near local thermal equilibrium that scales quadratically in the Lorentz factor. However, this was later shown to be due to an improper approximation in the plasma temperature and velocity used in Ref. \cite{BarrosoMancha:2020fay} such that such a scaling no longer appears when one includes a non-homogeneous temperature background \cite{Ai:2021kak}. Moreover, before the wall enters the ultrarelativistic regime, the friction does not grow monotonically as there is a peak dominantly caused by the local thermal equilibrium friction~\cite{Laurent:2022jrs,Ai:2023see}. One has to check that this peak is not strong enough to stop the further acceleration of the bubble wall~\cite{Ai:2024shx}.

A simple estimate of the gravitational wave spectrum arising from the collision term can be calculated using the envelope approximation \cite{Caprini:2007xq,Huber:2008hg,Caprini:2009fx,Jinno:2016vai}. Here, it is assumed that the bubble walls are infinitesimally thin, and just after the nucleation, the energy-momentum tensor not only vanishes everywhere in the space except on these walls but also in the sections of the wall that are contained inside other bubbles after the overlap. This leaves only the envelope surrounding the region of true vacuum.  

A spherically symmetric system does not change the gravitational field and hence can never act as a source of GWs. This suggests that a spherically symmetric expanding bubble can only generate GW after its collision with other bubbles which results in the breaking of spherical symmetry on the highly energetic walls of the bubble. Assuming that the bubble nucleation takes place exponentially and the thin-wall and envelope approximation remains valid, the GW spectrum that should be seen today can be written as \cite{Jinno:2016vai}

\begin{equation}
  \Omega_{\text{co}} (f)h^2=1.67\times10^{-5}\bigg(\frac{g_\star}{100}\bigg)^{-1/3}\bigg(\frac{\kappa_{\text{co}}\alpha}{1+\alpha}\bigg)^2\bigg(\frac{\beta}{H_\star}\bigg)^{-2}S_{\text{co}}(f)\Delta,
  \label{gw_collision}
\end{equation}
where the spectral form is given by
\begin{equation}
    S_{\text{co}}(f)=\bigg[0.064\bigg(\frac{f}{f_{\text{co}}}\bigg)^{-3}+0.456\bigg(\frac{f}{f_{\text{co}}}\bigg)^{-1}+0.48\bigg(\frac{f}{f_{\text{co}}}\bigg)\bigg]^{-1}
\end{equation}
with the peak frequency
\begin{equation}
    f_{\text{co}}=1.65\times10^{-5}\text{Hz}\bigg(\frac{g_\star}{100}\bigg)^{1/6}\bigg(\frac{T_\star}{100~\text{GeV}}\bigg)\bigg(\frac{f_\star}{\beta}\bigg)\bigg(\frac{\beta}{H_\star}\bigg).
\end{equation}
Here, $f_\star/\beta$ denotes the peak frequency at creation in the Hubble units and may depend on the bubble wall velocity $v_w$ whereas $H_\star/\beta$ represents the nucleation rate relative to the Hubble rate. A more careful analysis was performed recently in ref \cite{Lewicki:2020azd} that included the contributions to the stress-energy tensor after the collision. In this case, the resulting spectrum still follows a broken power law. The envelope predicts an infrared spectrum scaling as $f^3$ and a UV spectrum scaling as $f^-1$ whereas this analysis found the power law of the infrared and UV spectrum could vary in an approximate range of $(f^{[1-3]},f^{[2-2.5]})$ respectively. Finally, $\kappa _{\rm co}$ is the fraction of energy stored in the wall and $\Delta$ captures the dependence of the amplitude upon the wall velocity,
\begin{equation}
    \Delta = \frac{0.48 v_w^3}{1+5.3 v_w^2+5 v_w^4} 
\end{equation}

\subsubsection{Turbulence}
The aftershock or the turbulence developed in the plasma following the bubble collisions also leads to the generation of the GWs. It is interesting to point out that while the turbulent motion can generate the GWs on its own, turbulent motion can also impact the sound wave source even if the GWs from turbulence are negligible. This is because turbulence transfers energy from bulk motion on large scales to small scales. The turbulence can be characterized by irregular eddy motions and is typically modeled by Kolmogorov's theory~\cite{kolmogorov1995turbulence,Kosowsky:2001xp,Caprini:2009yp}. The expansion and the collision of bubbles tend to stir the background plasma resulting in eddies that can be of the order of the bubble radius. Now, in order to develop a stationary, homogeneous, and isotropic turbulence, the plasma needs to be stirred continuously which is only possible if the stirring time is larger than the circulating time of the largest eddies. Early work on turbulence was able to obtain an analytic handle on the non-stationary case and consider a finite source \cite{Caprini:2009yp}. Beyond this, it is an enormous challenge to get an analytic handle on turbulence and so far we must rely largely on simulations, which we summarize in section \ref{sec:simulations}. Here we briefly summarize the gravitational wave spectrum from Kologorov turbulence with a finite time source as outlined in ref \cite{Caprini:2009yp}.

The spectrum is given by,
\begin{equation}
  \Omega_{\text{tur}} (f)h^2=3.35\times10^{-4}\bigg(\frac{g_\star}{100}\bigg)^{-1/3}\bigg(\frac{\kappa_{\text{tur}}\alpha}{1+\alpha}\bigg)^{3/2}\bigg(\frac{\beta}{H_\star}\bigg)^{-1}S_{\text{tur}}(f)v_w,
  \label{gw_tur}
\end{equation}
where $\kappa_{\text{tur}}$ denotes the efficiency of conversion of latent heat into the turbulent flows and the spectral shape of the turbulent contribution is given by,
\begin{equation}
    S_{\text{tur}}(f)=\bigg(\frac{f}{f_{tur}}\bigg)^3\bigg(\frac{1}{[1+(f/f_{tur})]^{11/3}(1+8\pi f/h_\star)}\bigg)^{7/2},
\end{equation}
with $h_\star$ being the Hubble rate at $T_\star$,

\begin{equation}
    h_\star=16.5\times10^{-6}\text{Hz}\bigg(\frac{g_\star}{100}\bigg)^{1/6}\bigg(\frac{T_\star}{100~\text{GeV}}\bigg),
\end{equation}
and $f_{\text{tur}}$ being the peak frequency and is almost three times larger than the one obtained from the sound wave,
\begin{equation}
    f_{\text{tur}}=27\times10^{-6}\text{Hz}\frac{1}{v_w}\bigg(\frac{g_\star}{100}\bigg)^{1/6}\bigg(\frac{T_\star}{100~\text{GeV}}\bigg)\bigg(\frac{\beta}{H_\star}\bigg).
\end{equation}

\subsection{Theoretical issues with thermal field theory}

\begin{figure}
    \centering
    \includegraphics[width=0.65\textwidth]{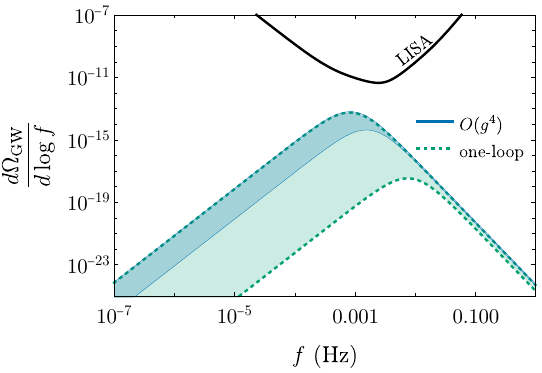} \caption{Theoretical uncertainties for a singlet extension to the Standard model are multiple orders of magnitude, with the more common one loop band not even capturing the range of values predicted at next to leading order. Figure taken from ref. \cite{Gould:2021oba}. Note that the uncertainties in this plot are purely from the thermal field theory and uses the RMS fluid velocity in predicting the sound shell source.}
    \label{fig:thunc}
\end{figure}

\begin{figure}
    \centering
    \includegraphics[width=0.8\textwidth]{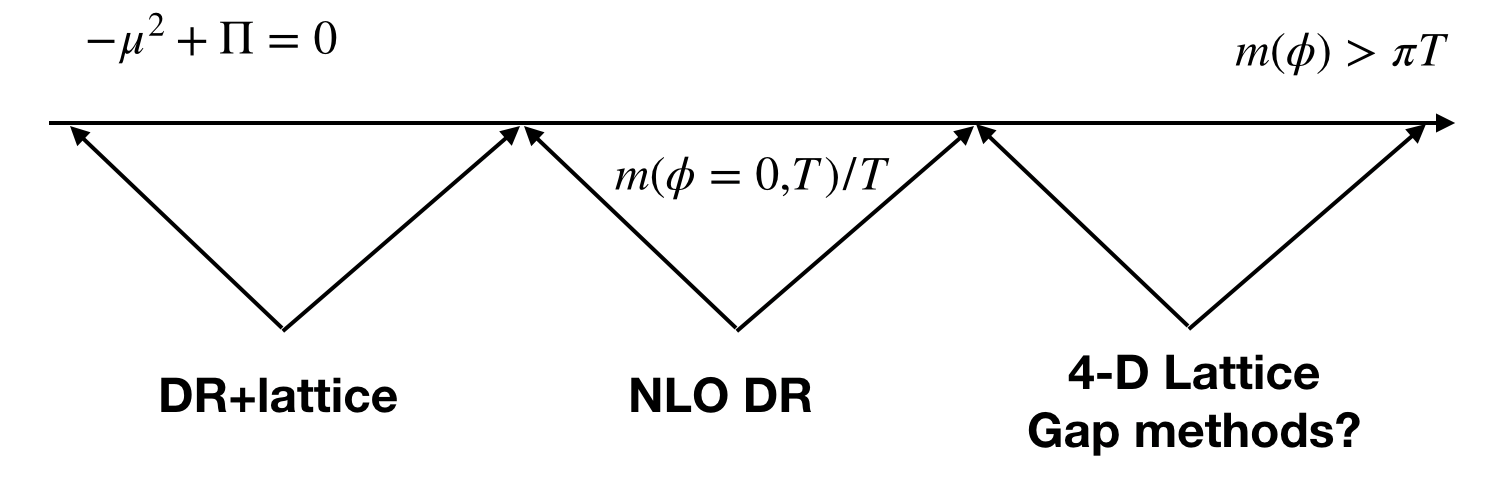}
    \caption{Perturbation theory is expected to only be valid in a narrow window where the masses are small enough that the high-temperature regime is valid but large enough to avoid infrared divergences near the critical temperature. For lighter masses, dimensional reduction allows the numerical simulation to be more tractable. For larger masses a large 4-dimensional simulation, although perhaps gap equation techniques show some early promise.}
    \label{fig:validityschematic}
\end{figure}

Achieving a phase transition requires that the thermal loop corrections to the tree level potential modify it drastically enough to substantially alter its shape and qualitative features. Having loop corrections being dominant is cause for alarm if we are to use perturbation theory. A second issue is that perturbation theory might actually diverge \cite{Linde:1980ts}. Consider the expansion parameter in the loop expansion for bosonic virtual states at zero temperature compared to its finite temperature counterpart
\begin{equation}
    \frac{g}{\pi^2} \to \frac{g}{\pi ^2} f\sim \frac{g}{\pi^2}\frac{T}{m} \ .
\end{equation}
Here $f$ is the bosonic distribution function and we have used the high-temperature expansion. Note that for some bosons the mass scales with the field value, $m\propto \phi$. However, if we are describing corrections to the potential we need to consider corrections when $\phi \to 0$ and $T/m$ diverge. A partial fix to this issue is to resume our theory to modify the masses of all particles running in loops by their loop correction,
\begin{equation}
    \frac{g}{\pi^2}\frac{T}{m} \to \frac{g}{\pi^2}\frac{T}{\sqrt{m^2 + \Pi}}
\end{equation}
where $\Pi$ is the thermal correction to the mass. The downside of this is that the mass of the dynamical scalar field is usually tachyonic at the origin. It is in principle possible for the thermal mass to cancel the tachyonic mass, again resulting in an infrared divergence. Indeed this is what occurs in a second-order transition which is why even sophisticated uses of perturbation theory badly estimate the critical mass at which the electroweak phase transition changes from crossover to weakly first order \cite{dine1992towards,Csikor:1998eu}. Fortunately, this does not appear to be an issue for phase transitions strong enough to detect \cite{Gould:2022ran}.
While Linde's infrared problem presents a catastrophic failure of perturbation theory at high orders, one can still use perturbation theory to understand the behaviour of strong first-order transitions (and recent work that Linde's infrared problem may not be as numerically important as previously thought \cite{Ekstedt:2022zro}. However, we still have an issue of loop corrections needing to dominate to ensure that temperature corrections qualitatively modify the nature of the effective potential. This calls for a reorganization of perturbation theory so that instead of a loop expansion, we ought to organize the theory in order of the size of each contribution. Consider the most simple scalar field theory with a potential
\begin{equation}
    V = \frac{m^2}{2} \phi^2 + \frac{g^2}{4!} \phi ^4
\end{equation}
The loop corrections, assuming the validity of the finite temperature expansion, are 
\begin{eqnarray}
    V_{\rm 0-T} &=& \frac{(m^2 + \frac{g^2}{2}\phi ^2)^2}{64 \pi ^2} \left( \log \left[\frac{(m^2 + \frac{g^2}{2}\phi ^2)}{\bar{\mu} ^2} - \frac{3}{2} \right] \right) \\ 
    V_{\rm 1-T} &=& \frac{g^2}{24}\phi ^2 T^2 - \frac{1}{12\pi}{(m^2 + \frac{g^2}{2}}\phi ^2)^{3/2} T \ ,
\end{eqnarray}
where $\bar{\mu}$ is the renormalization scale in the $\bar{MS}$ scheme.
A method of testing the size of higher-order corrections is to measure the scale dependence. If perturbation theory is working well, the implicit scale dependence from RG running should approximately cancel the explicit scale dependence in the loop correction, with any residual scale dependence indicating the approximate size of the neglected higher-order terms. The renormalization group equations for this minimal theory is
\begin{eqnarray}
    \bar{\mu} \frac{\partial m^2}{\partial \bar{\mu}} &=& \frac{g^2m^2}{16 \pi ^2}  \\ 
    \bar{\mu} \frac{\partial g^2}{\partial \bar{\mu}} &=& 3 \frac{g^4}{16 \pi ^2}
\end{eqnarray}
from which we can derive the scale dependence of each contribution to the potential to $O(g^4)$
\begin{eqnarray}
    \bar{\mu} \frac{\partial V}{\partial \bar{\mu} } &=& \frac{m^2 g^2}{32 \pi ^2} \phi ^2 +\frac{g^4}{128 \pi ^2}\phi ^4 \\
    \bar{\mu} \frac{\partial V_{\rm 0-T}}{\partial \bar{\mu}} &=& -\frac{m^4}{32 \pi ^2} - \frac{m^2 g^2}{32 \pi ^2} \phi ^2-\frac{g^4}{128 \pi ^2} \phi ^4 \\ 
    \bar{\mu} \frac{\partial V_{\rm 1-T}}{\partial \bar{\mu}} &=& \frac{g^4}{256 \pi ^2}T^2 \phi ^2 \ .
\end{eqnarray}
Apart from the running of a constant, field-independent term, the scale dependence from the temperature-independent pieces cancels to this order. On the other hand, there is nothing to cancel the temperature-dependent piece indicating that perturbation theory is not giving us the correct hierarchy of terms (from largest to smallest). Including the thermal mass correction in the loop partially solves our issue by yielding an additional piece
\begin{equation}
V_{\rm 0-T} =  \frac{(m^2 + \frac{g^2}{2}\phi ^2+\Pi)^2}{64 \pi ^2} \left( \log \left[\frac{(m^2 + \frac{g^2}{2}\phi ^2)+\Pi}{\bar{\mu}  ^2} - \frac{3}{2} \right] \right) \to \bar{\mu} \frac{\partial V_{\rm 0-T}}{\partial \bar{\mu} } \supset  -\frac{g^4}{48 \times 16 \pi ^2 }T^2 \phi ^2.
\end{equation}
So we partially solve our issue by resumming the mass. However, there is a 2-loop sunset diagram that has a term of the same order and completes the cancellation of the scale dependence of the thermal quadratic contribution to the effective potential
\begin{equation}
    V_{\rm sun}= -\frac{1}{12} g^4 \phi ^2 \frac{3 T^2}{32 \pi ^4}\left( \log \left[\frac{\bar{\mu}^2}{m^2 + \frac{g^2}{2}\phi ^2}\right]+2 \right) \left( \frac{\pi^2}{6} - \frac{\pi M}{2T} \right) \ . 
\end{equation}
Taking the derivative on the logarithm will yield a term $\sim - g^4T^2 \phi ^2/\pi^2$ which is precisely what we require to cancel the scale dependence. If only we were done. If we fish around in thermal field theory for all terms $O(g^4)$ we find more terms. Even in this unrealistically simple theory, it is quite cumbersome to collect all the terms needed to define the theory to accuracy $g^n$. Unlike zero temperature perturbation theory, a naive finite temperature loop expansions does not naturally organize perturbation theory into a series in terms of a small parameter $(g/\pi)^n$. 
Fortunately, there is a method to naturally organize perturbation theory. In imaginary time, the time domain is compact of size $1/T$. Thus the theory looks like a three-dimensional theory with a compactified extra dimension with the tower of heavy Kaluza Klein modes known as Matsubara modes. These modes are at the scale $\pi T$ so they can be integrated out as long as the high-temperature expansion is valid resulting in an effective three-dimensional theory \cite{Ginsparg:1980ef,Appelquist:1981vg,Nadkarni:1982kb,Landsman:1989be,Farakos:1994kx,Braaten:1995cm,Braaten:1995jr,Kajantie:1995dw}. Integrating out these heavy modes results in the mass terms being shifted by their thermal masses. The temporal modes of vector bosons in the effective theory have masses of the scale $O(g T)$ due to the thermal corrections to their masses (known as Debye masses). This, usually is large compared to the masses of the dynamical scalar as the Debye mass partially cancels against the tachyonic mass. This means that it might be appropriate to make an additional step in integrating the temporal modes of vector bosons. Further corrections to the dimensionally reduced theory arise from the matching relations between the 4-dimensional and 3-dimensional theories, which for the general field has the form
\begin{equation}
    \phi ^2 _{\rm 3D} = \frac{1}{T}(1 + \Pi _{\phi \phi }^\prime ) \phi ^2 _{\rm 4D} \ ,
\end{equation}
where $\Pi ' =\frac{d}{dk^2}\Pi$ is the derivative of the self-energy evaluated at zero momentum. These self-energies can be reabsorbed into the definition of all parameters in the theory. Unlike a loop expansion, dimensional reduction naturally organizes the theory into powers of the coupling $g_2$. For instance, in the standard model, we have
\begin{eqnarray}
V_3 &\supset & \frac{1}{2} \mu^2 _{h,3} h_3 ^2 + \frac{1}{4} \lambda _3 h_3^4 +\cdots \\    \mu ^2 _{h,3} &=& \mu _h^2 (\Pi _{hh}) - (\mu _h ^2 - \Pi _{hh})\Pi^\prime _{hh} \\
    \lambda _3 &=& T(\lambda - \frac{1}{2}\Gamma _{hhhh} - 2 \lambda \Pi ^\prime _{hh})
\end{eqnarray}
where $\Gamma _{hhhh}$ is the four-point function calculated in the full, four-dimensional theory. In the above, the self energies are themselves dependent on gauge couplings whose definitions have themselves been redefined by the matching relations. One can then use either one or two loop matching relations in order to write the theory to the desired order in $g_2$. Of course, even the Standard Model has many more parameters than $g_2$, so it is customary to consider each parameter as approximately equal to some power of $g_2$ when deciding which terms are small enough to leave out. 
The process of dimensional reduction for the case of the Standard Model is shown in table \ref{tab:dr:smeft}.
\begin{table}
\centering

\begin{tabular}{cccccc}
  \multicolumn{6}{l}{{\sl Start: {\bf $(d+1)$-dimensional SM}}} \\
  \hline
  \textbf{Scale} &
  \textbf{Validity} &
  \textbf{Dimension} &
  \textbf{Lagrangian} &
  \textbf{Fields} &
  \textbf{Parameters} \\
  \hline
  {\sl Hard} & $\pi T$ & $d+1$ &
  ${\cal L}_{\rm {SM}}$ &
  $A_{\mu},B_{\mu},C_{\mu},\phi,\psi^{ }_{i}$ &
  $m_h^{2},\lambda,
  g_1,g_2,g_s,y_t $ \\
  &&\multicolumn{4}{l}{$\Big\downarrow$ {\sl Integrate out $n\neq 0$ modes and fermions}} \\
  {\sl Soft} & $g T$ & $d$ &
  $\mathcal{L}_{3d}$ &
  $A_{r},B_{r},C_{r},$ &
  $\mu_{h,3}^{2},\lambda^{ }_{3},
  g^{ }_{3},m_D^{ },$\\
  &&&&
  $A^{ }_{0},B^{ }_{0},C^{ }_{0},\phi$ &
  $g_{1,{3}},m_D',g^{ }_{s,3},m_D''$
  \\
  &&\multicolumn{4}{l}{$\Big\downarrow$ {\sl Integrate out temporal adjoint scalars $A_{0},B_{0},C_{0}$}} \\
  {\sl Ultrasoft} & $g^{2}T/\pi$ & $d$ &
  $\bar{\mathcal{L}}_{3d}$ &
  $A_{r},B_{r},C_{r},\phi$ &
  $\bar{\mu}_{h,3}^{2}, \bar{\lambda}^{ }_{3},  \bar{g_2}^{ }_{3},\bar{g_1}_{3},\bar{g_{s,3}}$
  \\\hline
  \multicolumn{6}{l}{{\sl End: {\bf $d$-dimensional Pure Gauge}}} \\
\end{tabular}
\caption[Dimensional reduction]{
  Dimensional reduction of $(d+1)$-dimensional SM into
  effective $d$-dimensional theories based on the scale hierarchy at high temperatures.
  The effective couplings are functions of the couplings of their
  parent theories and temperature and are determined by a matching procedure.
  The first step integrates out
  all hard non-zero modes.
  The second step integrates out
  the temporal adjoint scalars $A^{ }_{0},B^{ }_{0},C^{ }_{0}$
  with soft Debye masses $m_D^{ },m_D',m_D''$.
  At the ultrasoft scale, only ultrasoft spatial gauge fields $A_{r},B_{r},C_{r}$
  (with corresponding field-strength tensors $G_{rs},F_{rs},H_{rs}$)
  remain along with a light Higgs that undergoes the phase transition. Table adapted from ref. \cite{Croon:2020cgk}}
\label{tab:dr:smeft}
\end{table}
The process is admittedly complicated. However, as shown in Fig. \ref{fig:thunc} the theoretical uncertainties of ordinary perturbation theory is multiple orders of magnitude 
\cite{Croon:2020cgk,Gould:2021oba}, even for simple extensions of the Standard Model. This can often be tamed if one calculates the dimensionally reduced theory at next to the leading order, as shown in Fig. \ref{fig:thunc}.
However, Dimensional reduction assumes a hierarchy of scales so that we can integrate out the Matsubara modes. However, for sufficiently strong phase transitions, this hierarchy disappears. For stronger phase transitions, it is plausible that techniques that work in QCD could be repurposed. In particular, numerically solving the Dyson Schwinger equations is a technique that has borne fruit in solving strongly coupled theories \cite{Gao:2015kea}. Such a technique has been recently developed for thermal field theory, so far ignoring the momentum dependence of the gap equations \cite{Curtin:2016urg,Curtin:2022ovx}.  A schematic of the current situation is shown in Fig \ref{fig:validityschematic} and, as we will see in the next section, simulations do seem to reproduce the above heuristic arguments.

\subsection{Simulations}\label{sec:simulations}
Cosmic phase transitions are such complicated systems, far from equilibrium, that it is necessary to check our analytic methods against simulations. Current numeric results seem to indicate that the theory community has a long way to go in both steps, understanding finite temperature perturbation theory and modeling the gravitational wave spectrum.
Recent Monte-Carlo analysis treatments of thermal field theory demonstrated that, in line with expectations, there is indeed a Golilocks zone where perturbation theory yields correct answers when the phase transition is strong enough that infrared catastrophes aren't numerically important near the critical temperature, yet not so strong that expansions in the coupling constant have slow convergence. However, they found convergence only when expanding in $g^2/\lambda$ rather than the traditional expansion in $g^2$, where $g$ is the SU(2) gauge coupling constant and $\lambda$ is the Higgs quartic \cite{Gould:2022ran}. 

Recent simulations of the sound shell model indicate that the sound shell model overestimates the gravitational wave spectrum due to energy lost to vorticity modes \cite{Cutting:2019zws}. Generally, this is a small suppression unless the velocity is small and the trace anomaly large. As for the shape, numerical treatment of the soundshell model does appear to agree fairly well with simulations \cite{Hindmarsh:2015qta,Hindmarsh:2017gnf}.

The gravitational wave spectrum from scalar shell collisions differs quite dramatically in shape, size, and peak frequency when one compares the predictions of lattice and the envelope approximation \cite{Child:2012qg,Cutting:2018tjt,Cutting:2020nla}. A significant issue is probably that the contribution to the stress-energy tensor from the bubble wall is ignored after the collision. There has been recent work to try and model this effect and it remains to be seen how much this closes the gap between simulations and theory \cite{Jinno:2017fby,Konstandin:2017sat,Jinno:2019jhi,Lewicki:2020azd,Jinno:2020eqg,Di:2020kbw}.

Of all sources of gravitational waves during a phase transition, our understanding of turbulence is perhaps the most remedial. Recent work considered the gravitational wave background produced a period of freely decay vortical turbulence \cite{Auclair:2022jod}. They were able to analytically fit the predicted gravitational wave spectrum, however this eventually needs to matched with a sufficient understanding of how first-order phase transitions source turbulence.

\section{Gravitational waves from topological defects}\label{sec:top}


Cosmic phase transitions are not the only source of gravitational waves that can be produced during an event of cosmological symmetry breaking. One possibility is that the vacuum manifold is sufficiently interesting to allow massive configurations of fields that are at least metastable, known as topological defects \cite{Kolb:1990vq,vilenkin2000cosmic,Maggiore:2018sht}. What makes topological defects a particularly compelling source of gravitational waves is that one does not need to live in a corner of the parameter space to produce such defects. In other words, these topological defects can be formed independently of the order of the phase transition and can generate GWs. This is in contrast to gravitational waves from phase transitions, which require a strong first-order phase transition. Second, and also in contrast to cosmic phase transitions, a gravitational wave signal from such defects can become easier to detect if the scale of physics involved is higher. This is assured directly in the case of cosmic strings as the waveform is relatively flat over many decades of frequency with an amplitude that grows with the symmetry-breaking scale. In the case of domain walls, there is an interplay between the symmetry-breaking scale which determines the power of the gravitational waves emitted, and the lifetime of the wall. In this section, we will give a brief introduction to how to use topology to characterize topological defects, what we know about the gravitational wave signals in the case of each defect as well as hybrids before giving an overview of some recent applications. 
\subsection{Categorization by homotopy group}

During an epoch of cosmological spontaneous symmetry breaking, if the topology of the vacuum manifold is non-trivial, the transition leaves traces known as topological defects. Depending on the nature of the symmetry under consideration, these defects can be global or local and the phenomenology differs in each case. The full set of possible global or local defects includes domain walls (DW), cosmic strings (CS), monopoles, and textures. To describe the topology of the vacuum manifold, let us use the notation $\mathcal{M}=\mathcal{G}/\mathcal{H}$, where $\mathcal{G}$ and $\mathcal{H}$ are, respectively, the symmetry groups before and after the symmetry breaking. The topological properties of the manifold, $\mathcal{M}$, can be diagnosed using homotopy theory. The $n^{\text{th}}$ homotopy group $\Pi_n(\mathcal{M})$ classifies qualitatively distinct mapping from the $n-$ dimensional sphere $S^{n}$ into the manifold $\mathcal{M}$. Below we classify different topological defects based on their vacuum
manifold properties and homotopy group,

\begin{enumerate}
    \item Domain Walls: They are formed when the vacuum manifolds are disconnected, $i.e~ \Pi_0(\mathcal{M})\neq \mathcal{I}$, where $\mathcal{I}$ denotes a trivial homotopy group. In other words $\Pi_0(\mathcal{M})$ is a homotopy group that counts disconnected components. For example, let us consider a discrete symmetry $Z_2$ spontaneously broken to an $\mathcal{I}$. In this case, the vacuum manifold is given by $\mathcal{M}=Z_2/\mathcal{I}$ and $\Pi_0(\mathcal{M})=Z_2\neq\mathcal{I}$.
    
     \item Cosmic strings: CS are formed when the vacuum manifolds are not simply connected, $i.e~ \Pi_1(\mathcal{M})\neq \mathcal{I}$. Considering a $U(1)$ symmetry that is spontaneously broken to  
     $\mathcal{I}$, the vacuum manifold can then be written as $\mathcal{M}=U(1)/\mathcal{I}$ and $\Pi_1(\mathcal{M})=U(1)\neq\mathcal{I}$.
    \item Monopoles: These point like defects are formed when $\Pi_2(\mathcal{M})\neq \mathcal{I}~i.e.$ when the manifold of degenerate vacua contains non-contractible two-surfaces
    \item Textures: Textures appear in the scenarios where the vacuum manifold has a non-trivial third homotopy group, $\Pi_3(\mathcal{M})\neq\mathcal{I}$. For example, if a $O(4)$ is spontaneously broken to $O(3)$, the vacuum manifold is given by $\mathcal{M}=O(4)/O(3)$ and $\Pi_3(\mathcal{M})=O(4)/O(3)\neq\mathcal{I}$.
\end{enumerate}

\subsection{Gravitational waves from Domain walls} 
\label{subsec:DW}
A stochastic GW background can be sourced from different topological defects. As a first example, let us consider domain walls \cite{Kolb:1990vq,vilenkin2000cosmic,Saikawa:2017hiv}. These are sheet-like topological defects that form if a discrete symmetry is spontaneously broken in the early Universe. Once created, their energy density scales inversely with the scale factor. Therefore, they can soon dominate the total energy density of the Universe and later alter the products of BBN as well as the Cosmic microwave background (CMB).

Several solutions exist to overcome domain wall domination~\cite{vilenkin2000cosmic}.  For example, if the formation of DW takes place before the inflation, the walls can be inflated away far beyond the present Hubble radius. On the other hand, wall domination can also be avoided in a model where a discrete symmetry that is broken at a very scale is restored at a lower temperature. In this scenario, wall tension (discussed later) is temperature dependent and walls are always over-damped ($\rho_{\text{DW}}/\rho<<1$), hence they never dominate the Universe. Another way out is if the DWs are unstable \cite{vilenkin2000cosmic}. If domain walls arise from the spontaneous breaking of a global discrete symmetry, one can introduce a small bias to the scalar potential from an explicit symmetry-breaking term that allows the walls to annihilate. Finally, a wall-dominated universe can be avoided if the discrete symmetry responsible for the generation is embedded in a continuous symmetry group and a prior step in the symmetry-breaking chain admits cosmic strings. In such a case, the walls can be bounded by strings and these hybrid defects \cite{vilenkin2000cosmic,Dunsky:2021tih} can decay before they dominate the Universe.   

We initiate our discussion by briefly describing the dynamics of DW followed by how it generates the SGWB \cite{Saikawa:2017hiv,Huang:1985tt,Nakayama:2016gxi,Gelmini:2020bqg,Blasi:2022ayo,Blasi:2023sej,Bhattacharya:2023kws} that can be detected by the current and future GW detectors. To do that we consider a simplistic scenario of a real scalar field $\phi$ that remains invariant under a discrete $Z_2$ symmetry ($\phi\to-\phi$). The $Z_2$ symmetry is spontaneously broken, once the scalar obtain a non-zero vacuum expectation value (vev), $i.e.~\langle \phi\rangle=\pm v$. A simple example of such a potential is,
\begin{equation}
   V(\phi)=\frac{\lambda}{4}(\phi^2-v^2)^2, 
\end{equation}
\noindent where $\lambda$ is a quartic coupling. Notice, that the scalar has two degenerate minima at $\phi=\pm v$. As a result of this spontaneous symmetry breaking (SSB), two domains can appear and walls are produced at the boundary. Assuming a static planar doamin wall lying perpendicular to the $x-$axis in a Minkowski space, $\phi(x)$, the analytical solution\footnote{This solution is only valid when the thickness of wall is smaller than $H^{-1}/\sqrt{2}$ \cite{Dolgov:2016fnx}.} to the equation of motion for this scalar is,
\begin{equation}
    \phi(x)=v~\text{tanh}\bigg(\sqrt{\frac{\lambda}{2}}v x\bigg).
\end{equation}
Here a local \emph{kink} is observed at $x=0$, that takes $\phi$ from $-v$ at $x=-\infty$ to $+v$ at $x=+\infty$. The surface tension, $\sigma$, is controlled by the symmetry breaking scale,
\begin{equation}
   \sigma=\int_\infty^\infty dx\bigg(\frac{1}{2}\bigg[\frac{\partial\phi(x)}{dx}\bigg]^2+V(\phi(x))\bigg)=\sqrt{\frac{8\lambda}{9}}v^3. 
\end{equation}

Once the DWs are formed they can eventually dominate the energy budget of the Universe as their energy density dilutes much slower than radiation. Their dynamics can be described by considering them as a system of isotropic gas~\cite{vilenkin2000cosmic} with an equation of state,

\begin{equation}
 p_{\text{DW}}=\bigg(u^2-\frac{2}{3}\bigg)\rho_{\text{DW}},  
\end{equation}
where $p_{\text{DW}}$ and $\rho_{\text{DW}}$ denotes the pressure and energy density of the DWs while $u$ represents the wall's velocity. While in the case of highly relativistic walls, $u\to 1$, a usual equation of state for a relativistic gas is obtained $i.e.~ p_{\text{DW}}=\frac{1}{3}\rho_{\text{DW}}$, the non-relativistic regime ($u<<1$) gives $i.e.~ p_{\text{DW}}=-\frac{2}{3}\rho_{\text{DW}}$. Shortly after forming, the domain walls are expected to have a large mass compared to the background temperature, implying that they will be non-relativistic.

Assuming the Universe to be homogeneous on a scale much greater than the wall separation, it can be approximately described by an FRW metric,
\begin{equation}
    ds^2=dt^2-a^2(t)dx^2,
\end{equation}
with the equation of state $p=w\rho$, the scale factor is then given by,
\begin{equation}
    a(t)\propto t^{\frac{2}{3(w+1)}}.
    \label{scalefactor_DW}
\end{equation}
The DW energy density can then be related to the scale factor as $\rho_{\text{DW}}\propto a^{-3(1+w)}$. For non-relativistic DWs while $a(t)\propto t^2$, the energy density is $\rho_{\text{DW}}\propto a^{-1}$.  This slow dilution makes a stable domain wall network cosmologically dangerous and conflicting with observation. At this stage it is important to point out that after their formation, the DW dynamics is governed by two different kinds of forces (i) the tension force $p_{\rm T}\sim \sigma/R_{\rm w}$ that helps flatten the wall where $\sigma$ denotes the surface tension and $R_{\rm w}$ represents the curvature radius of the walls, and (ii) friction force $p_{\rm F}\sim vT^4$ that appears if the field composing the core of domain wall interacts with the particles in the thermal bath with $v$ being the velocity of DWs. These two forces are balanced $i.e.~ p_{\rm T}\sim p_{\rm F}$. Using this condition and replacing $T^4\sim M_{\rm Pl}^2/t_r^2$ (assuming the radiation-domination) one can easily show that $\rho_{\rm DW}\propto t_r^{-2}$ or $\rho_{\rm DW}\propto a^{-1}$ (as also discussed above) where $t_r$ denotes the time when DWs become relativistic. At late times, the friction force is exponentially damped (when the temperature of the radiation bath becomes less than the mass of particles that interact with domain walls). Once $p_{\rm F}$ becomes irrelevant, the DW dynamics is dominated by the
tension force $p_{\rm T}$, which stretches DWs up to the horizon size. Numerical studies \cite{Press:1989yh, Garagounis:2002kt, Oliveira:2004he, Avelino:2005kn, Leite:2011sc, Leite:2012vn, Martins:2016ois} have shown that the evolution of DW in this regime can be described by the scaling solution. Here, their energy density evolves according to the simple scaling law $\rho_{\rm DW}\propto t^{-1}$, and their typical size is given by the Hubble radius $ \sim t^2$. In the scaling regime, the energy density of the DW can be expressed as,
\begin{eqnarray}
    \rho_{\rm DW}= \sigma\frac{\mathcal{A}}{t},
    \label{DW_scaling}
\end{eqnarray}
with $\mathcal{A}\simeq 0.8\pm 0.1$ being the area parameter~\cite{Hiramatsu:2013qaa}. Even if their energy density is subdominant today, they may still produce excessive density perturbations observable in the CMB epoch if their surface energy density is above $\mathcal{O}(\text{MeV}^3)$, this bound is also known as the Zel’dovich-Kobzarev- Okun bound~\cite{Zeldovich:1974uw}.

As we mentioned, one way to resolve this issue is to make the DWs unstable and allow the network to decay. Let us consider a scenario where the degeneracy of the vacuum is broken by introducing an energy bias in the scalar potential~\cite{Vilenkin:1981zs,Gelmini:1988sf,Larsson:1996sp,Bhattacharya:2023kws}. It is important to point out that, although an energy bias~\cite{Chiang:2020aui,King:2023ayw,Lu:2023mcz,King:2023ztb} is required to make the DWs unstable, a very large energy difference from the beginning will not allow the formation of DWs at the first place~\cite{Gelmini:1988sf}. So there must be a hierarchy between the bias terms and the rest of the potential. If the discrete symmetry is approximate, a bias can be established by introducing an explicit symmetry-breaking (in this case $Z_2$) term ~\cite{Saikawa:2017hiv}, 
\begin{equation}
    \Delta v(\phi)=\epsilon v\phi\bigg(\frac{1}{3}\phi^2-v^2\bigg),
\end{equation}
where $\epsilon$ is a dimensionless constant. Although this potential has a minima at $\pm v$, there also exists an energy difference between them,
\begin{equation}
    V_{\text{bias}}=V(-v)-V(+v)=\frac{4}{3}\epsilon v^4.
\end{equation}
As a result of this energy difference, the false vacuum tends to shrink $i.e.$ there exists a volume pressure force acting on the walls whose magnitude can be estimated as $p_V\sim V_{\text{bias}}$. DW collapses when this pressure force becomes greater than the tension force $p_T\sim\sigma\frac{\mathcal{A}}{t}$. Their annihilation time can be estimated by comparing these two forces,
\begin{equation}
    t_{\text{ann}}=C_{\text{ann}}\frac{\sigma\mathcal{A}}{V_{\text{bias}}},
    \label{t_ann}
\end{equation}
where $C_{\text{ann}}$ is a coefficient of $\mathcal{O}(1)$. Assuming the annihilation occurs in the radiation-dominated era, the temperature at $t=t_{\text{ann}}$ becomes proportional to the inverse square root of Eq. \ref{t_ann}, 
\begin{equation}
T_{\rm ann} = 3.41 \times 10^{-2}\,\mathrm{GeV}\,C_{\rm ann}^{-1/2}\mathcal{A}^{-1/2}\left(\frac{g_*(T_{\rm ann})}{10}\right)^{-1/4}\left(\frac{\sigma}{\mathrm{TeV}^3}\right)^{-1/2}\left(\frac{V_{\rm bias}}{\mathrm{MeV}^4}\right)^{1/2}, \label{T_ann_general}
\end{equation}
 The annihilation of DWs may produce GWs, which are potentially observable today. The Gravitational wave spectrum produced by a DW at cosmic time $t$ can be characterized by the ratio of the rate of change of GW energy density $\rho_{\rm GW}$ with respect to the frequency of the GW $f$ to the critical energy density of the Universe $\rho$,

\begin{equation}
    \Omega_{\text{GW}}(t,f)=\frac{1}{\rho_c(t)}\frac{d\rho_{\rm GW}(t)}{d \ln f}
    \label{GW_spec}
\end{equation}
where $\rho_{\rm GW}$ and $ f$ denote the energy density and the frequency of the GW respectively. The main feature of the GW spectrum from DW is that it follows a broken power law, where the breaking point has a frequency determined by the annihilation time (when $p_V\sim p_T$) and the peak amplitude is determined by the energy density in the domain walls (which in turn determines the gravitational power radiated). The peak amplitude is therefore dependent upon the fraction of energy radiated into gravitational waves, $\epsilon$, the surface tension, $\sigma$, and the radius of the domain wall which is the Hubble radius at the collapse time. Finally, it will depend upon an $\mathcal{O}(1)$ factor $\mathcal{A}$ that captures the precise details of a simulation. Assuming an instantaneous annihilation of the DWs at $t=t_{\rm ann}$, the peak amplitude is then~\cite{Saikawa:2017hiv},
\begin{equation}
     \Omega_{\text{GW}}(t_{\rm ann})_{\rm peak}=\frac{1}{\rho_c(t_{\rm ann})}\bigg(\frac{d\rho_{\rm GW}(t_{\rm ann})}{d \ln f}\bigg)_{\rm peak}=\frac{8\pi\tilde{\epsilon}_{\rm GW}G^2\mathcal{A}^2\sigma^2}{3H^2(t_{\rm ann})},
\end{equation}
where the parameter $\tilde{\epsilon}_{\rm GW}$ is estimated to be $\tilde{\epsilon}_{\rm GW}\simeq 0.7\pm0.4$~\cite{Hiramatsu:2012sc} and $G$ is the Newton’s gravitational constant. Once the DWs have entered the scaling regime \cite{Avelino:2005kn}, they decay, $i.e$ when their energy density evolves according to the simple scaling law $\rho_{\rm DW}\propto t^{-1}$, and their size is typically given by the Hubble radius $\sim t$ and hence the peak amplitude at present time $t_0$ is,
\begin{equation}
    \Omega_{\text{GW}}h^2(t_{0})_{\rm peak}=7.2\times10^{-18}\tilde{\epsilon}_{\rm GW}\mathcal{A}^2\bigg(\frac{\sigma}{\rm TeV^3}\bigg)^2\bigg(\frac{T_{\rm ann}}{10^{-2}\rm GeV}\bigg)^{-4}.
\end{equation}
The redshifted peak frequency can be determined by
\begin{eqnarray}
 f_{\rm peak}\simeq \bigg(\frac{a(t_{\rm ann})}{a(t_0)}\bigg)H(t_{\rm ann})\\
    f_{\rm peak}\simeq 1.1\times10^{-9}~{\rm Hz}\bigg(\frac{T_{\rm ann}}{10^{-2}\rm GeV}\bigg).
\end{eqnarray}

To depict the GW spectrum, the following parametrization for a broken power-law spectrum~\cite{Caprini:2019egz,Ferreira:2022zzo} can be adopted
\begin{eqnarray}
\Omega^{}_{\rm GW}h^2_{}  =  \Omega_{\text{GW}}h^2_{\rm peak} \frac{(a+b)^c}{\left(b x^{-a / c}+a x^{b / c}\right)^c} \ ,
\label{eq:spec-par}
\end{eqnarray}
where $x := f/f_p$, and $a$, $b$ and $c$ are real and positive parameters. Here the low-frequency slope $a = 3$ can be fixed by causality, while numerical simulations suggest $b \simeq c \simeq 1$~\cite{Hiramatsu:2013qaa}. While these simulations agree with a scenario where the $Z_2$ symmetry is broken, the scenarios with larger discrete symmetries require careful treatment.

\begin{figure}
    \centering
    \includegraphics[width=0.7\textwidth]{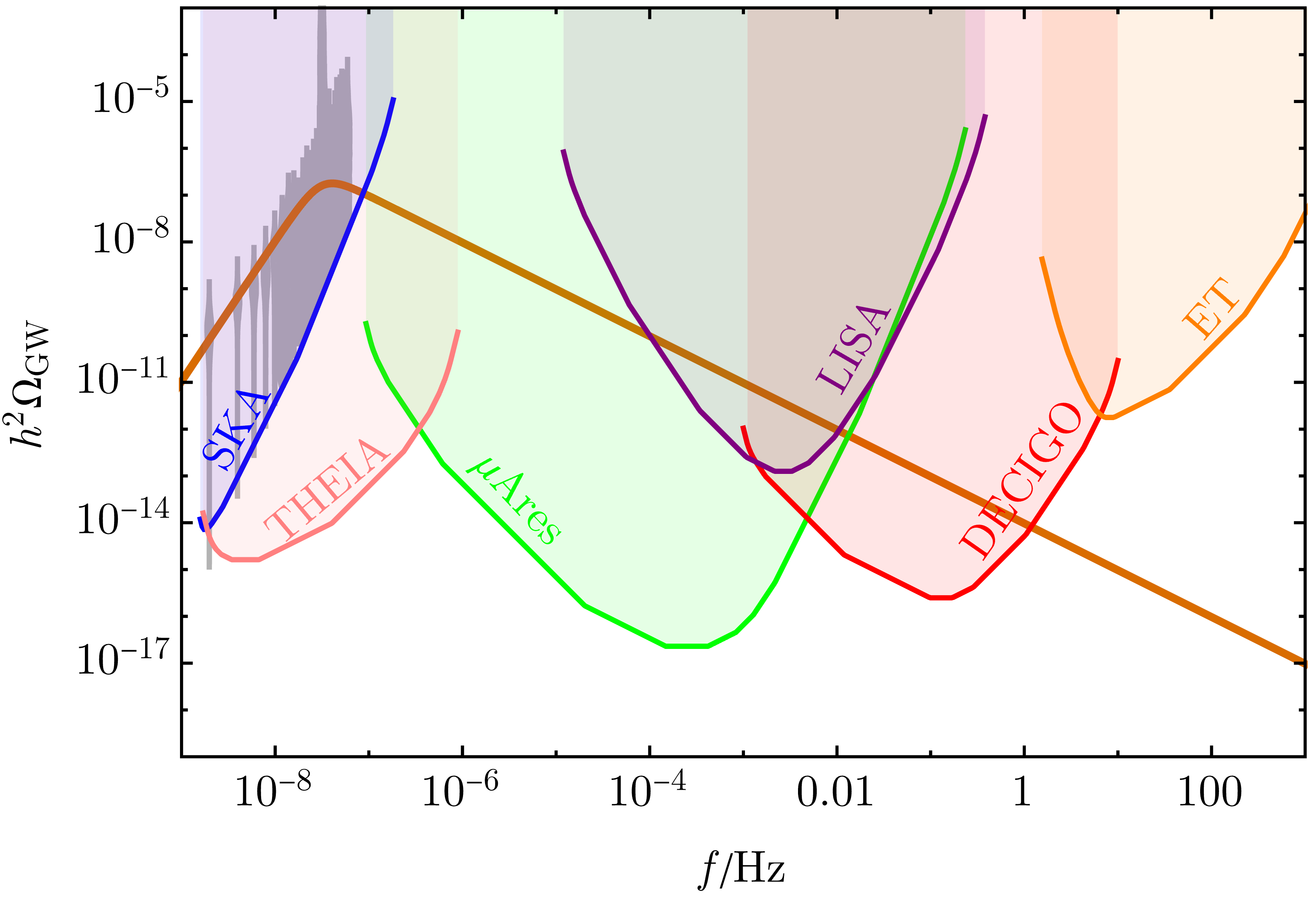}
    \caption{Shape of gravitational wave spectra as a function frequency generated from DW annihilation. Figure taken from ref \cite{King:2023ztb} }
    \label{fig:GW_DW}
\end{figure}

In the case of $Z_N$ symmetry for $N>2$, the UV power law can be modified \cite{Hiramatsu:2012sc,Ferreira:2022zzo,Wu:2022stu}. In this case, multiple degenerate vacua appear, hence leading to a domain wall network that is more complicated to the $Z_2$ case. Again, the degeneracy among the vacua can be broken by introducing a bias term in the scalar potential, and eventually, a domain with the lowest energy dominates over others, which causes the annihilation of the walls and the production of GW. It is interesting to point out that GW signals observed in these scenarios can deviate from what is observed in the case of $Z_2$ domain walls as also pointed out by a recent study \cite{Hiramatsu:2012sc}, that suggests that $b$ in Eq.\ref{eq:spec-par} decreases with the increasing value of $N$.



\subsection{Gravitational waves from Cosmic Strings}
\label{subsec_CS}

Let us now turn our attention to cosmic strings (CS). CSs are one-dimensional defects that appear in many extensions of the SM, e.g., field theories with a spontaneously broken U(1) symmetry (gauge or global) \cite{Kibble:1976sj,Nielsen:1973cs,Vachaspati:1984dz,vilenkin2000cosmic,King:2020hyd,Huang:2020bbe,Huang:2020mso,Lazarides:2021uxv,Maji:2023fhv,Saad:2022mzu,Ahmed:2023pjl,Antusch:2023zjk,Ahmed:2023pjl,Ahmed:2023rky,Ahmed:2022rwy,Afzal:2022vjx}, and the fundamental and/or composite strings in superstring theory \cite{Copeland:2003bj,Dvali:2003zj,Polchinski:2004ia,Jackson:2004zg,Tye:2005fn}. On the other hand, more complicated strings, such as $Z_N$-strings, can also be encountered during the phase transitions followed by a remnant discrete symmetry. For example a $Z_2$ string can appear in a GUT model discussed in \cite{Kibble:1982ae}

\begin{equation}
    SO(10)\longrightarrow SU(5)\times Z_2
\end{equation}

\noindent In a situation where  $N > 2$, $Z_N$-strings can form networks with vertices from which several strings can emerge.

Just like the DWs, once the strings are formed they can also eventually dominate the energy density of the Universe. The effective equation of state for the strings is then given by,

\begin{equation}
 p_{\text{S}}=\bigg(\frac{2}{3}u^2-\frac{1}{3}\bigg)\rho_{\text{S}},  
\end{equation}
where $p_{\text{S}}$ and $\rho_{\text{S}}$ denotes the pressure and energy density of the strings while $u$ represents the string's velocity. For a non-relativistic strings ($u<<1$) gives $i.e.~ p_{\text{S}}=-\frac{1}{3}\rho_{\text{S}}$.  Following Eq.~\ref{scalefactor_DW}, the CS energy density can then be related to the scale factor as $\rho_{\text{S}}\propto a^{-3(1+w)}$. For non-relativistic string while $a(t)\propto t$, the energy density is $\rho_{\text{S}}\propto a^{-2}$.  This suggests that strings can also act as a cosmological disaster. However, the evolution of the string network is much more complicated than this. One should include the two important effects 
\begin{itemize}
    \item The intercommutation of intersecting string segments leads to the formation of loops of different sizes.
    \item The decay of smaller loop by radiating GWs.
\end{itemize}

 At the time of the CS formation, the Universe might have been filled with a surplus amount of strings per horizon. This results from the fact that the correlation length of the field associated with the strings is often much less than the horizon size. Their abundance is then reduced by their self-intersection. String friction with the background plasma, if present, tends to freeze the string on the length scale which is larger than the friction length scale,
\begin{equation}
  R_{\rm fric} = v_{\rm drag} \times t,  
\end{equation}
where $v_{\rm drag}$ is the drag speed of the string in the plasma due to friction. This friction retards the rate at which strings reach scaling. However, during this time, strings can still self-intersect and reduce their abundance slowly until typically one string exists per horizon, thereby reaching the scaling regime. At this stage, one string per horizon is maintained as the strings typically intersect whenever
\begin{equation}
    t_{\rm intersect} = L/v_{\rm string} = {\rm Hubble \  time}
\end{equation}
is satisfied where $L$ denotes the average inter-string separation scale. This is an attractor solution since the strings will not dilute less than one per horizon since if they dilute a bit more they become superhorizon, and they are stretched as $a(t) \sim t^{1/2} (t^{2/3})$ in the RD (MD) era which is slower than the horizon expansion of $t$. As a result of this, the horizon always ``catches" up with the superhorizon strings as long as the scale factor grows slower than $t$.

There exist several studies in the literature that have calculated the SGWB from evolving cosmic string networks \cite{Hindmarsh:1997qy,Vincent:1997cx,Matsunami:2019fss,Hindmarsh:2021mnl,Blanco-Pillado:2023sap}.  Nevertheless, there exists a technical challenge that persists for pure numerical simulation to track the network’s evolution over the entire relevant cosmic history. The issue with the CS network is that at the micro level, we use a field-theoretic treatment. Simulations of the field-theoretic treatment seem to suggest that very limited energy radiated into GW. Once the string is long enough we use a classical Nambu-Goto action string theory treatment. Here, most of the energy is dumped into GWs. It is hard to tell which treatment is right due to the enormous hierarchy of scales involved $i.e.$ inverse string length vs symmetry-breaking scale. 
Therefore it is difficult to make predictions for observations today. The debate has been ongoing for quite some time, with new insights on both sides of the argument appearing in the last couple of years (see for instance \cite{Matsunami:2019fss,Hindmarsh:2021mnl,Blanco-Pillado:2023sap}) leaving little confidence that the last word has been uttered. \par For the purposes of this review we will proceed with the assumption that the Nambu-Goto treatment is valid. We will use the Velocity-dependent One-Scale (VOS) model \cite{Martins:1996jp,Martins:2000cs} that captures the essential physics, and use it to study and predict the evolution of the string network over a long range of time while calibrating the input model parameters with data points for early time evolution that have been made available by simulations. Next, we briefly review the VOS model.

Following Eq.~\ref{GW_spec}, one can calculate the GW spectrum produced by evolving CS. The idea here is to add up the GW emission from all the loops throughout the history of the Universe. Now, the first basic ingredient for calculating the GW spectrum from the CS is the GW power $P_{\rm GW}$. The total power emitted in gravitational waves by strings loops can be estimated from the quadrupole formula, $ P_{\rm GW} \approx \frac{G}{45} \sum_{i,j} \langle \dddot{Q}_{ij} \dddot{Q}_{ij} \rangle$. For a string of length $l$, the quadrupole moment is $Q\sim \mu l^3$ where $\mu$ denotes the energy per unit length of a string or the string tension. Now, the oscillation frequency of a relativistically oscillating loop is $\omega\sim 1/l$, so $\dddot{Q}\sim Q\omega^3\sim\mu$. This gives $ P_{\rm GW}  \propto G \mu^2$.   
Note that, since the loop also moves relativistically, one cannot use the quadrupole formula directly. An oscillating string loop of length $l$ emits a discrete set of frequencies $\omega_n=2\pi f_n$. Hence, the total power radiated is
\begin{eqnarray}
    P_{\rm GW}=\sum_n P_n,
\end{eqnarray}
where $P_n=G\mu^2p_n$, so

\begin{eqnarray}
    P_{\rm GW} (f,l)= G \mu^2 \sum_n p_n,\\
    = G \mu^2 l\int_0^\infty df P(fl),
\end{eqnarray}
where 
\begin{eqnarray}
  P(x)\equiv\sum_n p_n\delta(x-x_n).  
\end{eqnarray}

\noindent In a continuous limit,
\begin{eqnarray}
    \int_0^\infty dx P(x)=\Gamma,
\end{eqnarray}
where GW emission efficiency, $\Gamma \sim 50$ \cite{Blanco-Pillado:2013qja, Blanco-Pillado:2017oxo, Vilenkin:1981bx, Blanco-Pillado:2011egf} is determined by Nambu-Goto (NG) simulations.
Finally, the total power emitted is then expressed as,
\begin{eqnarray}
   P_{\rm GW} = \Gamma G \mu^2.
   \label{pow_cs}
\end{eqnarray}

Following this, the observed energy density of GW at a particular frequency $f$ today is obtained by adding the amount of energy produced at each moment of cosmic evolution for loops of all sizes. Since the $\rho_{\rm GW}$ redshifts as $(a(t)/a_0)^{-4}$ and hence $\rho_{\rm GW}(t)=(a(t)/a_0)^{-4}\rho_{\rm GW}(t_0)$. On the other hand, the emitted frequency is related to the observed frequency today as $f_e=(a(t)/a_0)^{-1}f_0$, so
\begin{eqnarray}
    \frac{d\rho_{\rm GW}(t)}{df_e}=\bigg(\frac{a(t)}{a_0}\bigg)^{-3} \frac{d\rho_{\rm GW}(t_0)}{df_0}.
\end{eqnarray}
\noindent  Finally, including the result of redshifting, the emission for the moment of emission to today is,
\begin{eqnarray}
\label{eqn:energyI}
   \frac{d\rho_{\rm GW}}{df}(t_0,f) = G\mu^2\int_0^{t_0}{dt \left(\frac{a(t)}{a_0}\right)^3 \int_0^\infty{dl ~ l~ n(l,t)~ P \left(\frac{a_0}{a(t)} f l\right)}}\,,
\end{eqnarray}
where $a(t)$ is the scale factor which takes the value $a_0$ today and $n(l,t)$ is the loop number density (see Eq. \ref{loop_number_density} ). Note that we have just used $f$ to denote $f_0$ in Eq. \ref{eqn:energyI}. 

The second important ingredient for calculating the GW spectrum from the CS is the number density $n(l,t)$ of non-self-intersecting, sub-horizon, CS loops of invariant length $l$ at cosmic time $t$. To extrapolate the results from simulations (which runs only over a finite time interval) to any moment in the history of the network, the scaling of loops is important as it suggests that,
\begin{eqnarray}
    n(l,t)=t^{-4}n(x),
\end{eqnarray}
where $x=l/t$ is the ratio of the loop size to roughly the horizon scale. 

To obtain $ n(l,t)$ one approach is to determine the loop production function $\mathbf{f}(l,t)dl$  namely the number density of non-self-intersecting loops of lengths between $l$ and
$l + dl$ produced per unit time, per unit volume, which in scaling satisfies,
\begin{eqnarray}
    \mathbf{f}(l,t)=t^{-5}\mathbf{f}(x).
\end{eqnarray}
The loop number density can be obtained by solving the Boltzmann equation for the loops. The Boltzmann equation incorporates the production of loops from the infinite string network as described by $\mathbf{f}(l,t)dl$, dilution of loops due to the expansion of the Universe, loss of energy through the GWs radiation. The loop number density thus can be calculated by integrating the loop production function \cite{Auclair:2019wcv}
\begin{eqnarray}
    n(l,t)  = \int_{t_{i}}^{t}{dt' \mathbf{f}(l',t') \left(\frac{a(t')}{a(t)}\right)^3}\,,
    \label{loop_number_density}
\end{eqnarray}
where the effect of the expansion is explicitly seen through the dependence of the scale factor $a(t)$, and $l'$ (which is given below) contains information on the evolution of the length of the loop due to its gravitational decay from the time of formation $t'$ to the observation time $t$. 

Finally, we can relate the loop production function $\mathbf{f}(l,t)$ with the long string network energy density $\rho_\infty$ by assuming that the production of loops is the dominant energy loss mechanism of the long string network, which is given by
\begin{eqnarray}
    \label{eqn:cc}
   \frac{d \rho_{\infty}}{dt} = -2H(1+\bar{v}^2)\rho_{\infty} - \frac{d \rho_{\infty}}{dt} \bigg{|}_{\rm loop}\,,
\end{eqnarray} 
where $\bar{v} = \sqrt{\langle v^2 \rangle}$ is the root-mean-squared (RMS) velocity of the long strings strings. The first term in this equation describes the dilution of the long string energy density in an expanding Universe, while the second, describes energy loss into loop formation. The energy loss in loop formation is proportional to the loop production function, loop length, and string tension,
\begin{eqnarray}
  \frac{d \rho_{\infty}}{dt} \bigg{|}_{\rm loop} \equiv \mu\int_0^{\infty}{l\mathbf{f}(l,t)dl}.
  \label{energy_loss}
\end{eqnarray}
\noindent Loop production is essential to achieve the linear scaling of long strings, see e.g.~\cite{vilenkin2000cosmic}.

 To determine $n(l,t)$, it is necessary to have a handle on the evolution of the long string energy density $\rho_\infty$ and the root mean velocity $\bar{v}$ of the long string appearing in Eq. \ref{eqn:cc}. To do that one can use the VOS as it describes both the scaling evolution of the long string together with non-scaling evolution through the radiation-matter transition. The evolution can be obtained by solving the coupled differential equations (VOS equation of motion) involving the rate of change of RMS velocity of the long strings and the characteristic length between the strings,
\begin{eqnarray}
    \frac{d\bar{v}}{dt} & = & \left(1-\bar{v}^ 2\right)\left[\frac{k(\bar{v})}{L}-2H\bar{v}\right]\,,\nonumber\\
\frac{dL}{dt} & = & \left(1+\bar{v}^ 2\right)HL+\frac{\tilde{c}}{2}\bar{v}\,, 
\label{eqn:vosL}
\end{eqnarray}
where $L\equiv (\mu/\rho_\infty)^{1/2}$ denotes the characteristic length that measures the average distance between long strings and $\tilde{c}$ quantifies the efficiency of the loop-chopping mechanism. Note that the second equation of Eq. \ref{eqn:vosL} is nothing but Eq. \ref{eqn:cc} written in terms of $L$ and $\tilde{c}\bar{v}$ where
\begin{eqnarray}
   \tilde{c}\bar{v} = \mu \frac{L}{\rho_\infty}\int_0^\infty l\mathbf{f}(l,t)dl.
   \label{tildec}
\end{eqnarray} 
The function $k(\bar{v})$ accounts for the effects of small-scale structure (namely,  multiple kinks) on long strings \cite{Martins:2000cs}, 
\begin{eqnarray} \label{eqn:curavture}
k(\bar{v})=\frac{2\sqrt{2}}{\pi}\left(1-\bar{v}^2\right)\left(1+2\sqrt{2}\bar{v}^3\right)\frac{1-8\bar{v}^6}{1+8\bar{v}^6}\,,
\end{eqnarray}
This reproduces the expected asymptotic behavior of $k(\bar{v})$ both in the relativistic and non-relativistic limits. 
The linear scaling of the long string network ( of a form  $L=\xi_s t$ and $\bar{v}=$ constant \cite{Sousa:2013aaa}) in the radiation- and matter-dominated backgrounds follows directly from Eq.~\ref{eqn:vosL} since the particular solutions are
\begin{eqnarray}\label{eqn:scaling}
\frac{L}{t}=\sqrt{\frac{k(\bar{v})(k(\bar{v})+\tilde{c})}{4\nu(1-\nu)}} \equiv \xi_s \qquad\mbox{with}\qquad \bar{v}=\sqrt{\left(\frac{k(\bar{v})}{k(\bar{v})+\tilde{c}}\right)\left(\frac{1-\nu}{\nu} \right)}  \equiv \bar{v}_s \,,
\end{eqnarray}
where the subscript $s$ stands for `scaling', are attractor solutions of these equations for $a\propto t^{\nu}$ and $0<\nu<1$ with $\xi_s,\bar{v}_s$ being constant. The scaling solution is attainable only for a constant expansion exponent $\nu$, so, such a regime can only be maintained deep into radiation or matter eras \cite{Sousa:2013aaa}. More generally, Eq.~\ref{eqn:vosL} can be solved throughout any cosmological era, including the radiation-to-matter and matter-to-dark-energy transitions, and hence one can trace the evolution of cosmic string networks in a realistic cosmological background~\cite{Sousa:2013aaa}.

The second step in determining $n(l,t)$ is to relate the loop production function and the long string network as described by the VOS model. Now defining a new variable $\xi\equiv \frac{L(t)}{t}$ which describes the length parameter and as before, $x \equiv \frac{l}{t}$, it follows from the second equation of Eq.~\ref{eqn:vosL} and Eq. \ref{tildec} that the loop production function satisfies,
\begin{eqnarray}
    \label{eqn:energyconservation}
    \int_0^\infty{ x \mathbf{f}(x) ~dx} = \frac{2}{\xi^2}\left[ 1- \nu (1+\bar{v}^2)\right] = \tilde{c} \frac{\bar{v}}{\xi^3}\equiv C_{\rm eff}\,.
\end{eqnarray}
Now, one can make the following assumption\footnote{There exist two other models to calculate the loop number density. In the first one \cite{BlancoPillado:2011dq,Blanco-Pillado:2013qja,Auclair:2019wcv} one can use loop production functions for non-self-intersecting loops directly from NG simulations of cosmic string and in the second one \cite{Lorenz:2010sm,Ringeval:2005kr,Auclair:2019wcv} the loop production function is not the quantity inferred from the simulation but the distribution of non-self-intersecting scaling loops is directly extracted from the simulation.} {\it throughout cosmic history, all loops are assumed to be created with a length $l$ that is a fixed fraction of the characteristic length of the long string network}, namely $l = \alpha_L L$, with $\alpha_L <1$. Thus
\begin{eqnarray}
\mathbf{f}(x) =  {\tilde C} \delta\left(x - \alpha_L\xi\right) .
\label{eq:ffirst}
\end{eqnarray}
From Eq.~(\ref{eqn:energyconservation}) we obtain,
\begin{eqnarray}
\tilde C =  \frac{\tilde{c}}{\alpha_L}\frac{\bar{v}}{\xi^4}\,
\label{c1}
\end{eqnarray}
with $\bar{v}$ and $\xi=L/t$ being the solutions of the VOS Eq. \ref{eqn:vosL}. Note that the value of $\tilde C$ in the above equation is an upper limit since Eq. \ref{eqn:energyconservation} does not carry the information that some of the energy of the long string network goes redshifting of the peculiar velocities of loops. Introducing $f_r\sim\sqrt{2}$ which denotes the reduction factor that accounts for the energy loss into redshifting of the peculiar velocities of loops and $\mathcal{F}$ that denotes the fraction of loops formed with size $\alpha_L$ as not all the loops created need to be of the same size. Taking these corrections into account, the loop production function can be written as
\begin{eqnarray}
\mathbf{f}(x) = \left(\frac{\mathcal{F}}{f_r}\right) {\tilde C} \delta\left(x - \alpha_L\xi\right)\equiv A ~ \delta\left(x - \alpha_L\xi\right)\, ,
\label{eq:ffirst}
\end{eqnarray}
This expression is valid throughout cosmic history, even when the cosmic string network is not in a linear scaling regime (in this case $x$ and $\tilde{C} $ will be time-dependent). 

Finally, substituting Eq.~\ref{eq:ffirst} in Eq.~\ref{loop_number_density} one can obtain loop number density for all times including during the radiation-to-matter and matter-to-dark-energy transitions but this requires solving the VOS equation of motion. During the {\it radiation era} ($\nu=1/2$), the long string network is scaling and described by the VOS solutions (Eq. \ref{eqn:scaling}), namely $\xi_r=0.271$ and $v_r=0.662$, hence it follows that the loop distribution takes a simple form 
\begin{eqnarray}
\label{eqn:VOSrad}
n_r(x)=\frac{A_r}{\alpha}\frac{\left(\alpha+\mathrm{\Gamma} G\mu \right)^{3/2}}{\left(x+\mathrm{\Gamma} G\mu \right)^{5/2}}\,,
\end{eqnarray} 
with $A_r= 0.54$ (fixing ${\cal F}=0.1$), and where $\alpha = \alpha_L \xi_r$. As noted above, this expression is only valid for $x\leq \alpha$. 
In a matter-only universe ($\nu=2/3$), the VOS scaling solutions (Eq. \ref{eqn:scaling}) give $\xi_m=0.625$ and $v_m=0.583$ and the loop distribution is
\begin{eqnarray}
\label{eqn:VOSmat}
n_m(x)=\frac{A_m}{\alpha_m}\frac{\alpha_m+\mathrm{\Gamma} G\mu}{\left(x+\mathrm{\Gamma}G\mu\right)^2}\,,
\end{eqnarray} 
where $\alpha_m=\alpha_L \xi_m$, $A_m=0.039$ and $x\leq \alpha_m$. 

We are now at a point where we must take divergent paths - the evolution of the string network is qualitatively different if the strings arise from a global or a local symmetry. We will refer to each case as a global and local string network respectively and separately describe their evolution including the above effects. 

\subsubsection{Cosmic strings from global $U(1)$ symmetry breaking}
Let us first consider the case of global strings. Although we will introduce some key concepts that are common for both global as well as local strings.
We will consider a simple setup with a global $U(1)$ symmetry that is spontaneously broken by a complex scalar $\phi$. The model features global strings that result from this breaking. The setup can be described by the Lagrangian,
\begin{eqnarray}
    \mathcal{L}=\partial_\mu\phi^*\partial^\mu\phi-V(\phi),\\
    V(\phi)=\frac{1}{2}\lambda(|\phi|^2-\frac{1}{2}\eta^2)^2,
\end{eqnarray}
where $\eta$ denotes the vacuum expectation value of $\phi$. This spontaneous breaking of global $U(1)$ also leads to the formation of a massless Nambu-Goldstone boson. Depending on the magnitude of $\eta$, these global strings can either decay dominantly to the Goldstone bosons or the GWs (see the discussion below Eq. \ref{Et} ). In this scenario, the GWs are mostly suppressed and their amplitude mildly falls with frequency for most of the spectrum of interest.  Although, this makes their detection a bit difficult, works like \cite{Chang:2019mza, Chang:2021afa, DiBari:2023mwu,Gorghetto:2021fsn} suggest that they can still be detected by GW experiments like LISA, AEDGE, DECIGO, and BBO by considering the breaking scale $\eta$ above $10^{14}$ GeV.  

Once formed, the loop oscillates and emits its energy in the form of Goldstone (first term) and gravitational waves (second term) as shown in the equation below \cite{Vilenkin:1986ku}
\begin{eqnarray}
\frac{dE}{dt} = - \Gamma_a \eta^2 -\Gamma G\mu^2, \label{Et}.
\end{eqnarray}
The parameter $\Gamma_{(a)}$ depends only on the loop trajectory \cite{vilenkin2000cosmic,Battye:1997jk},  and hence we expect that the value of the Goldstone radiation constant $\Gamma_a$ must be close to the value of the GW radiation constant $\Gamma \sim \Gamma_a$ \cite{Battye:1997jk}.  The Goldstone radiation constant is found to be $\Gamma_a \sim 65$ \cite{vilenkin2000cosmic, Battye:1997jk}. The absence of the gauge field in the breaking of the global $U(1)$ symmetry implies the presence of Goldstone (as discussed earlier) with logarithmically divergent gradient energy, 
\begin{eqnarray}
 \mu(t) = 2\pi \eta^2 \log{\frac{L(t)}{\delta}},
 \label{mu_gs}
\end{eqnarray} 
The string tension depends on the ratio of two scales, a macroscopic scale $L(t)$ (close to the Hubble scale), and a microscopic scale $\delta(t) \propto 1/\eta$ representing the width of the string core.  It is interesting to point out that even though the $\Gamma_{(a)}$ are similar in size, the GW radiation has a relative suppression of $ \eta ^2 / M_{\rm p}^2$. However, this suppression becomes less severe once the symmetry-breaking scale $\eta$ gets closer to $M_{\rm p}$. For this reason, next-generation gravitational wave detectors are only sensitive to global strings formed during symmetry breaking occurring near the GUT scale.
Due to the emission of the Goldstones together with the GW,  a loop of initial length $\ell_i = \alpha t_i$ at a later time reduces, 
 \begin{eqnarray}
\ell (t) \simeq \alpha t_i - \Gamma G \mu (t-t_i) - \frac{\Gamma_a}{2\pi} \frac{t-t_i}{\log{\frac{L(t)}{\delta}}}, \label{ell}.
\end{eqnarray}
where the second and third terms denote the decrease in loop size for gravitational wave emission and Goldstone emission, respectively. 

A string loop emits the Goldstone and the GW from normal mode oscillations with frequencies \cite{Vachaspati:1984gt,Allen:1991bk,Maggiore:2018sht},  
\begin{eqnarray}
    \tilde{f}_k = 2k/\tilde{l},
\end{eqnarray}
where $k=1,2,3, \ldots$, and $\tilde{l} \equiv l (\tilde{t})$ is the instantaneous size of a loop when it radiates at $\tilde{t}$.
The distribution of power in the different harmonics is determined by the small-scale structure of the loops. It was shown \cite{Vachaspati:1984gt,Allen:1991bk} that, in the large $k$ limit, the power emitted in each mode scales as $k^{-4/3}$ if the loop contains points where the velocity is locally 1 known as cusps; as $k^{-5/3}$, if it has kinks (which are discontinuities in the tangential vector introduced by the intercommutation process); and $k^{-2}$ when kink-kink collisions occur. In principle, when computing the SGWB generated by cosmic string loops, one should consider the contribution of all the harmonic modes of emission. 
The radiation parameters can be decomposed as $\Gamma = \sum_k \Gamma^{(k)}$ and $\Gamma_a = \sum_k \Gamma_a^{(k)}$, with
\begin{eqnarray}
\Gamma^{(k)} = \frac{\Gamma k^{-n}}{\sum_{j=1}^{\infty} j^{-n}}, \quad \text{and\quad} \Gamma_a^{(k)} = \frac{\Gamma_a k^{-n}}{\sum_{j=1}^{\infty}j^{-n}},
\label{gamma_k}
\end{eqnarray}
and the normalization factor is approximately $\sum_{j=1}^{\infty}j^{-4/3} \simeq 3.60$. 

Taking into account the redshifted  frequency $f_k = \frac{a(\tilde{t})}{a(t_0)} \tilde{f}_k$, with $t_0$ denoting the present time and $a(t_0) \equiv 1$ being scale factor today, the gravitational wave amplitude is summed over all normal modes
\begin{eqnarray}
\Omega_{\rm GW}(f) = \sum_k \Omega_{\rm GW}^{(k)} (f) = \sum \frac{1}{\rho_c}\frac{d\rho_{\rm GW}}{d \log{f_k}}.
\label{omega_CS}
\end{eqnarray}
Now, the contribution from an individual $k$ mode can be obtained by taking an integral over the evolving number density, 
\begin{eqnarray}
\Omega_{\rm GW}^{(k)} (f) = \frac{\mathcal{F}_\alpha F_a}{\alpha \rho_c} \frac{2k}{f} \int_{t_f}^{t_0} d\tilde{t} \frac{C_{\rm eff}\left(t_i^{(k)}\right)}{t_i^{(k)4}} \frac{\Gamma^{(k)}G\mu^2}{\alpha + \Gamma G \mu + \frac{\Gamma_a}{2\pi \frac{L(t)}{\delta}}} \left[\frac{a(\tilde{t})}{a(t_0)}\right]^5 \left[\frac{a\left(t_i^{(k)}\right)}{a(\tilde{t})}\right]^3 \theta(\tilde{\ell}) \theta(\tilde{t}-t_i), \nonumber \\ \label{GWk}
\end{eqnarray}
while $t_f$ denotes the time of network formation, the first Heaviside theta functions guarantee causality (i.e the size of the source is limited to the horizon size), and the second Heaviside theta function suggests that the GWs are only emitted once the loops are formed.   Following Eq. \ref{ell}, one can easily find that a loop that emits GW at time $\tilde{t}$ leading to an observed frequency $f$ was formed at the time
\begin{eqnarray}
t_i^{(k)} = \frac{\tilde{\ell}(\tilde{t},f,k) + \left(\Gamma G \mu + \frac{\Gamma_a}{2\pi \frac{L(t)}{\delta}}\right)\tilde{t}}{\alpha + \Gamma G \mu + \frac{\Gamma_a}{2\pi \frac{L(t)}{\delta}}},
\end{eqnarray} 
where the loop size can be written as $\tilde{\ell} = 2ka(\tilde{t})/f$.

The GW spectrum can be divided into three regions. The spectrum in these three region can be approximated by \cite{Chang:2021afa}

\begin{equation}
    {\Omega_{\rm GW}(f)h^2} \simeq  \left\{ \begin{array}{cc}
         5.1\times10^{-15}\left( \frac{\eta}{10^{15}{\rm GeV}} \right)^4\left( \frac{f}{f_\eta} \right)^{-1/3}, & {\rm for}~f_\eta<f \\
     8.8\times10^{-18}\left( \frac{\eta}{10^{15}{\rm GeV}} \right)^4\log^3\left[ \left(\frac{2}{\alpha f}\right)^2\frac{\eta}{t_{\rm eq}}\frac{1}{z_{\rm eq }^2\sqrt{\xi}}\Delta_{R}^{1/2}(f) \right]\Delta_R(f), & {\rm for}~f_{\rm eq}<f<f_\eta \\
      2.9\times10^{-12}\left( \frac{\eta}{10^{15}{\rm GeV}} \right)^4\left( \frac{f}{f_{\rm eq}} \right)^{-1/3}, & {\rm for}~f_0<f<f_{\rm eq} \\
      0,  & {\rm for}~f<f_0
    \end{array} \right.
\end{equation}
where $t_{\rm eq}$ and $z_{\rm eq}$ are the time and red-shift at the epoch of matter-radiation equality, $\Delta_R(f)$ takes care of the effect of varying the number of relativistic degrees of freedom, $g_\star$ and $g_{\star S}$, over time 
\begin{eqnarray}
    \Delta_R(f)=\frac{g_\star(f)}{g_\star^0}\bigg(\frac{g_{\star S}^0}{g_{\star S}(f)}\bigg)^{4/3},
\end{eqnarray}
with superscript $0$ denoting the values at present times. In the first region where the frequency is very high, $f_{\eta}<f$ \cite{Battye:1996pr}, the signal falls off, and the exact shape depends on the initial conditions and very early stages of the string network evolution not fully captured by the VOS model~\cite{Martins:1996jp, Martins:2000cs, Martins:2003vd, Martins:2016wqq, Martins:2018dqg, Correia:2019bdl} discussed above. $f_{\eta}$ is related to the time when the Goldstone radiation becomes significant. The radiation-dominated era that corresponds to the second region, $f_{\rm eq}<f<f_{\eta}$, the spectrum fall as $~\log^3{(1/f)}$. $f_{\rm eq}$ indicates the frequency corresponding to the emission around the time of matter-radiation equality. Finally, in the last region that corresponds to the matter-domination $f_0<f<f_{\rm eq}$, the spectrum falls as $f^{-1/3}$. $f_{\rm eq}$ is related to the time of matter-radiation equality, and $f_0$ is related to the emission at the present time. These characteristic frequencies are given by 
\begin{eqnarray}
f_{\eta} &\sim \frac{2}{\alpha t_n} \frac{a(t_{\eta})}{a(t_0)} \sim 10^{10}\ \text{Hz}, \\
f_0 &\sim \frac{2}{\alpha t_0} \sim 3.6 \times 10^{-16}\ \text{Hz}, \\
f_{\rm eq} &\sim 1.8 \times 10^{-7}\ \text{Hz}. 
\end{eqnarray}

\begin{figure}
    \centering
    \includegraphics[width=0.7\textwidth]{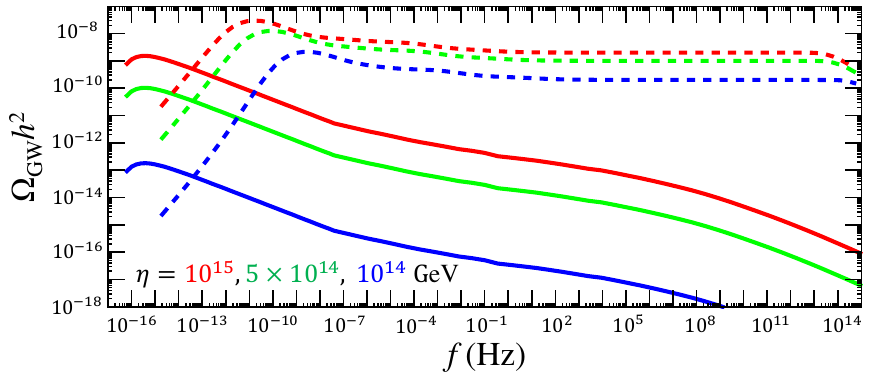}
    \caption{Shape of gravitational wave spectra as a function frequency generated from global (solid) and local (dashed) cosmic strings for different values of $\eta=10^{15}~\rm{GeV (red/upper)}, 5\times 10^{14}~\rm{GeV (green/middle)},10^{14}~\rm{GeV (blue/lower)}$. Figure taken from ref \cite{Chang:2021afa} }
    \label{fig:GW_CS_1}
\end{figure}

\subsubsection{Cosmic strings from gauged $U(1)$ symmetry breaking}

Unlike global strings, local strings or the Nambu-Goto strings are formed by gauging a $U(1)$ symmetry. In this scenario, the Goldstones are eaten by $U(1)$ vector bosons. Once formed, these Nambu-Goto strings radiate their energy primarily via loop generation and emission of GW which has a flat spectrum \cite{Gouttenoire:2019kij,Lazarides:2022spe,Borah:2022vsu,Lazarides:2022ezc,Borah:2023iqo}. $10\%$ of the loops formed are large and the remaining are highly boosted smaller loops \cite{Vanchurin:2005pa,Martins:2005es,Olum:2006ix,Ringeval:2005kr,Blanco-Pillado:2017oxo}. The loop formation from the long strings can again be described using the VOS model. Contrary to the case of global strings where energy per unit length, $\mu$ has a logarithmic dependence (see Eq.~\ref{mu_gs}), the energy is radiated at a constant rate during the oscillation of the larger loops for Nambu-Goto strings as $\mu$ remains constant, 
\begin{eqnarray}
    \frac{dE}{dt}=-\Gamma G\mu^2.
\end{eqnarray}
In general, for local strings $\mu\sim\mathcal{O}(\eta)$. Since there is no emission of Goldstone, Eq.~\ref{ell} can now be written as,
\begin{eqnarray}
\l (t) \simeq \alpha t_i - \Gamma G \mu (t-t_i) . \label{ell_local}    
\end{eqnarray}
Integrating over the emission time, the gravitational wave amplitude of the $k$-th mode is given by
\begin{eqnarray}
    \Omega_{\rm GW}^{(k)} (f) = \frac{1}{\rho_c}\frac{2k}{f} \frac{\mathcal{F}_\alpha \Gamma^{(k)}G\mu^2}{\alpha (\alpha + \Gamma G\mu)} \int_{t_F}^{t_0} d\tilde{t} \frac{C_{\rm eff} (t_i^{(k)})}{{t_i^{(k)}}^4} \left[ \frac{a(\tilde{t})}{a(t_0)} \right]^5 \left[\frac{a(t_i^{(k)})}{a(\tilde{t})}\right]^3\ \Theta(t_i^{(k)} - t_F)\,, \label{GWCS}
\end{eqnarray}
where $\Gamma^{(k)}$ can be obtained from Eq.~\ref{gamma_k}, $t_i^{(k)}$ is the formation time of loops contributing to the $k$-th mode and is given by
\begin{eqnarray}
    t_i^{(k)}(\tilde{t}, f) = \frac{1}{\alpha + \Gamma G \mu}\left[ \frac{2k}{f} \frac{a(\tilde{t})}{a(t_0)} + \Gamma G \mu \tilde{t} \right]. 
\end{eqnarray}
Summing over all modes, we get the total amplitude of the gravitational waves
\begin{eqnarray}
    \Omega_{\rm GW} (f) = \sum_k \Omega_{\rm GW}^{(k)} (f)\,,
\end{eqnarray}
where the sum can be easily evaluated using 
\begin{eqnarray}
    \Omega_{\rm GW}^{(k)} (f) = \frac{\Gamma^{(k)}}{\Gamma^{(1)}} \Omega_{\rm GW}^{(1)}(f/k) = k^{-4/3}\ \Omega_{\rm GW}^{(1)} (f/k)\,.
\end{eqnarray}

\noindent A typical shape of a GW signal generated by Nambu-Goto strings is a rising spectrum at a low frequency and thereafter a plateau is observed at higher frequencies. The height of the plateau is proportional to the symmetry-breaking scale $\eta$. This gentle dependence (linear) of $\Omega $ on $\eta$ in Fig. \ref{fig:GW_CS_1} means next-generation detectors can probe a large range of symmetry-breaking scales.

\subsection{Gravitational waves from Monopoles and Textures} 
Monopoles or localized defects are point-like defects that arise from the SSB of some larger spherical symmetry. The existence of a monopole solution was first discussed by 't Hooft and Polyakov in a gauge  $SO(3)$ model with a scalar triplet field. The setup can be described by the Lagrangian,
\begin{equation}
    {\cal L}= \frac{1}{2}D^\mu \phi D_\mu \phi  - \frac{1}{4} B_{\mu \nu} B^{\mu \nu} - \frac{\lambda}{4} \left(\phi ^2 - v^2_\phi \right)^2 \ .
\end{equation}
where $\phi$ is a scalar triplet. The `t Hooft-Polyakov monopole \cite{tHooft:1974kcl,Polyakov:1974ek} has the solution
\begin{eqnarray}
  \phi &=& \hat{r} \frac{h(v_\phi e r)}{er} \nonumber \\ 
  W^i _a &=& \epsilon _{aij} \hat{x} ^j \frac{1-f(v_\phi er)}{er} \nonumber \\
  W_a^0 &=& 0 \, ,
\end{eqnarray}
where, using the shorthand $\xi = v_\phi e r$ for the product of $v_\phi$ with the gauge coupling constant $e$ and radial coordinate $r$, the functions $f$ and $h$ are solutions to the equations \cite{tHooft:1974kcl,Polyakov:1974ek}. \begin{eqnarray}
  \xi ^2 \frac{d^2f}{d \xi^2} &=& f(\xi) h(\xi )^2 + f(\xi) (f(\xi)^2 -1) \\
  \xi^2  \frac{d^2 h}{d \xi ^2} &=& 2 f(\xi )^2 h(\xi) + \frac{\lambda}{e^2} h(\xi) (h(\xi)^2 - \xi ^2) \ .
\end{eqnarray}
The boundary conditions satisfy $\lim _{\xi \to 0} f(\xi)-1= \lim _{\xi \to 0}h(\xi ) \sim {\cal O}(\xi )$ and $\lim _{\xi \to \infty}f(\xi) = 0 $, $\lim _{\xi \to \infty} h(\xi )  \sim \xi$. The monopole mass again comes from solving the equations of motion and then calculating the static Hamiltonian,
\begin{eqnarray}
  E=m&=& \frac{4 \pi v_\phi}{e} \int _0 ^\infty \frac{d \xi}{\xi ^2 }\left[ \xi^2 \left( \frac{d f}{d\xi} \right)^2  +\frac{1}{2} \left( \xi \frac{dh}{d \xi} - h \right)^2 \right.  \nonumber \\ 
  && \left. + \frac{1}{2}(f^2-1)^2+ f^2h^2 + \frac{\lambda}{4 e^2}(h^2- \xi ^2)^2 \right] \ .
  \label{eq:monopoleMass}
\end{eqnarray}
 It has the form
\begin{equation}
    m=\frac{4 \pi v_\phi}{e} f(\lambda / e^2) \ .
\end{equation}
The solution has been calculated numerically for multiple values, and one finds that for $0.1<\lambda /e^2<10^1$, $f(\lambda /e^2)$ is slowly varying ${\cal O}(1)$ function \cite{tHooft:1974kcl}.

It is well known that monopoles with an initial abundance of roughly one per horizon at formation as computed by Kibble~\cite{Kibble:1976sj}, have no gravitational wave amplitude. Following the Kibble-Zurek mechanism~\cite{Zurek:1985qw}, it was recently shown in ~\cite{Dunsky:2021tih} that a relativistic monopole-bounded string network can emit a pulse of GW before decaying assuming the friction is low. Such a scenario falls under the category of hybrid defects and is discussed in section~\ref{hydef}.

Textures, on the other hand, are defects in three spatial directions \cite{vilenkin2000cosmic}. Sometimes they are also referred to as non-singular soliton, as the scalar field is nowhere topologically constrained to rise from the minimum of the scalar potential. They can arise either from the breaking of a global symmetry or the breaking of a gauge symmetry. After their formation, the global texture releases their energy during their collapse in the form of Goldstone bosons and GWs. It produces a flat spectrum with its peak amplitude controlled by the symmetry-breaking scale $i.e.~\Omega_{\rm{GW}}h^2 \sim v^4$, where $v$ is the breaking scale of global symmetry. The produced gravitational wave spectrum is scale-invariant as the amplitude does not depend on the frequency. There is no dependence on the self-coupling of the symmetry-breaking field. This is because the effective theory of the Goldstone modes, responsible for the creation of the GWs, is a nonlinear $\sigma$ model, and the coupling disappears when the scalar field mode is integrated out \cite{Durrer:1998rw,Figueroa:2012kw,Fenu:2009qf}.  On the other hand, gauged textures also can generate GW once produced from the breaking of local symmetry. The GW generated from the gauged texture is not scale invariant, this is because the gauge field configuration cancels the gradient of the scalar field on the large scale. In this case, there exists a cutoff on the spectrum $f_0\sim 10^{11}~\rm{Hz}$ \cite{Dror:2019syi}. To test the GW generated from the gauged textures one requires future GW experiments that are sensitive to a very high frequency.


\subsection{Hybrid defects}
\label{hydef}


\begin{figure}
    \centering
    \includegraphics[width=0.7\textwidth]{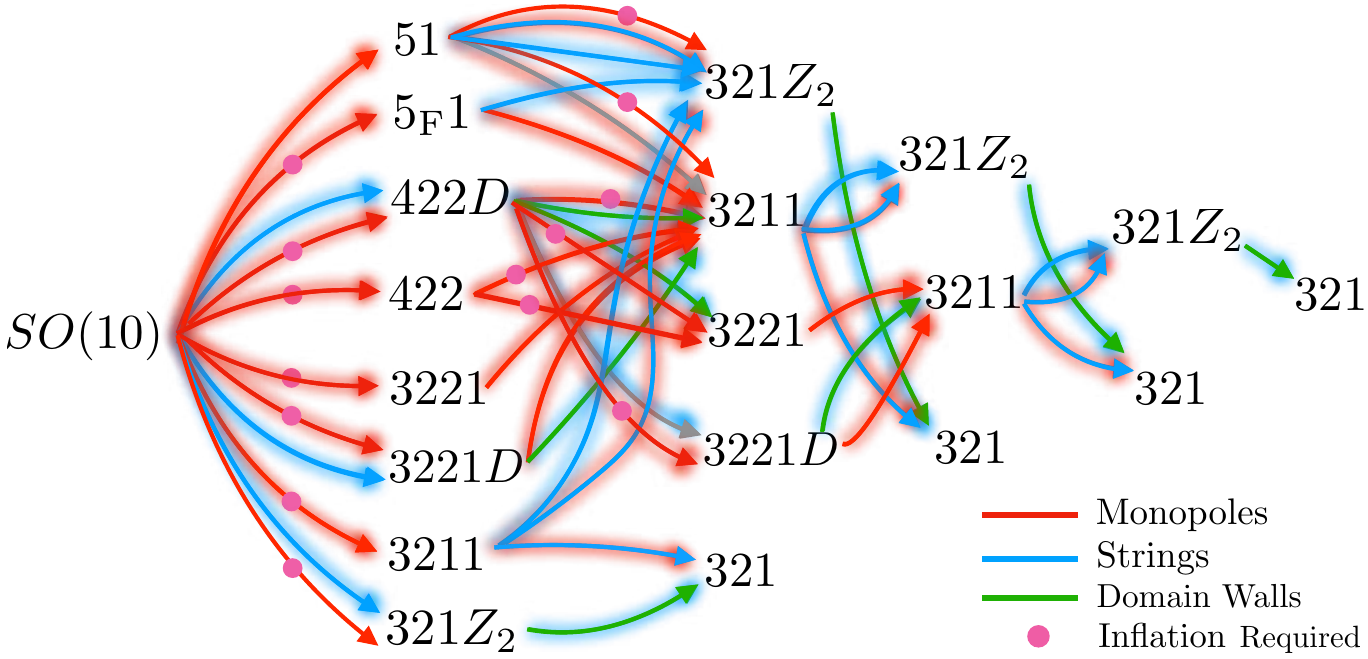}
    \caption{A simple schematic of $SO(10)$ symmetry breaking chain to SM that produces hybrid defects. Arrows with different colors represent different topological defects. The figure is taken from \cite{Dunsky:2021tih} }
    \label{fig:GW_CS}
\end{figure}

Hybrid defects \cite{Langacker:1980kd,Lazarides:1981fv,vilenkin1982cosmological,vilenkin2000cosmic,Dunsky:2021tih,Buchmuller:2023aus} are generated due to a chain of symmetry breaking events, $\mathcal{G}\to\mathcal{H}\to \mathcal{K}$, where the defects of different dimensionalities get clubbed together. For example, a Hybrid effect can be generated where cosmic strings are attached to DWs or monopoles are attached to a string. Such hybrid defects are generally unstable with one defect eating the other via the conversion of its rest mass into the other's kinetic energy. The leftover defect decays further resulting in the production of GW. For the purpose of demonstration, we show a simple schematic of $SO(10)$ symmetry breaking chain to SM that produces hybrid defects in Fig. \ref{fig:GW_CS}, the figure is taken from \cite{Dunsky:2021tih}. We now, briefly summarize some of these hybrid defects and the GW spectrum generated as a result of their collapse. It was proven in ref.~\cite{Dunsky:2021tih} that if a defect involving an $n$ dimensional object appears in the first symmetry breaking step, any defects from the next symmetry breaking step must have a dimension lower than $n$, so we can write the full list of possible hybrid defects as follows, 
\begin{itemize}
    \item  \textbf {String-Monopole network:} The hybrid defects associated with the string-monopole system \cite{Langacker:1980kd,Lazarides:1981fv,vilenkin1982cosmological,Dunsky:2021tih,Buchmuller:2023aus,Lazarides:2023ksx,Lazarides:2022jgr} can be categorized by either monopoles nucleating and eating a string network or strings attaching to and eating a preexisting monopole network. In the first case, a vacuum manifold $\mathcal{H}/\mathcal{K}$ is not simply connected to a full vacuum manifold $\mathcal{G}/\mathcal{K}$. Then $\pi_1(\mathcal{H}/\mathcal{K})\neq I$ results in the formation of topologically unstable strings that manifest themselves by nucleation of magnetic monopole pairs that eat the strings. Such kinds of monopoles must always arise from the earlier phase transition $\mathcal{G}/\mathcal{H}$ (generally before inflation) so that $\pi_2(\mathcal{G}/\mathcal{H})\neq I$. Consider a scenario where in the first phase a spherical symmetry (non-abelian gauge group) is broken to a $U(1)$ leading to a formation of monopoles, and in the second phase when $U(1)$ is broken, strings are formed. If the $U(1)$ symmetry involved in the production of monopoles partially overlaps or coincides with the $U(1)$ symmetry responsible for the string formation, the strings become metastable as monopoles-antimonopole pairs nucleate along the strings (see Fig.~\ref{fig:mono_nuc_1}) through Schwinger nucleation by quantum tunneling. Conversion of the string's rest mass into the monopole's kinetic energy leads to relativistic oscillations of the monopoles before the system collapses via radiating GW and monopole annihilations~\cite{Martin:1996cp,Leblond:2009fq,Dunsky:2021tih,Buchmuller:2021mbb,Afzal:2022vjx,Afzal:2023kqs,Ahmed:2023pjl,Antusch:2023zjk,Lazarides:2022jgr,Lazarides:2023ksx}. 

    \begin{figure}
    \centering
    \includegraphics[width=0.7\textwidth]{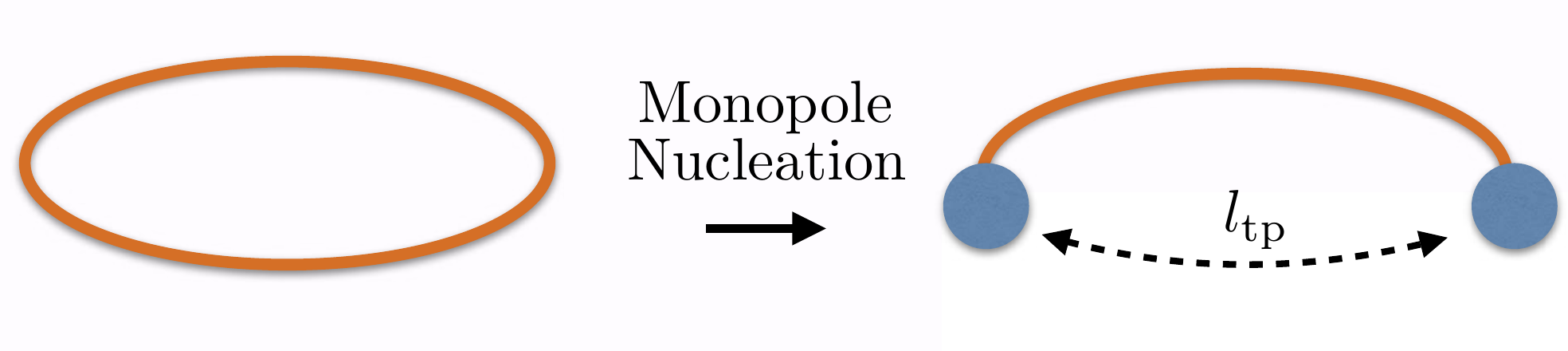}
    \caption{A cartoon showing a monopole nucleation process where a string is eaten and replaced by a monopole-antimonopole pair. The figure is taken from \cite{Dunsky:2021tih} }
    \label{fig:mono_nuc_1}
\end{figure}
   
The total power emitted in the form of GW by string loops before nucleation or by the relativistic monopoles post-nucleation can be calculated using the gravitational power as given in Eq.~\ref{pow_cs}.  The power emitted by the string loops or monopole-bounded strings should be comparable since the kinetic energy of the relativistic monopoles originates from the rest mass of the string. Here, $\Gamma \approx 50$ for string loops before nucleation and $\Gamma \approx 4 \ln \gamma_0^2$ for relativistic monopoles bounded to strings post-nucleation \cite{Leblond:2009fq}, whereas $\gamma_0 \approx 1 + \mu l /2m$ is the monopole Lorentz factor arising from the conversion of string rest mass energy to monopole kinetic energy. The power emitted by gravitational waves reduces the string length (see Eq.~\ref{ell_local})  with a loop lifetime of order $\alpha t_i/\Gamma G \mu$.
The string length ($l$) and harmonic number $n$ is set by the emission frequency, $f' = n/T = 2 n/l$, where $T= l/2$ is the period of any string loop \cite{Weinberg:1972kfs,vilenkin2000cosmic}. The redshifted frequency observed today is, 
\begin{equation}
    f = \frac{2n}{l} \frac{a(t_0)}{a(t)},
\end{equation}
where $t_0$ is the present time.

The number density spectrum of string loops is as before but modified by a term that captures the rate of cutting strings by monopoles, 
\begin{eqnarray}
 \mathcal{N}(l,t)_{\rm Schwinger} &\equiv \frac{dn}{dl}(l,t) \approx \frac{dn}{dt_i}\frac{dt_i}{dl} e^{-\Gamma_m l (t-t_i)}  \nonumber \\
 &=
 \frac{\mathcal{F} C_{\rm eff}(t_i)}{t_i^4 \alpha(\alpha+\Gamma G\mu)}\Bigl(\frac{a(t_i)}{a(t)}\Bigr)^3 e^{-\Gamma_m l (t-t_i)}
, 
\label{eq:stringNumberDensitySpectrum}
\end{eqnarray}
where the tunneling rate per unit string length can be estimated from the bounce action formalism ~\cite{Kibble:1982dd, Preskill:1992ck} and is exponentially suppressed by powers of the ratio of the symmetry breaking scales ~\cite{Leblond:2009fq}
\begin{equation}
    \label{eq:monopoleNucleationRate}
    \Gamma _{m} = \frac{\mu }{2 \pi } {\rm exp} (- \pi \kappa_m ) \ ,
\end{equation}
with  $\kappa_m = m^2/\mu $. Typically, $m\sim v_m$ and $\mu\sim v_\mu^2$ where $v_\mu$ ($v_m$) is the breaking scale that leads to the formation of CS (monopole). To get a GW signal in this situation, one needs to have $v_m\sim v_\mu$ so that $\kappa_m$ is not extremely large otherwise the GW spectrum remains identical to the standard stochastic string spectrum as the string remains stable for a larger value of $\kappa_m$. Going back to Eq. \ref{eq:stringNumberDensitySpectrum}, the exponential factor on the right side of Eq. \ref{eq:stringNumberDensitySpectrum} is the monopole nucleation probability. This effectively cuts off loop production and destroys the loops with a length large enough to nucleate with significant probability. The probability of nucleation is negligible and the string network evolves like a standard, stable string network for $\Gamma_m l (t-t_i) \ll 1$.
Note that this cutoff is time-dependent, 
\begin{equation}
\Gamma_m\, l (t-t_i) =\Gamma_m \frac{2n}{f}\frac{a(t)}{a(t_0)} (t-t_i)\ .
\end{equation}

Following Eq.~\ref{eq:stringNumberDensitySpectrum} one finds that even though the number density of string loops decreases when nucleation occurs (exponential suppression), the number density of string-bounded monopoles still increases.
A loop that nucleates monopoles will continue to nucleate and fragment into many monopole-bounded strings for $l_{\rm max} \gg l_{\rm tp}$, with size of order $l \sim l_{\rm tp} \ll l_{\rm max}$ with $l_{\rm tp}$ denoting the critical length above which it is energetically favorable for the string to form a gap of length $l_{\rm tp}$, separating two monopole endpoints and $l_{\rm max}$ denotes the maximum string size. In this situation, the net energy density eventually deposited into gravitational waves is much less even if the the total energy density in these pieces is comparable to the original energy density of the parent string loop. This can be understood in the following manner, the lifetime of the string-bounded monopoles $\sim \mu l_{\rm tp}/\Gamma G \mu^2$ is much smaller than the parent loop because their power emitted in gravitational waves is similar to pure loops while their mass is much smaller. The net energy density that is transferred into gravitational waves is, to a good approximation, the energy density of the defect at the time of decay. Since these pieces decay quickly and do not redshift $\propto a^3$ for as long as pure string loops, their relative energy density compared to the background at their time of decay is much less than for pure string loops. As a result, the net energy density that goes into gravitational radiation by monopole-bounded string pieces compared to string loops is small, and hence their contribution to the spectrum can be neglected.

The gravitational wave energy density spectrum generated from a network of metastable cosmic strings, including dilution and redshifting due to the expansion history of the Universe can be obtained using Eq. \ref{eqn:energyI}  and replacing, 

   \begin{equation}
     \frac{dn}{dl}(l,t') =  \mathcal{N}(l,t')_{\rm Schwinger},~
    \\~\text{and}
    \frac{dP(l,t')}{df'} = \Gamma G \mu^2 l \, g\left(f \frac{a(t)}{a(t')} l\right),
    \label{GW_ED_hydef I}
\end{equation}
with $t'$ being the emission time, $f' = a(t)/a(t')f$ being the emission frequency, and $f$ is the redshifted frequency observed at time $t$.  The normalized power spectrum for a discrete spectrum is \cite{vilenkin2000cosmic,Sousa_2013}
\begin{equation}
    \label{eq:g(x)Strings}
    g(x) = \sum_n \mathcal{P}_n \delta(x - 2 n) \,
\end{equation}
which ensures the emission frequency is $f' = 2n/l$. $\mathcal{P}_n = n^{-q}/\zeta(q)$ is the fractional power radiated by the $n$th mode of an oscillating string loop where the power spectral index, $q$, is found to be $4/3$ for string loops containing cusps \cite{Auclair:2019wcv,PhysRevD.31.3052}. The GW spectrum from a metastable string can be obtained by including the Schwinger production term, $\mathcal{N}_{\text{Schwinger}}$ in EQ. \ref{omega_CS}, 
    \begin{eqnarray}
        \Omega_{\text{GW}}(f)=\frac{8\pi}{3H_0}(G\mu)^2\sum_{n=1}^\infty\frac{2n}{f}\int_{t_{\text{form}}}^{t_0}dt
        \bigg(\frac{a(t)}{a(t_0)}\bigg)^5\times\mathcal{N}_{\text{Schwinger}}\bigg(l=\frac{2n}{f}\frac{a(t)}{a(t_0)},t\bigg)\mathcal{P}_n. 
    \end{eqnarray}

 \begin{figure}
    \centering
    \includegraphics[scale=0.3]{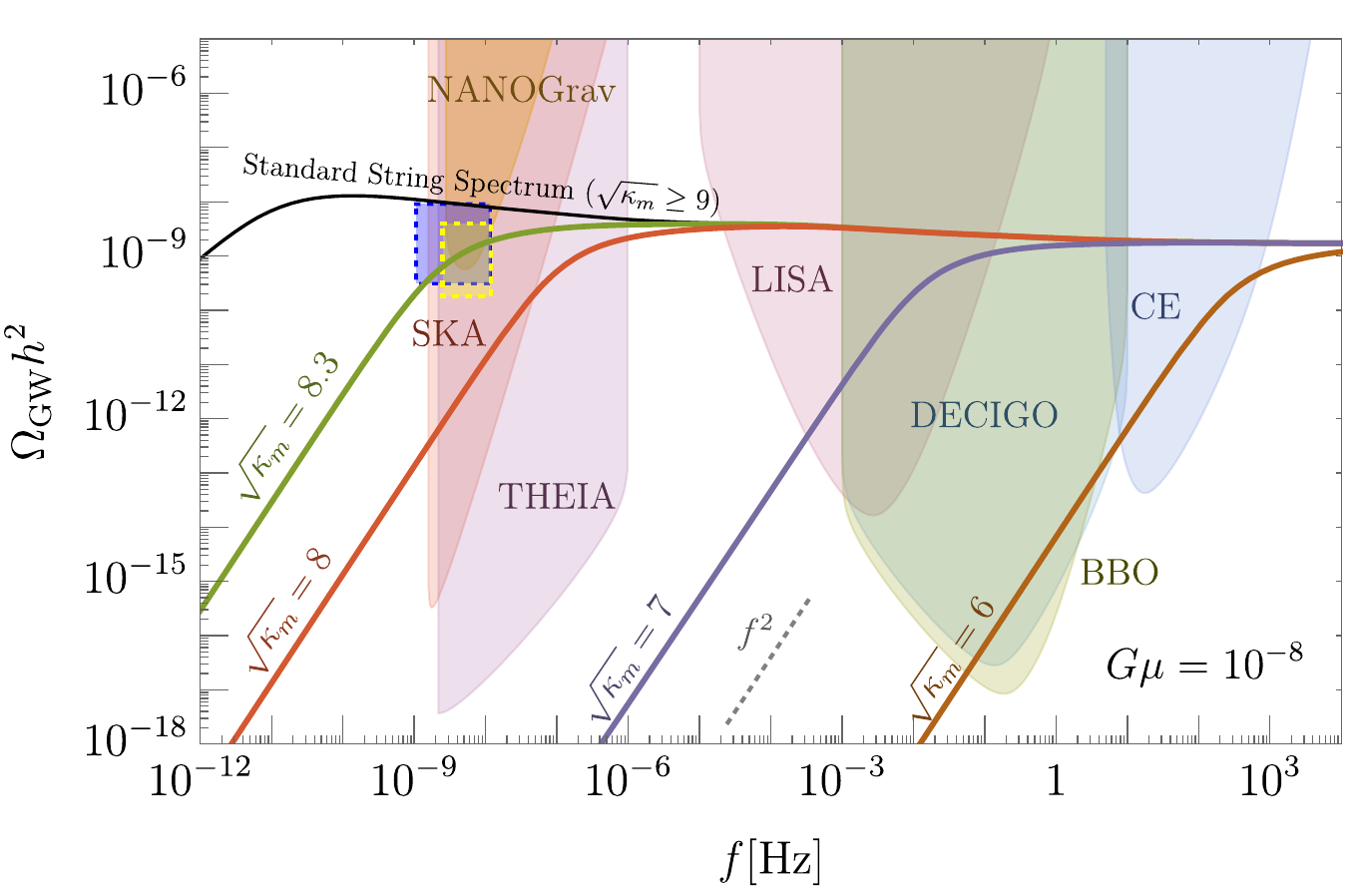}
    \includegraphics[scale=0.33]{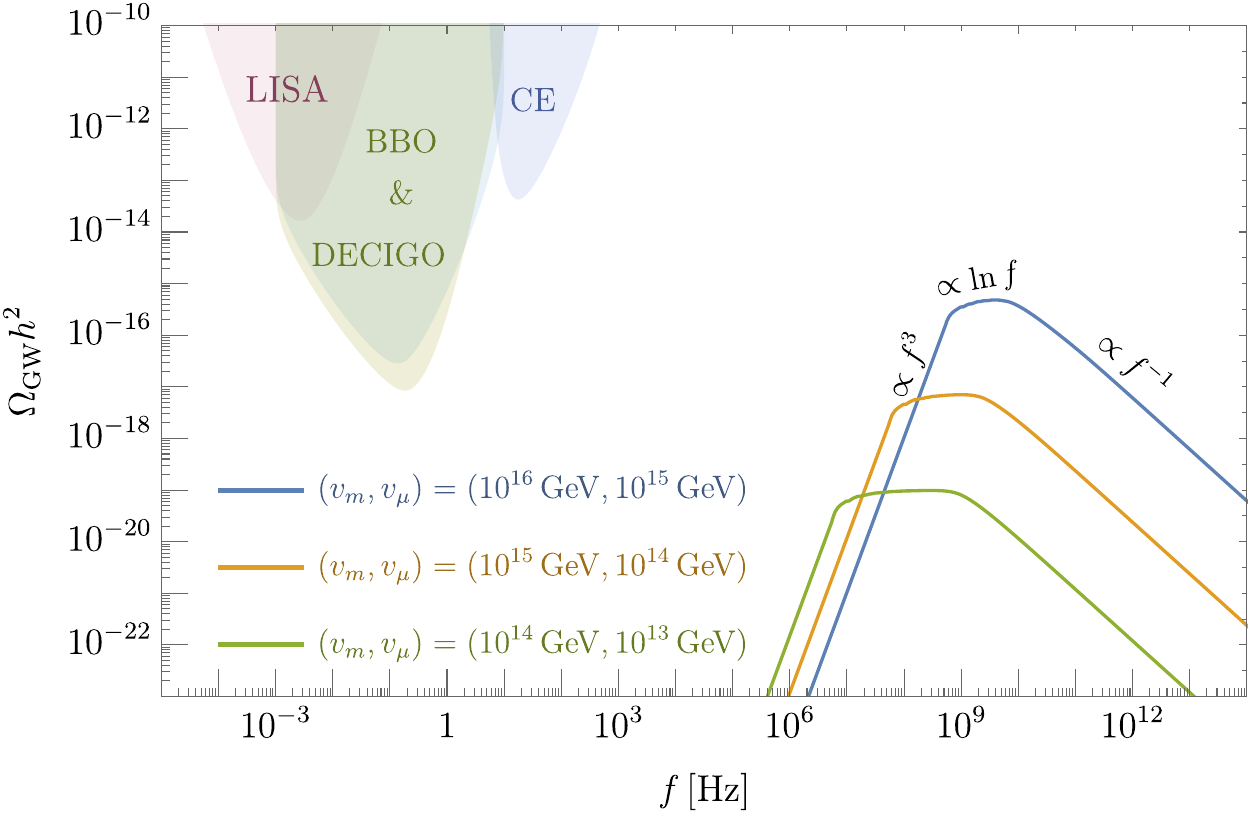}
    \caption{Left panel: GW spectrum emitted by strings that are eaten up by the nucleation of monopoles for a fixed $G\mu=10^{-8}$ and different values of $\kappa_m$. Right Panel:  GW spectrum emitted by monopoles that are eaten up by the string for three different values of combinations $(v_m,v_\mu)=(10^{16}~\rm{GeV},~10^{15}~\rm{GeV})~\rm{in~ blue/upper},~(10^{15}~\rm{GeV},~10^{14}~\rm{GeV})~\rm{in~ orange/middle},~(10^{14}~\rm{GeV},~10^{13}~\rm{GeV})$ in green/lower. The figures are taken from \cite{Dunsky:2021tih} }
    \label{fig:mono_nuc_2}
\end{figure}
The left panel of Fig. \ref{fig:mono_nuc_2} shows the GW spectra emitted by the strings eaten up by the monopole nucleation for different values of $\kappa_m$ and a fixed value of $G\mu=10^{-8}$.  One notices that by increasing the value of $\kappa_m$, the nucleation rate decreases, which corresponds to a cosmologically stable CS. Larger loops, corresponding to lower frequencies and later times of formation, vanish because of Schwinger production of monopole-antimonopole pairs and hence the gravitational wave spectrum is suppressed at low frequencies, scaling as an $f^2$ power law in the infrared.  The slope is therefore easily distinguishable from the one obtained from strings without monopole pair. production
 \begin{figure}
    \centering
    \includegraphics[width=0.7\textwidth]{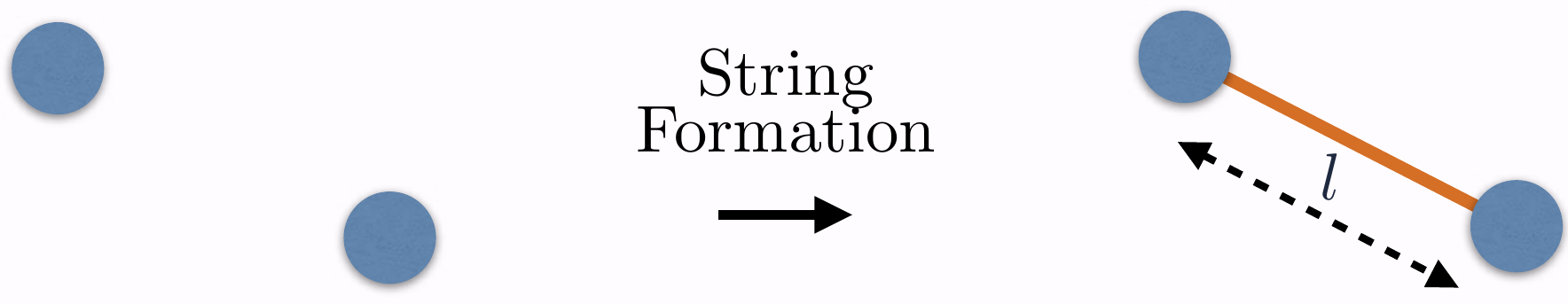}
    \caption{A illustration of monopole squeezing into flux tubes or strings. The figure is taken from \cite{Dunsky:2021tih} }
    \label{fig:mono_nuc_3}
\end{figure}

    In the second case, one can consider a similar breaking chain but with a subtle difference, here the inflation occurs before the formation of monopole and string. The magnetic field of monopoles squeezes into flux tubes (CS) connecting pairs of monopoles and antimonopoles \cite{Lazarides:1981fv,Vilenkin:1984ib}. The strings gradually shrink and lead to the annihilation of monopole and antimonopole pairs. Fig.~\ref{fig:GW_CS}, shows an example of a chain where this gastronomy scenario occurs is $3221\to 3211\to 321$. While $3221\to 3211$ results in the production of monopoles, the second breaking $3211\to 321$ connects the monopoles to strings. 
    
    In a system where a string is attached to a pre-existing monopole \cite{Lazarides:1981fv,Holman:1992xs,Martin:1996ea,vilenkin2000cosmic}, the monopole's initial abundance is considered to be one per horizon at the formation as computed by Kibble~\cite{Kibble:1976sj} and as a result, there is no GW amplitude. In a recent work~\cite{Dunsky:2021tih}, it was shown that if the Kibble-Zurek mechanism~\cite{Zurek:1985qw} is considered and the monopole-antimonopole freeze-out occurring in between monopole and string formation is taken into account, the monopole-antimonopole pairs annihilate in less than a Hubble time with a non-relativistic velocity. This situation again does not generate any GWs. On the other hand, if some of the monopoles with mass $m$ and string scale $v_\mu$ exist, then the monopole-bounded strings can be relativistic. They can generate GWs before decaying if the friction is not large.


Following the Kibble--Zurek mechanism, once formed, the monopole-antimonopole pairs start annihilating.  Their freeze-out abundance at any temperature depends on the critical temperature $T_c$,\cite{Preskill:1979zi}
\begin{eqnarray}
   \label{eq:monopoleNumDensity}
    \frac{n_m(T)}{T^3} = \left[\frac{T_c^3}{n_m(T_c)} + \frac{h^2}{\beta_m} \frac{C M_{Pl}}{m}\left(\frac{m}{T} - \frac{m}{T_c}\right)\right]^{-1}
\end{eqnarray}
With $n_m$ being the number density of monopoles, $C = (8\pi^3 g_*/90)^{-1/2}$,  critical temperature $T_c$  corresponds to a temperature at which the vacua have degenerate energy and

\begin{eqnarray}
   \beta_m \simeq \frac{2 \pi}{9} \sum_i b_i \left(\frac{h e_i}{4\pi}\right)^2 \ln{\Lambda}
\end{eqnarray}
counts the particles of charge $e_i$ in the background plasma that the monopole scatters off. The magnetic coupling is $h = 2\pi/e$, where $e$ is the $U(1)$ gauge coupling constant, $\Lambda \sim {1/(g_* e^4/16 \pi^2)}$ is the ratio of maximum to minimum scattering angles of charged particles in the plasma, and $b_i = 1/2$ for fermions and $1$ for bosons \cite{vilenkin2000cosmic,goldman1981gravitational}. With $e \sim 0.3$ and a comparable number of electromagnetic degrees of freedom as in the Standard Model, $\beta_m \sim 20$. 

At a scale below $v_\mu$, the magnetic fields of the monopoles squeeze into flux tubes. The string length here is set by the typical separation distance between monopoles,
\begin{eqnarray}
    \label{eq:kibbleStringLength}
    l \approx \frac{1}{n_m(T = v_\mu)^{1/3}}. 
\end{eqnarray}
Eq.~\ref{eq:kibbleStringLength} is valid when the correlation length of the string Higgs field denoted by $\xi_\mu$ follows the condition $\xi_\mu \geq l$ \cite{vilenkin2000cosmic}.  Since the string correlation length grows fast enough with time $\propto t^{5/4}$ \cite{Kibble:1976sj,vilenkin2000cosmic}, the string-bounded monopole becomes straightened out within roughly a Hubble time of string formation and ends up with a length close to Eq.~\ref{eq:kibbleStringLength}. For $T_c = v_m \lesssim 10^{17} \, \rm GeV$, $l$ is far below the horizon scale. Consequently, $l$ is not conformally stretched by Hubble expansion and only can decrease with time by energy losses from friction and gravitational waves.

Because the string rest mass is converted to monopole kinetic energy, the initial string length Eq. \ref{eq:kibbleStringLength} determines whether or not the monopoles can potentially move relativistically. Relativistic monopoles can emit a brief pulse of gravitational radiation before annihilating while non-relativistic monopoles will generally not. This can be achieved in a situation where $\beta\sim 1$ and $T^2\sim \mu$ (string tension) (see \cite{Dunsky:2021tih} for details). Here, the pulse of energy density emitted by relativistic monopoles in the form of gravitational waves is well approximated by the product of the power emitted ($P_{\rm GW} = \Gamma G\mu^2$)  by oscillating monopoles connected to strings,  the number density ($n_m(v_\mu)$) of monopoles at $T=v_\mu$ and  the lifetime of the string-bounded monopoles $\tau$, 
\begin{eqnarray}
    \rho_{\rm GW, burst} \approx  n_m(v_\mu)P_{\rm GW} \, \tau.
    \label{GW_burst}
\end{eqnarray}
The GW amplitude from the burst is simply given by Eq. \ref{GW_burst} and redshifted to the spectrum we see today,
\begin{eqnarray}
    \label{eq:monopoleBurstAmplitude}
    \Omega_{\rm GW, burst} &= \frac{ \rho_{\rm GW, burst}}{\rho_c(v_\mu)} \Omega_r \left(\frac{g_{*0}}{g_*(v_\mu)}\right)^{ \scalebox{1.01}{$\frac{1}{3}$}}  
    \nonumber
    \\
    &\approx \frac{30 \pi^2}{g_*(v_\mu)\beta}\Gamma G \mu \left(\delta \frac{m}{M_{\rm Pl}}\right)^{ \scalebox{1.01}{$\frac{2}{3}$}}  , 
\end{eqnarray}
where 
\begin{eqnarray}
   \label{eq:delta}
   \delta = \frac{1}{C\beta_m h^2}\left(\frac{4\pi}{h^2}\right)^2\text{Max}\left\{1, \, \frac{v_\mu}{m} \left(\frac{\beta_m h^2}{4\pi}\right)^2 \right\}.
\end{eqnarray}
and $\rho_c(v_\mu)$ is the critical energy density of the Universe at string formation, which is assumed to be in a radiation-dominated era. 
The amount of monopole-antimonopole annihilation that occurs before string formation at $T = v_\mu$ is characterized by the argument of the `Max' function of Eq. \ref{eq:delta}. For a small $v_\mu/m$, the freeze-out annihilating monopoles finishes before string formation, and the max function of Eq. \ref{eq:delta} saturates at $1$. In this situation, $\delta \approx 10^{-4}\beta_m^{-1}(e/0.5)^4$.

Similarly, the peak frequency for the burst is controlled by the inverse length which in turn is controlled by the symmetry-breaking scale, 
\begin{eqnarray}
    f_{\rm burst} \sim \frac{1}{l}\frac{a(t_\mu)}{a(t_0)}
    &\approx   10^8 \, {\rm Hz} \left(\frac{v_m}{10^{14} \, \rm GeV} \, \frac{\delta}{10^{-4}} \, \frac{106.75}{g_*(v_\mu)}\right)^{ \scalebox{1.01}{$\frac{1}{3}$}}
\end{eqnarray}
where $a(v_\mu)$ and $a(t_0)$ are the scale factors at string formation and today, respectively.

With the understanding of the monopole burst spectrum, one can compute $\Omega_{\rm GW}$ for a situation where $\beta_m \sim 1$. Applying the chain rule, one can replace $\frac{dn(l,t')}{dl}$ in Eq.\ref{GW_ED_hydef I} with 
\begin{equation}
    \frac{dn}{dl}(l,t') = \frac{dn}{dt_k}\frac{dt_k}{dl}.
\end{equation}

where primed coordinates refer to the time of emission and unprimed refer to the present so that gravitational waves emitted from the monopoles at time $t'$ with frequency $f'$ will be observed today with frequency $f = f' a(t')/a(t)$. Further, $t_k$ is the formation time of monopole-bounded strings of length $l(t_k)$,
\begin{eqnarray}
    \frac{dn}{dt_k} \simeq n_m(t_k)\delta(t_k - t_\mu) \left(\frac{a(t_k)}{a(t)} \right)^3
\end{eqnarray}
is the string-bounded monopole production rate, which is localized in time to the string formation time, $t_\mu \simeq C M_{Pl}/v_\mu^2 $. $dt_k/dl$ is found by noting that the energy lost by relativistic monopoles separated by a string of length $l$ which is given by by the rate of change in string length and monopole mass, 
\begin{eqnarray}
   \frac{dE}{dt} = \frac{d}{dt}(\mu l + 2m) \approx -\beta_m v^2 \mu \,.
\end{eqnarray}
A relativistic monopole can be achieved when $\mu l \gg 2m$ \cite{Dunsky:2021tih}. As a result,  monopole-bounded strings that form at time $t_k$ with initial size $l(t_k)$ decrease in length according to 
\begin{eqnarray}
   l(t) \simeq l(t_k) - \beta_m v^2 (t - t_k)
\end{eqnarray}
so that
\begin{eqnarray}
   \frac{dt_k}{dl} \simeq \frac{1}{\beta_m v^2} \approx 1 .
\end{eqnarray}
The normalized power spectrum for a discrete spectrum can be decomposed into the product of the delta function and fractional power radiated by the $n^{th}$ mode of an oscillating string loop, 
\begin{eqnarray}
   \label{eq:g(x)Monopoles}
    g(x) = \sum_n \mathcal{P}_n \delta(x - n \xi) \qquad \xi \equiv \frac{l}{T}
\end{eqnarray}
ensures the emission frequency of the $n$th harmonic is $f' = n/T$, where $T$ is the oscillation period of the monopoles. For pure string loops, $T = l/2$ ($\xi = 2$, reducing to Eq.~\ref{eq:g(x)Strings}), whereas for monopoles connected to strings, $T = 2m/\mu + l \simeq l$ \cite{Leblond:2009fq,Martin:1996cp} ($\xi \approx 1$).  $\mathcal{P}_n \approx n^{-1}$ is found \cite{Leblond:2009fq,Martin:1996cp} for harmonics up to $n \approx \gamma_0^2$, where $\gamma_0 \simeq (1 + \mu l/2m)$, is the Lorentz factor of the monopoles. For $n > \gamma_0^2$, $P_n \propto n^{-2}$,$\Gamma \approx 4 \ln \gamma_0^2$.


    In this situation, a string eats the monopole, and the emitted  GW spectrum looks like,

    \begin{eqnarray}
        \Omega_{\text{GW}}=\sum_n\frac{8\pi(G\mu)^2}{3H_0^2}\bigg(\frac{a(t_\mu-l_*)}{a(t-0)}\bigg)^5\bigg(\frac{a(t_\mu)}{a(t_\mu-l_*)}\bigg)^3\times\Gamma\mathcal{P}_n\frac{\zeta n}{f}\frac{n_m(t_\mu)}{\beta_m v^2}
    \end{eqnarray}
    where
    \begin{eqnarray}
        l_*=\frac{\frac{\zeta_n n}{f}\frac{a(t_\mu)}{a(t_0)}-n_m(t_\mu)^{-1/3}}{\beta_m v^2}
    \end{eqnarray}
\noindent The GW spectrum for this scenario is shown in the right panel of Fig. \ref{fig:mono_nuc_2}. Here, at high frequencies, the spectral shape goes as $f^{-1}$, prior to which a small plateau is observed where the spectrum goes as $\text{ln}f$ and at low frequencies decays as $f^3$. 
    \item {\bf String-Wall network:} Analogous to a string-monopole system, the hybrid defects associated with a string-wall network \cite{Dunsky:2021tih,Maji:2023fba,Lazarides:2023ksx,Gelmini:2021yzu,Gelmini:2022nim,Gelmini:2023ngs,Maji:2023fba,Li:2023gil} can also be categorized as either the destruction of a domain wall network by the nucleation of string-bounded holes on the wall that expand and eat the wall, or the collapse and decay of a string-bounded wall network by walls that eat the strings. If a vacuum manifold $\mathcal{H}/\mathcal{K}$ is disconnected but $\mathcal{G}/\mathcal{K}$ is connected, DWs are formed at transition from  $\mathcal{H}\to\mathcal{K}$ and $\Pi_0(\mathcal{H}/\mathcal{K})\neq \mathcal{I}$. The DWs of this nature are topologically unstable as $\Pi_0(\mathcal{G}/\mathcal{K})=\mathcal{I}$. This instability is caused by the nucleation of string-bounded holes on the wall that expand and destroy the wall completely. As a consequence of Schwinger nucleation, a wall (formed after inflation) gets bounded by a string (formed before inflation) if both defects are related to the same discrete symmetry. Here, the wall's rest mass energy is transformed into the string's kinetic energy leading to its rapid expansion. This rapid expansion causes the collapse of the string wall system and produces GWs in the process.

\begin{figure}
    \centering
    \includegraphics[width=0.7\textwidth]{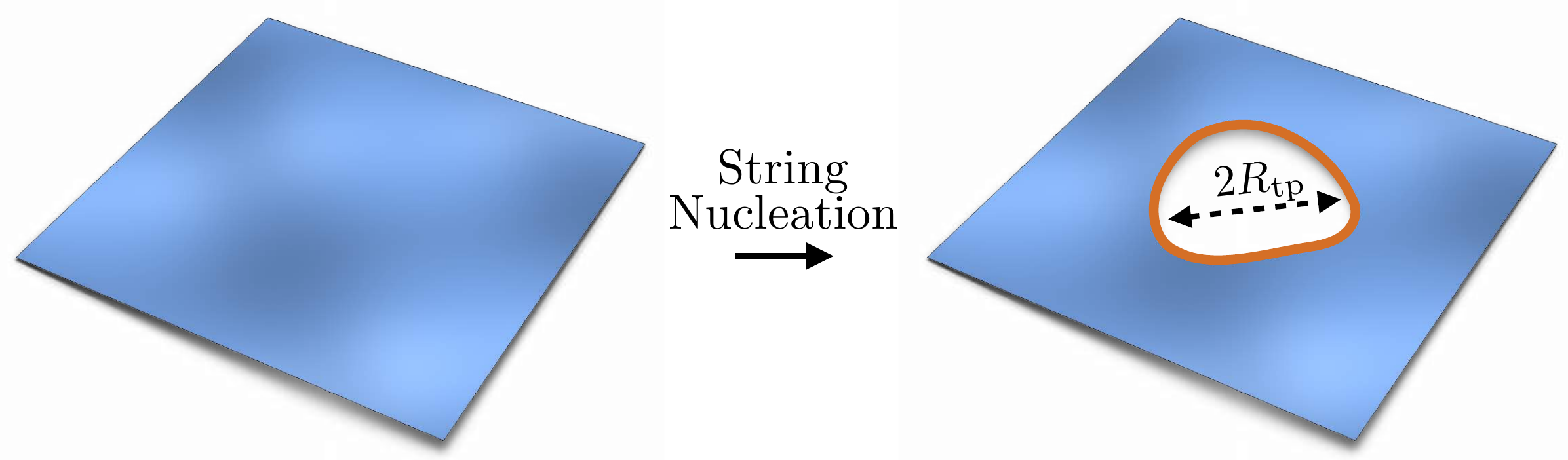}
    \includegraphics[width=0.7\textwidth]{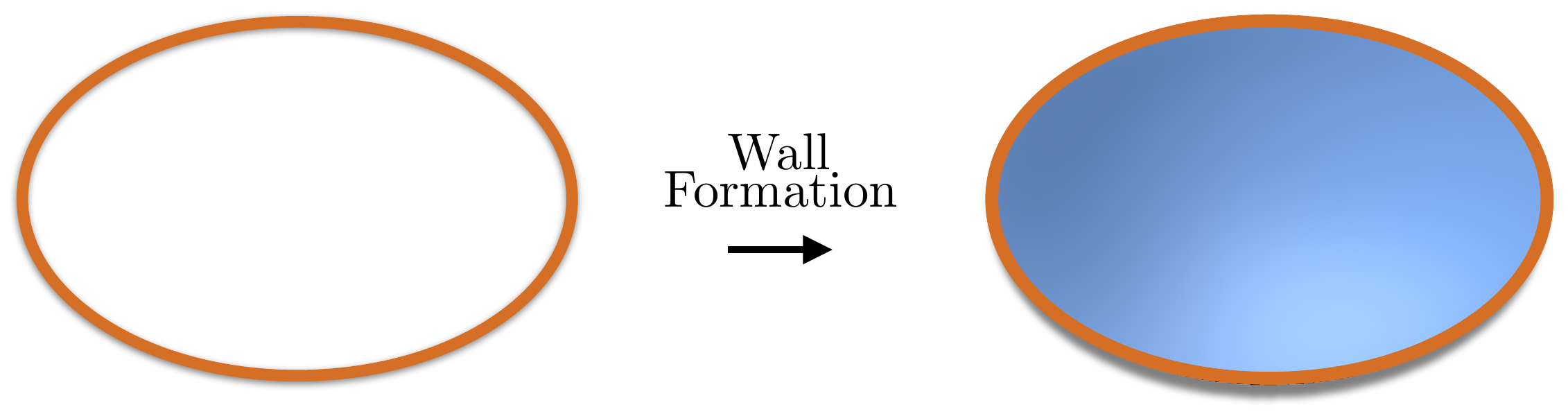}
    \caption{A cartoon showing a string nucleation process in the top panel and wall formation in the bottom panel. The figure is taken from \cite{Dunsky:2021tih} }
    \label{fig:mono_nuc}
\end{figure}

To understand this, one can solve the Boltzmann equation \cite{Dunsky:2021tih} for the total energy density, $\rho_{\rm GW}$ in the gravitational wave background, 
\begin{eqnarray}
    \label{eq:GWNucleation}
    \frac{d\rho_{\rm GW}}{dt} + 4H \rho_{\rm GW}  = \mathcal{A}\mathcal{B}\frac{G \sigma^2}{t}\theta(t_\Gamma - t) - x \frac{d \rho_{\rm DW}}{dt} \theta(t - t_\Gamma).
\end{eqnarray}
While the first term on the RHS of Eq. \ref{eq:GWNucleation} corresponds to the rate of energy density lost into GWs by DWs at time $t$ in the scaling regime and prior to the nucleation, the second term shows the energy transferred into GWs after string began nucleating and eating the wall with $x \in [0,1]$ denoting an efficiency parameter characterizing the fraction of the energy density of the wall transferred into gravitational waves at time, 
\begin{eqnarray}
    t_\Gamma \sim \frac{1}{\sigma A} e^{S_E} \sim \frac{1}{\sigma^{1/3}}\exp{\frac{16 \pi \kappa_s}{9}}.
\end{eqnarray}
Here, we take the wall area, $A$ at time $t_{\Gamma}$ to be $\sim t_{\Gamma}^2$ following the scaling regime and $\kappa_s=\mu^3/\sigma^3$ where $\sigma$ denotes the surface tension of the DW that depends on the breaking scale $v_\sigma$ that leads to the formation of the DW (see section \ref{subsec:DW} for details). Also one needs to have $\mu^3\sim\sigma^3$, otherwise, the DW remains long-lived.  When the strings begin nucleating at $t_\Gamma$, they quickly expand from an initial radius $R_{\rm tp} = 2 R_c$ with $R_c\equiv \mu/\sigma$ according to \cite{Dunsky:2021tih}
\begin{eqnarray}
   \label{eq:stringExpansionWall}
    R_s(t) = \sqrt{4 R_c^2 + (t-t_{\Gamma})^2} .
\end{eqnarray}
In the above, $R_{\rm tp}$ denotes the critical string radius that is obtained when the balance between string creation and domain wall destruction is balanced. Above this, it is energetically favorable for the string to nucleate and continue expanding and consuming the wall as shown in the top panel of Fig. \ref{fig:mono_nuc}. 
As a result, the strings rapidly accelerate to near the speed of light as they `eat' the wall. The increase in string kinetic energy arises from the devoured wall mass. Thus, shortly after $t_\Gamma$, most of the energy density of the wall is transferred to strings and string kinetic energy.

Assuming a conservative limit $x=0$ \cite{Dunsky:2021tih}, the solution to Eq.~\ref{eq:GWNucleation} during an era with scale factor expansion $a(t) \propto t^\nu$ is then

\begin{eqnarray}
        \rho_{\rm GW}(t) &= \Bigg{\{} \begin{array}{cc}
          \mathcal{A}\mathcal{B}\frac{ G \sigma^2}{4 \nu}\left(1 - \left(\frac{t_{\rm scl}}{t}\right)^{4\nu}\right) &\; t \leq t_{\Gamma}  \\
      \left(\rho_{\rm GW}(t_\Gamma) +  x  \mathcal{A}\frac{\sigma}{{t_\Gamma}}\right) \left(\frac{a(t_{\Gamma})}{a(t)}\right)^{4} &\; t > t_{\Gamma} 
    \end{array} 
    \label{eq:gwEnergyDensityNucleation}
    \end{eqnarray}

Eq.~\ref{eq:gwEnergyDensityNucleation} demonstrates that the gravitational wave energy density background quickly asymptotes to a constant value after reaching scaling at time $t_{\rm scl}$ and to a maximum at the nucleation time $t_\Gamma$. We thus expect a peak in the gravitational wave amplitude of approximately \cite{Dunsky:2021tih},

    \begin{eqnarray}
    \Omega_{\text{GW, max}}=\frac{16\pi}{3}[(G\sigma t_\Gamma)^2+2xG\sigma t_\Gamma]\Omega_r\bigg(\frac{g_{*0}}{g_{*}(t_\Gamma)}\bigg)^{1/3}.
     \label{eq:peakAmplitudeNucleation}
\end{eqnarray}

where we take $t_\Gamma > t_{\rm scl}$, $\mathcal{A} = \mathcal{B} = 1$, and a radiation dominated background at the time of decay with $\nu = \frac{1}{2}$. $\Omega_r = 9.038 \times 10^{-5}$ is the critical energy in radiation today \cite{Aghanim:2018eyx}.

The first term in Eq. \ref{eq:peakAmplitudeNucleation}, the contribution to the peak amplitude from gravitational waves emitted prior to nucleation, agrees well with the numerical results of \cite{Hiramatsu:2013qaa} if $t_\Gamma$ maps to the decay time of unstable walls in the authors' simulations. The second term in Eq. \ref{eq:peakAmplitudeNucleation}, the contribution to the peak amplitude from GW emitted after nucleation, has not been considered in numerical simulations. The post-nucleation contribution dominates the pre-nucleation contribution if $x \gtrsim G \sigma t_{\Gamma}$, which may be important for short-lived walls. The complex dynamics of string collisions during the nucleation phase motivate further numerical simulations. 

The frequency dependence on the gravitational wave amplitude may be extracted from numerical simulations of domain walls in the scaling regime. The form of the spectrum was found in \cite{Hiramatsu:2013qaa} to scale as
 \begin{eqnarray}
        \Omega_{\text{GW}}(f)\simeq \Omega_{\text{GW, max}}\Bigg{\{} \begin{array}{cc}
         \left( \frac{f}{f_p} \right)^{-1} & f> f_p \\
     \left( \frac{f}{f_p} \right)^{3}    & f < f_p
    \end{array} 
    \end{eqnarray}

where 
\begin{eqnarray}
    f_{\rm p} \sim \frac{1}{t_{\Gamma}}\frac{a(t_\Gamma)}{a(t_0)}
\end{eqnarray}
The infrared $f^3$ dependence for $f < f_{\rm p}$ arises from causality arguments for an instantly decaying source \cite{Caprini:2009fx}. This behavior is shown in Fig. \ref{fig:nucleationSpectrum}. 

 \begin{figure}
    \centering
    \includegraphics[width=0.48\textwidth]{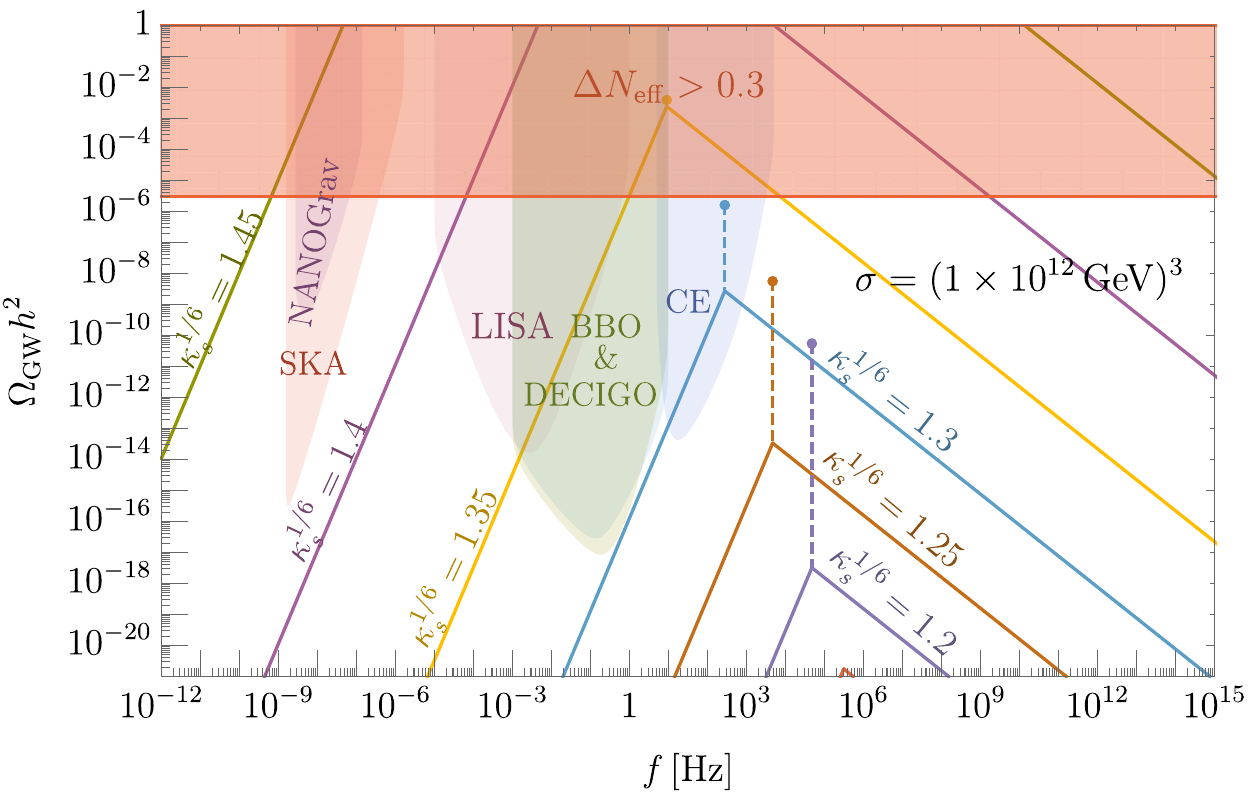}
    \includegraphics[width=0.48\textwidth]{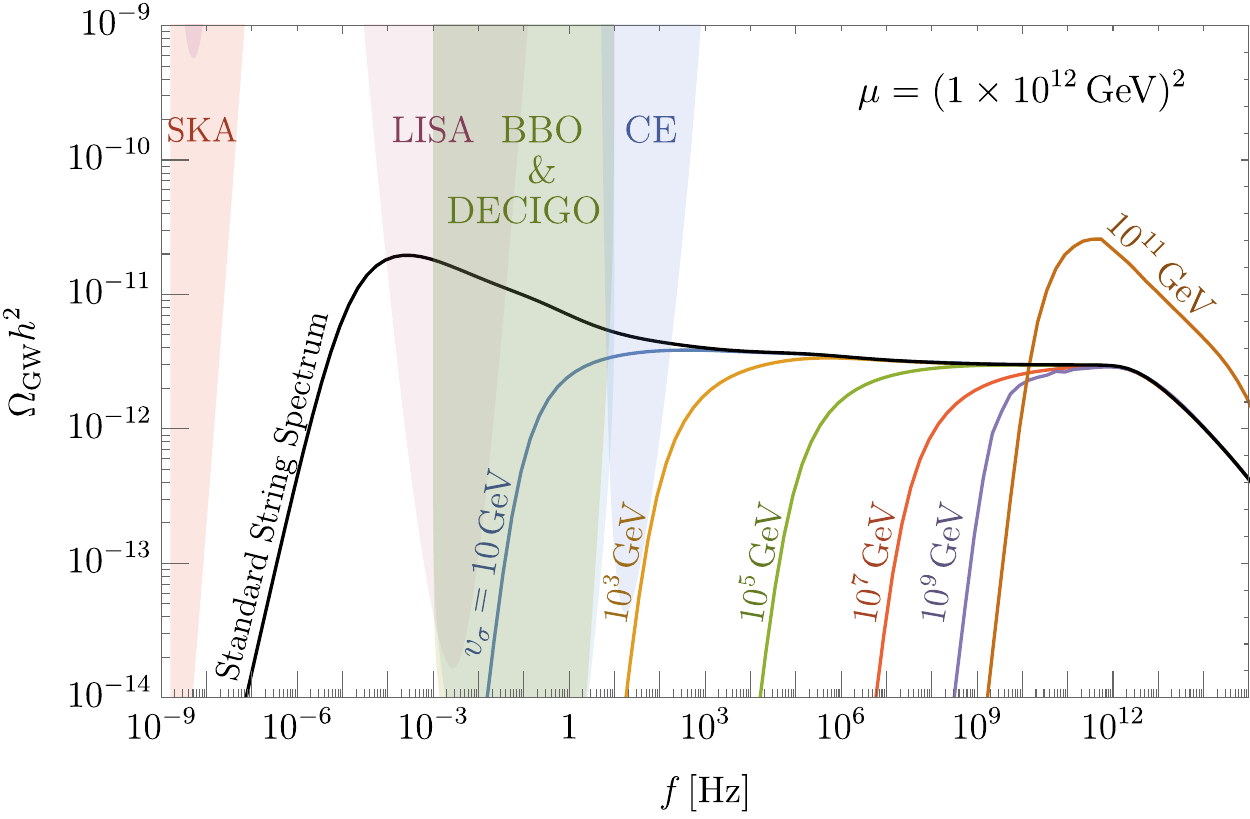}
    \caption{Left Panel: GW spectrum emitted by domain walls that are eaten by nucleation of strings for fixed $\sigma^{1/3} = 10^{12}\, {\rm GeV}$. Right Panel: GW spectrum emitted by strings that are eaten by domain walls for fixed $\sqrt{\mu} = 10^{12}\, {\rm GeV}$. The figure is taken from \cite{Dunsky:2021tih}.}
    \label{fig:nucleationSpectrum}
\end{figure}



 Finally, consider the case of a string wall network where DWs get attached to pre-existing strings \cite{Dunsky:2021tih}. Here, both strings and DWs are formed after inflation. In this scenario, the strings are formed before the wall formations and once the walls are in the picture they fill the spaces between the strings. After the DW-bounded string system enters the horizon, the system oscillates with constant amplitude resulting in the production of the GWs.

Considering a string boundary of a circular wall with radius $R$, it was shown that while the wall tension dominates when the radius $R>>R_c$ with $R_c\equiv\mu/\sigma$, the string dynamics reduces to the pure string loop motion for $R<<R_c$ where the string dominates the dynamics and the power is independent of loop size which is in agreement with the pure string case. Using the VOS model of an infinite string-wall network they showed that the walls pull their attached strings into the horizon when the curvature radius of the hybrid network grows above $R_c$. Once this network is inside the horizon, it oscillates and radiates GWs. As expected, for $R<<R_c$ (pure string limit), the power emitted in the GWs goes as $P_{\rm GW} \propto G\mu^2$. On the other hand, for $R >> R_c$, the domain wall dominates the dynamics and the power deviates from the pure string case, increasing quadratically
with $R/R_c$. Since $R_c \equiv \mu/\sigma$, this is equivalent to $P_{\rm GW} \propto G \sigma^2 R^2 \propto G \sigma M_{\rm DW}$, in agreement with the quadrupole formula expectation for gravitational wave emission from domain walls.

With the idea of GW power emitted by a string-bounded domain wall system (see \cite{Dunsky:2021tih} for detail) one can now calculate the gravitational wave spectrum from a network of circular string-bounded walls. Here we briefly discuss the analytical estimate of the expected amplitude and frequency of the spectrum. For the numerical computation, we refer readers to \cite{Dunsky:2021tih}.

To start with, first consider a pure string loop without walls that form at time $t_k$ with initial length $l_k = \alpha t_k$, where $\alpha \simeq 0.1$. Once these loops are inside the horizon, they oscillate, and their energy density redshifts $\propto a^{-3}$ because their energy $E = \mu l$ is constant in the flat space limit. The loops radiate 
 GW with power $P_{\rm GW} = \Gamma_s G \mu^2$ as discussed above, where $\Gamma_s \approx 50$, and eventually decay from gravitational radiation at time
\begin{eqnarray}
    t_{\Gamma} \approx \frac{\mu l_k}{\Gamma_s G \mu^2} \quad (\text{Pure string loop lifetime}).
    \label{life_time}
\end{eqnarray}
When the pure string loops form and decay in a radiation-dominated era, their energy density at decay is
\begin{eqnarray}
    \rho(t_\Gamma) \approx \mu l_k n(t_k)\left(\frac{t_k}{t_\Gamma}\right)^{3/2}
    \label{rho_tau}
\end{eqnarray}
with $n(t_k) \approx  \frac{1}{3}\frac{\mathcal{F}C_{\rm eff}}{\alpha t_k^3}$ being the initial number density of loops of size $l_k$ that break off from the infinite string network in a scaling regime \cite{Cui:2018rwi,Gouttenoire:2019kij,Sousa_2013}, $\mathcal{F} \approx 0.1$ \cite{Blanco-Pillado:2013qja} and $C_{\rm eff} \approx 5.4$ in a radiation dominated era \cite{Cui:2017ufi, Blasi:2020wpy}. 
Following Eq. \ref{life_time} and Eq. \ref{rho_tau}, the gravitational wave amplitude coming from these pure string loops is approximately
\begin{eqnarray}
    \label{eq:pureStringEstimation}
    \Omega_{\rm GW}^{(\rm str)} &\approx& \frac{\rho(t_\Gamma)}{\rho_c (t_{\Gamma})} \Omega_{\rm r} \left(\frac{g_{*0}}{g_*(t_{\Gamma})}\right)^{ \scalebox{1.01}{$\frac{1}{3}$}}  \\
    &=& \frac{32 \pi}{9}\mathcal{F}C_{\rm eff}\sqrt{\frac{\alpha G \mu}{\Gamma_s}}\Omega_{\rm r} \left(\frac{g_{*0}}{g_*(t_{\Gamma})}\right)^{ \scalebox{1.01}{$\frac{1}{3}$}}  \nonumber
\end{eqnarray}
where $\rho_c(t_\Gamma)$ is the critical energy density of the Universe at $t_\Gamma$. 

Before reaching $t = t_*$, the dynamics of the string-bounded walls are dominated by the strings and the spectrum must be approximately that of a pure string spectrum with $\Omega_{\rm GW}$ given approximately by Eq.~\ref{eq:pureStringEstimation}, independent of frequency. Now at time $t_k = t_*$, let us consider the formation of a near circular string-bounded wall with initial circumference $l_k = \alpha t_k$. While for $l_k \lesssim 2 \pi R_c$, the power emitted and total mass of the system is effectively identical to the pure string case and $\Omega_{\rm GW}$ remains the same as Eq. \ref{eq:pureStringEstimation}, for $l_k \gtrsim 2\pi R_c$, the power emitted and mass of the system is dominated by the DW contribution of the wall-string piece. In this situation, the wall-bounded string decays while radiating the GW at the time
\begin{eqnarray}
    t_{\Gamma} \approx \frac{\sigma l_k^2/4\pi}{\Gamma(l_k) G \mu^2} \approx \frac{1}{G \sigma}
\end{eqnarray}
For the wall-bounded strings forming and decaying in a radiation-dominated era, their energy density at decay is 
\begin{eqnarray}
    \rho(t_\Gamma) \approx \frac{\sigma  l_k^2}{4\pi} n(t_k)\left(\frac{t_k}{t_\Gamma}\right)^{3/2}
\end{eqnarray}
where $n(t_k) \approx  \frac{1}{3}\frac{\mathcal{F}C_{\rm eff}}{\alpha t_k^3}$ follows from the infinite string-wall network being in the scaling regime with $\mathcal{F}$ and $C_{\rm eff}$ expected to be similar to the pure string values right before the infinite network collapses at $t_*$. Hence the GW amplitude in this scenario is controlled by the energy density of the string-wall system at decay,  
\begin{eqnarray}
    \label{eq:wallStringEstimation}
    \Omega_{\rm GW} &\approx& \frac{\rho(t_\Gamma)}{\rho_c (t_{\Gamma})} \Omega_{r} \left(\frac{g_{*0}}{g_*(t_{\Gamma})}\right)^{1/3} \\
    &=& \frac{8}{9}\mathcal{F}C_{\rm eff} \alpha \sqrt{G \sigma t_k}\Omega_{r}\Big(\frac{g_{*0}}{g_*(t_{\Gamma})}\Big)^{ \scalebox{1.01}{$\frac{1}{3}$}} \,. \nonumber
\end{eqnarray}
The largest amplitude of Eq. \ref{eq:wallStringEstimation} occurs at the latest formation time $t_k$, which is $t_*$, the time of the collapse of the infinite network. As a result of this, a `bump' relative to the flat string amplitude appears if
\begin{eqnarray}
    \label{eq:omegaRatio}
    \frac{\Omega_{\rm GW}}{\Omega_{\rm GW}^{(\rm str)}} \approx \frac{1}{4\pi}\sqrt{\frac{\Gamma_s \alpha t_*}{R_c}} \approx 0.2 \left(\frac{\alpha}{0.1}\right)^{ \scalebox{1.01}{$\frac{1}{2}$}} \left(\frac{\Gamma_s}{50}\right)^{ \scalebox{1.01}{$\frac{1}{2}$}} \left(\frac{t_*}{R_c}\right)^{ \scalebox{1.01}{$\frac{1}{2}$}}
\end{eqnarray}
is greater than $1$ and at a frequency 
\begin{eqnarray}
    \label{eq:peakFreqEstimation}
    f_{\rm peak} \sim \frac{1}{l_k} \frac{a(t_\Gamma)}{a(t_0)}.
\end{eqnarray}
since the walls remain the same size once inside the horizon and dominantly emit at the frequency of the harmonic, $f_{\rm emit} \sim l_k^{-1}$. Here, $l_k \approx \alpha t_*$.

The estimation of Eq.~\ref{eq:omegaRatio} indicates that if $t_* >> R_c$, then $\Omega_{\rm GW}$ features a `bump' relative to the flat string spectrum before decaying. 
In this situation, the walls are massive and large enough to live much longer than the pure string of the same size. Due to this, they have an enhanced energy density before decaying in comparison to the shorter-lived pure string loops. In a situation with $t_* \approx R_c$ no such enhancement is observed because string-bounded walls are smaller in size and decay quickly.

With the qualitative features of the spectrum understood, we turn to a numerical computation of $\Omega_{\rm GW}$. The GW spectrum obtained from the numerical computation of \cite{Dunsky:2021tih} for this system is shown in the right panel of Fig. \ref{fig:nucleationSpectrum} for a fixed value of $\sqrt{\mu}=10^{12}$ GeV. It is also interesting to point out that, the spectrum here decays as $f^3$ (due to causality) rather than $f^2$ decay signal which was observed in the situation with the monopoles eating the strings.
\end{itemize}
\subsection{Applications of gravitational waves from topological defects}

Cosmic strings as discussed earlier, can play a vital role in understanding and testing the evolutionary history of the Universe. A string network continuously emits gravitational wave power for the entirety of its existence, and as the string network enters the scaling regime, the fraction of energy density in gravitational waves should remain constant for all frequencies unless there is some change in the energy budget of the Universe. GW signals emitted from the CS can therefore provide an unprecedented window into the evolution of the very early universe before BBN and the CMB. On the other hand, if a network of CS that was formed before or shortly after the start of inflation can regrow after getting diluted and generate GW signals that can be observed today. In this section, we will first cover how strings can be witnesses to a period of inflation before covering how they can be witnesses to dramatic changes in the Universe such as an early period of matter domination or kination. Finally, we will cover how topological defects can provide information about high-scale physics from leptogenesis to quantum gravity.

As is well known, primordial cosmological inflation dilutes all the relics that were generated before its commencement to a negligible level, hence one expects that a CS if produced before the onset of inflation will also be diluted. A recent study~\cite{Cui:2019kkd}, has shown that a network of cosmic strings diluted by inflation can replenish to a level that can result in the generation of GW signals that can be probed by the present and future GW detectors. This is due to the energy density diluting slower than radiation. Unlike the signals that result from an undiluted CS which are of the form of SGWB, the signal resulting from the diluted CS can be distinctive bursts of GWs. 

To understand this, we assume a constant Hubble parameter $H_I=V_I/3M_{\text{Pl}}^2$ during inflation that describes the inflationary energy density from the initial time $t_I$ to end time $t_E$ with $V_I=M^4$ ($M\leq 10^{16}$ GeV~\cite{Akrami:2018odb}), the cosmological scale factor grows as $a(t)\propto e^{H_It}$. Once the inflation is completed, the radiation-dominated era is initiated with temperature $T_{RH}\leq(30/\pi^2g_*)^{1/4}M$. Under the assumption that the temperature during the reheating remains below the symmetry breaking scale, the dilution of the pre-existing CS by thermal processes like symmetry restoration can be avoided.
Once again following the VOS model used for describing the horizon-length long strings during 
and after inflation~\cite{Guedes:2018afo}, one can study the evolution of the correlation length parameter $L$ and the velocity parameter $\bar{v}$ of a long string~\cite{Martins:1996jp,Martins:2000cs}. Starting with the initial conditions,
\begin{equation}
    L(t_F)\equiv L_F=\frac{1}{\xi H_I},
\end{equation}
where $\xi^2$ corresponds approximately to the number of long strings within the Hubble
volume at time $t_F$. $L(t)$ soon reaches an attractor solution during  inflation after $t_F$,
\begin{equation}
    L(t)=L_Fe^{H_I(t-t_F)},~ \bar{v}(t)=\frac{2\sqrt{2}}{\pi}\frac{1}{H_IL(t)}.
\end{equation}
For  $HL>>1$ and $\bar{v}<<1$, the above solution corresponds to the dilution of the long string network. As a result of this, the evolution of the string network after inflation
takes a very simple form while $HL>>1$ with $(L/a)$ approximately constant.  It follows that $HL$ decreases after inflation, corresponding to the gradual regrowth of the
string network. If the network is to produce a potentially observable signal in GWs, at least a few strings are needed  within our current Hubble volume corresponding to
$HL \leq 1$ today. The conditions under which there is enough string regrowth for $HL\to 1$ while also maintaining a sufficient amount of inflation can be obtained by comparing the evolution of $L$ before scaling to that of the curvature radius
$R=1/(H\sqrt{|\Omega-1|})$ which evolves in precisely the same way, independently of the details of inflation or reheating. The total number of inflationary e-foldings can then be written as $N_{tot}=H_I(t_E-t_I)\equiv N_F+\Delta N$ with $\Delta N\geq 0$ being the number of e-foldings between $t_I$ and $t_F$. Note that $\Delta N=0$ corresponds to the string forming before or at the start of inflation. Applying the curvature limit $|\Omega_0-1|=0.0007\pm0.0037$~\cite{Aghanim:2018eyx}, one finds,
\begin{equation}
    \Delta N + \ln\xi \geq
2.7  + \frac{1}{2}\ln(|\Omega_{I}-1|) 
+ \frac{1}{2}\ln\left[
\Omega_{\Lambda}(1+\tilde{z})^{-2}
+ \Omega_m(1+\tilde{z})
+ \Omega_r(1+\tilde{z})^2
\right] 
\end{equation}
where $\tilde{z}$ is the redshift at which $HL\to 1$, $\Omega_a$ are the fractional
energy densities in dark energy, matter, and radiation relative to critical today, $|\Omega_I-1|$ is the deviation from flatness at the start of inflation, $\ln(|\Omega_0-1|)/2\leq2.7$ is the bound from Planck~\cite{Aghanim:2018eyx}, and the last line describes the additional evolution
of $(\Omega-1)$ between  $\tilde{z}$ and now.

As discussed previously, the scaling solution produces larger string loops together with the smaller loops. The smaller loops are highly relativistic and most of the
energy transferred to the smaller loops is in the form of kinetic energy that simply redshifts away, and thus larger loops are expected to be the dominant source of GWs. For a Nambu-Goto string large loops oscillate, emit energy in the form of GWs, and gradually shorten
as shown in Eq.~\ref{ell_local}. Unlike the vanilla case discussed in section \ref{subsec_CS}, the GW in this situation is mostly emitted by short, violent, collimated bursts involving cusps or kinks on the string loops. Bursts emitted by a cosmic string network over its cosmic history that are not resolved contribute to the characteristic stochastic gravitational wave background of the network. 
For comparing the GW signals with current and future detectors, one can 
separate the contributions of the recent burst (with large amplitude) and earlier bursts (unresolved). Recent bursts can also be potentially observed as distinct, individual events. 
If a burst is to be resolved in a given frequency band $f$, 
it must produce a strain greater than the experimental sensitivity $h > h_{\rm exp}$ with a rate less than $f$.  
The rate of such events is~\cite{Siemens:2006vk,Auclair:2019wcv,Cui:2019kkd}
\begin{eqnarray}
R_{\rm exp}(f) = \int_{0}^{z_*} \!dz \int_{\max(h_{\rm min},h_{\rm exp})}^{h_{\rm max}} 
\!\!\!dh \; \frac{d^2R}{dz\, dh}(h,z,f)~~~~~ \ 
\end{eqnarray}
where $z_*$ (denotes the largest redshift contributing to the rate) enforces the rate condition and is given by
\begin{eqnarray}
\label{eq:astar}
f= \int_{0}^{z_*}\!dz \int_{h_{\rm min}}^{h_{\rm max}}\!dh \;  
\frac{d^2R}{dz\, dh}(h,z,f) \, .
\end{eqnarray}
Unresolved bursts contribute to the SGWB as~\cite{Siemens:2006vk,Auclair:2019wcv}
\begin{eqnarray} \label{eq:OmegaGW}
\Omega_{\rm GW}(f) =\frac{4\pi^2 f^3}{3H_0^2}
\int_{z_*}^{\infty}\!dz\,\int_{h_{\rm min}}^{h_{\rm max}}\!dh \; 
h^2\,
\frac{d^2R}{dz\, dh}(h,z,f) 
\end{eqnarray}
with $\Omega_{GW} = (f/\rho_c)\,d\rho_{\rm GW}/df$ and \begin{eqnarray}
\frac{d^2R}{dz\,dh} &=&
\frac{2^{3(q-1)}\,\pi\, G\mu\,  N_q}{(2-q)}
\frac{r(z)}{(1+z)^5H(z)}\,
\frac{n(l,z)}{h^2f^2}
\end{eqnarray}
where $l$ is a function of $h$, $f$, and $z$, and 
\begin{eqnarray}
h_{\rm min}=\frac{1}{(1+z)f^2}\frac{G\mu}{r(z)} \ ,~
h_{\rm max}=\frac{[\alpha L(z)]^{q-2}}{f^q(1+z)^{q-1}} 
\frac{\, G\mu}{r(z)} 
\end{eqnarray}
Since the total GW signal is expected to be dominated by bursts from
cusps~\cite{Ringeval:2017eww,Blanco-Pillado:2017oxo}, 
one can set $q=4/3$ and $N_q=2.13$ (to match $\Gamma = 50$~\cite{Auclair:2019wcv}) for the analysis purpose.

\begin{figure}
\centering
\includegraphics[width=0.48\textwidth]{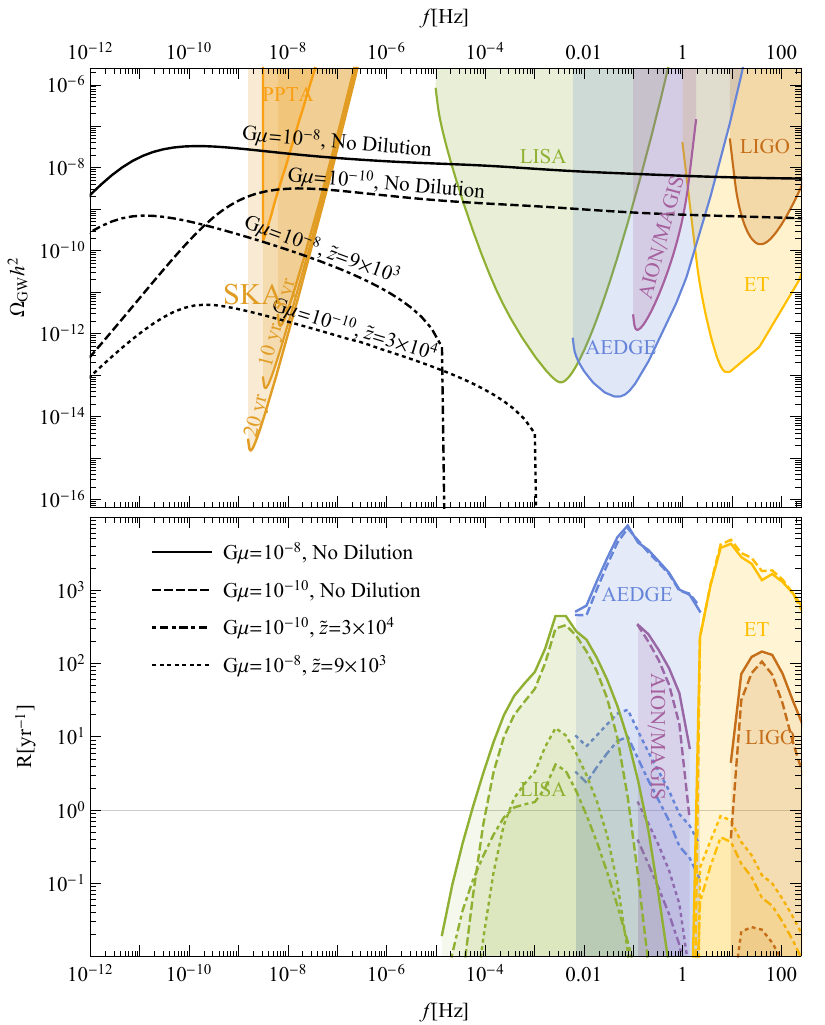}
\caption{
GW spectrum from diluted and undiluted CS
as a function of frequency observed today. While the top panel shows the
stochastic GW background, the lower panel gives the event rates of
resolved bursts.  In both panels, the curves are shown for $G\mu =
10^{-8},\,10^{-10}$ for undiluted networks as well as for two diluted
networks with $z = 9\times 10^{3}$ for $G\mu = 10^{-8}$, 
and $z = 3\times 10^{-4}$ for $G\mu = 10^{-10}$. The figure is taken from \cite{Cui:2019kkd}
\label{fig:StringRegrowthplot_1}
}
\end{figure}

The top (bottom) panel of Fig.~\ref{fig:StringRegrowthplot_1} shows the variation of the SGWB spectrum obtained from the unresolved (resolved) burst rate with frequency for diluted and undiluted string networks. The SGWB for the diluted networks falls off as $f^{-1/3}$ at high frequency due to the inclusion of the sufficient number 
of modes when computing the SGWB with the normal mode method (see \cite{Cui:2019kkd}) for the details. At higher frequencies, a sharper drop is observed from the burst 
method that results from the subtraction of infrequent bursts from the SGWB.

Let us now turn our attention to how the GW emitted from the CS can help us understand the cosmic history of the Universe after the inflation but at times well
before primordial nucleosynthesis and the cosmic microwave background where standard cosmology
has yet to be tested. Recent studies like \cite{Cui:2017ufi,Cui:2018rwi}, have demonstrated how the spectrum of GWs emitted by a Nambu-Goto cosmic string network depends on the energy content of the universe when they are produced. For example, if the Universe transitions from radiation domination at very high temperatures to matter domination and then back to radiation domination before the onset of BBN, a GW spectrum can be observed that is different from what is seen in the standard cosmology and hence can be distinguished. 

In a scenario like this, one can use the cosmological time $ t_\Delta$ (epoch of most recent radiation-domination era) to calculate the frequency $ f_\Delta$ where such a deviation would appear.  The first significant modification in the frequency spectrum will appear when the dominant emission time $\tilde{t}_M$ comes from loops created
at $t_i^{(k=1)} \simeq t_{\Delta}$.  This results in an approximate transition
frequency $f_{\Delta}$ as the solution of
\begin{eqnarray}
t_i(\tilde{t}_M(f_{\Delta})) = t_{\Delta} \ .
\end{eqnarray}
Under the approximation $a(t) \propto t^{1/2}$ at the radiation dominated era one can obtain \cite{Cui:2017ufi,Cui:2018rwi}
\begin{eqnarray}
f_{\Delta} &\simeq& 
\sqrt{\frac{8\,z_{\rm eq}}{\alpha\,\Gamma G\mu}}\,
\bigg(\frac{t_{eq}}{t_{\Delta}}\bigg)^{1/2}\,t_0^{-1} \\
&\simeq&
\sqrt{\frac{8\,}{z_{\rm eq} \alpha\,\Gamma G\mu}}\,
\bigg[\frac{g_*(T_\Delta)}{g_*(T_0)}\bigg]^{1/4}\bigg(\frac{T_{\Delta}}{T_0}\bigg)\,t_0^{-1}\nonumber
\end{eqnarray}
where $z_{\rm eq} \simeq 3387$ is the redshift at matter-radiation equality,
and $T_0 = 2.725\,\text{K}$ is the temperature today.
A more accurate dependence is obtained by fitting it to a full numerical 
the calculation that properly accounts for variations in $g_*$ gives \cite{Cui:2017ufi,Cui:2018rwi}
\begin{eqnarray} \label{eq:fdeltaforlargealpha}
f_{\Delta}= 
  (8.67\times 10^{-3} \, {\rm Hz})\, 
\bigg(\frac{T_\Delta}{\text{Gev}}\bigg)
\bigg(\frac{0.1\times 50\times 10^{-11}}{\alpha\,\Gamma\,G\mu}\bigg)^{1/2}
  \left(\frac{g_*(T_\Delta)}{g_*(T_0)}\right)^\frac{8}{6} \left(\frac{g_{*S}(T_0)}{g_{*S}(T_\Delta)}\right)^{-\frac{7}{6}} \ .
\end{eqnarray}
This is found to be accurate to about $10\%$. One can reinterpret this $f_\Delta$ as a frequency required to test cosmology up to temperature $T_\Delta$. This is because the measured  GW spectrum obtained from the CS will remain approximately flat till $f_{\Delta}$ for a given value of $\alpha$ and $\Gamma G\mu$ and hence would provide strong evidence for radiation domination up to the corresponding temperature $T_{\Delta}$.

\begin{figure}
\centering
\includegraphics[scale=0.5]{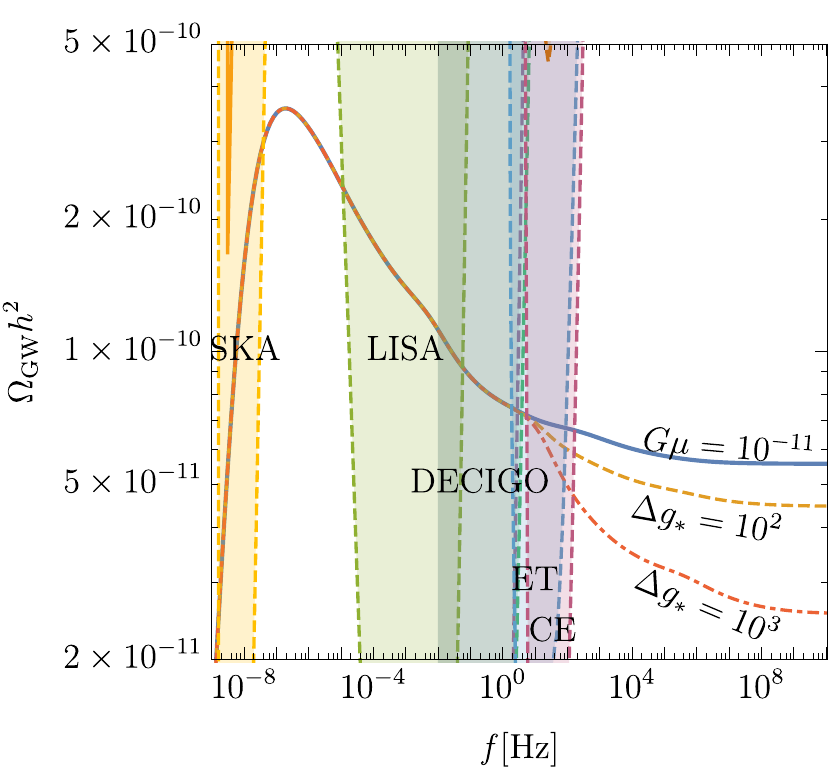}
\includegraphics[scale=0.33]{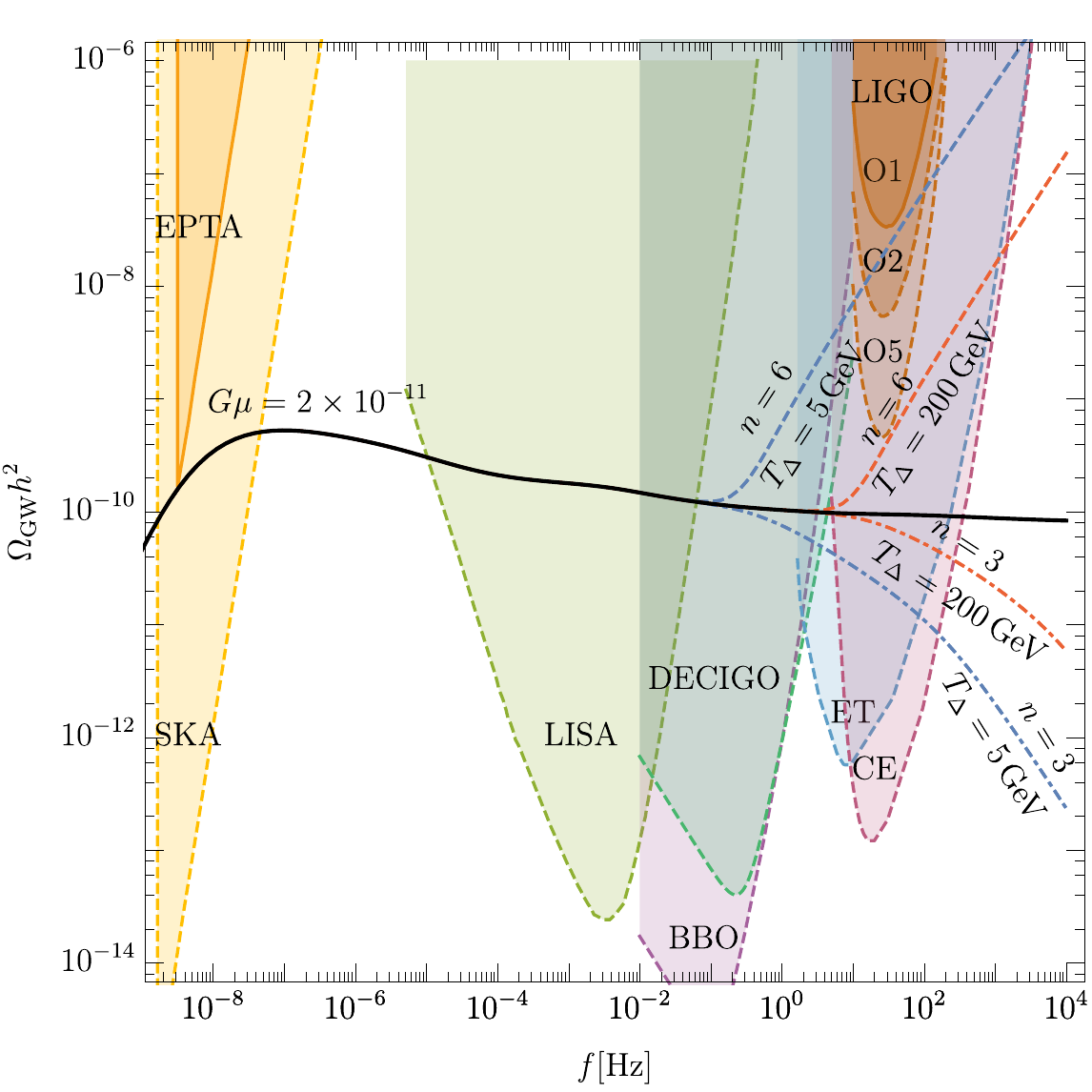}
\caption{
Left panel: Modification of the GW spectrum obtained from the CS by changing the number of DOFs. Here $G\mu=10^{-11}$ GeV and $T_\Delta=200$ GeV. Right panel: GW spectrum resulting from the early period of kination (n=6), matter (n=3) domination lasting until temperature $T_\Delta$. The figure is taken from \cite{Cui:2018rwi}
\label{fig:StringRegrowthplot}
}
\end{figure}

Such cosmological modification can be observed in a scenario having a very large number of additional degrees of freedom present in the spectrum at high energy. If additional degrees of freedom exist till temperature $T_\Delta$, the energy density of GW will be altered above $f_\Delta$ in comparison to what is expected in a standard cosmology with only SM degrees of freedom. In such a scenario, one obtains a standard $\Omega_{\rm GW}(f)$ vs.\ $f$ for cosmic strings behaviour up to $f_\Delta$ and thereafter fall-off is observed for $f>f_\Delta$ as shown in left panel of Fig. \ref{fig:StringRegrowthplot}. For a detailed discussion see \cite{Cui:2017ufi,Cui:2018rwi}.  

The refs. \cite{Cui:2017ufi,Cui:2018rwi} also discuss scenarios where GWs from cosmic strings evolve in a non-standard phase, either of an early matter domination phase $(n=3)$ or an early kination $(n=6)$ phase where $\rho\propto a^{-n}$ denotes the dilution of energy density of different dominant phases. The early matter domination can result from the presence of a new degree of freedom whose late decay brings the Universe back to the radiation domination era. Such slow decay should also satisfy BBN constraints. In other words, the universe transitions from radiation domination at very high temperatures to matter domination and then back to radiation domination (by the decay of said particles) before the onset of BBN. Another captivating way to generate such deviation is to consider a scalar $\phi$ oscillating into a potential of form $V(\phi)\propto \phi^N$ that gives $\rho\propto a^{-6N/(N+2)}$. In the limit $N\to\infty$ one finds a phase with $n=6$, the scalar can then decay. This leads to a cosmological history of very early radiation domination to kination domination (oscillation energy dominating) and back to radiation (by the decay of the moduli).
CS can act as a tool to probe these scenarios of alternate cosmologies, this is because they rapidly enter a scaling regime $i.e.$  their energy density scales with scale factor $a$ the same as the dominant energy density of the Universe. While for an early matter domination era, the CS will scale like $a^3$ during that phase, an early kination phase will make cosmic strings scale as $a^6$. The modified scaling behavior of CS  will alter the energy density of the GWs emitted through its non-standard redshifting. \cite{Cui:2017ufi,Cui:2018rwi} shows these effects quantitatively, which leads to a sharp fall-off in $\Omega_{\rm GW}(f)$ at high frequency $f$ (corresponding to the new phase era) if there is early matter domination and a sharp rise in $\Omega_{\rm GW}(f)$ if there is an early kination phase. This is shown in the right panel of Fig. \ref{fig:StringRegrowthplot}

\begin{figure}
\centering
\includegraphics[scale=0.5]{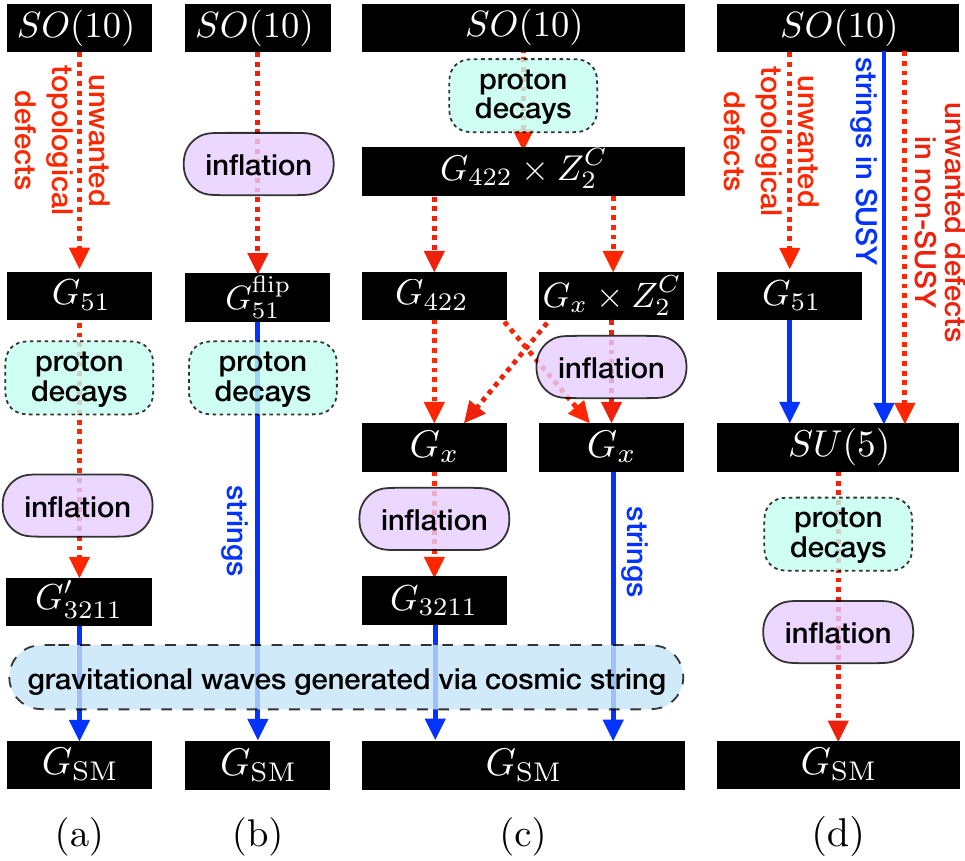}
\caption{
 The breaking chains of $SO(10) \to G_{\rm SM}$ are shown
along with their terrestrial and cosmological signatures where
$G_x$ represents either $G_{3221}$ or $G_{421}$. The figure is taken from \cite{King:2020hyd}
\label{fig:lepto}
}
\end{figure}

Apart from testing the scenarios of modified cosmologies, a GW emitted from the CS can also help in testing the other early Universe physics like why there exists a plethora of matter over anti-matter in our Universe. Arguably the most elegant way to understand this issue is through leptogenesis ~\cite{Fukugita:1986hr,Luty:1992un,Pilaftsis:1997jf}, using type I seesaw mechanism \cite{Minkowski:1977sc,Yanagida:1979as1,Yanagida:1979gs,GellMann:1980vs,Mohapatra:1979ia,Schechter:1980gr,Schechter:1981cv,Datta:2021elq,DuttaBanik:2020vfr,Datta:2021elq,Bhattacharya:2021jli,Bhattacharya:2023kws}\footnote{A recent work \cite{King:2023cgv} has also shown a new method towards distinguishing the Dirac vs Majorana nature of neutrino masses from the spectrum of GW associated with the neutrino mass genesis.}. Here an out-of-equilibrium decay of a heavy state generates an asymmetry in between leptons and their antiparticles. Once generated, the lepton asymmetry catalyzes a baryon asymmetry through sphaleron transitions \cite{Kuzmin:1985mm,Bento:2003jv,DOnofrio:2014rug}. Unfortunately, the scale of leptogenesis is much beyond the scales that can be tested on present-day earth-based detectors.  
The mass of the Right-handed neutrinos (RHN) responsible for explaining the light neutrino masses lies below the Planck or a possible GUT scale suggesting some symmetry that survives below these scales to protect the right-handed neutrino mass. Thermal leptogenesis demands the breaking of this symmetry to be below the scale of inflation or the asymmetry will be catastrophically diluted. Such a breaking can be observed through its predicted cosmological defects. In \cite{Dror:2019syi}, a $B-L$ that protects the RHN masses is spontaneously broken to generate the RHN masses, this process also results in the generation of CS that radiates a spectrum of stochastic gravitational waves. When embedding this symmetry into a grand unified group, one finds that the majority of viable symmetry-breaking paths admit cosmic strings. The GW signal is expected to be tested using future detectors that are expected to probe the entire mass range relevant to the paradigm of thermal leptogenesis.\footnote{If it is a global symmetry protecting the seesaw mass, it is still possible a signal is visible if the seesaw scale is very high \cite{Fu:2023nrn}} Symmetry breaking paths that do not admit cosmic strings could result in a signal from proton decay \cite{King:2020hyd,King:2021gmj,Saad:2022mzu,Fu:2023mdu}. Recent work has also considered the complementarity of gravitational waves and neutrino oscillation experiments within the concrete framework of an SO(10) model \cite{Fu:2022lrn}. A diagrammatic representation of the breaking chains of $SO(10) \to G_{\rm SM}$ is shown in Fig. \ref{fig:lepto} . 

\begin{figure}[t!]
  \centering
  \includegraphics[width=0.32\linewidth]{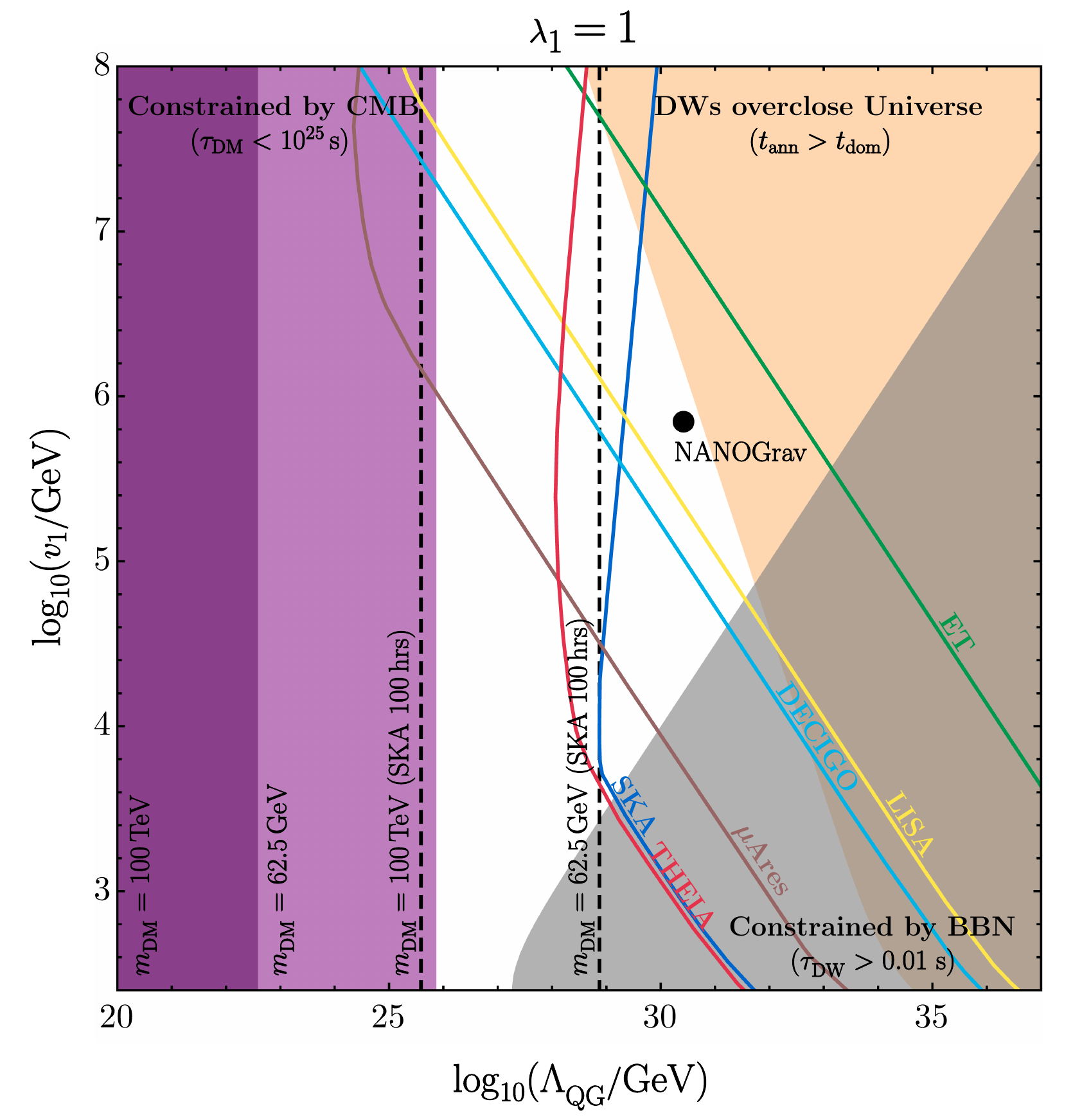}
  \includegraphics[width=0.32\linewidth]{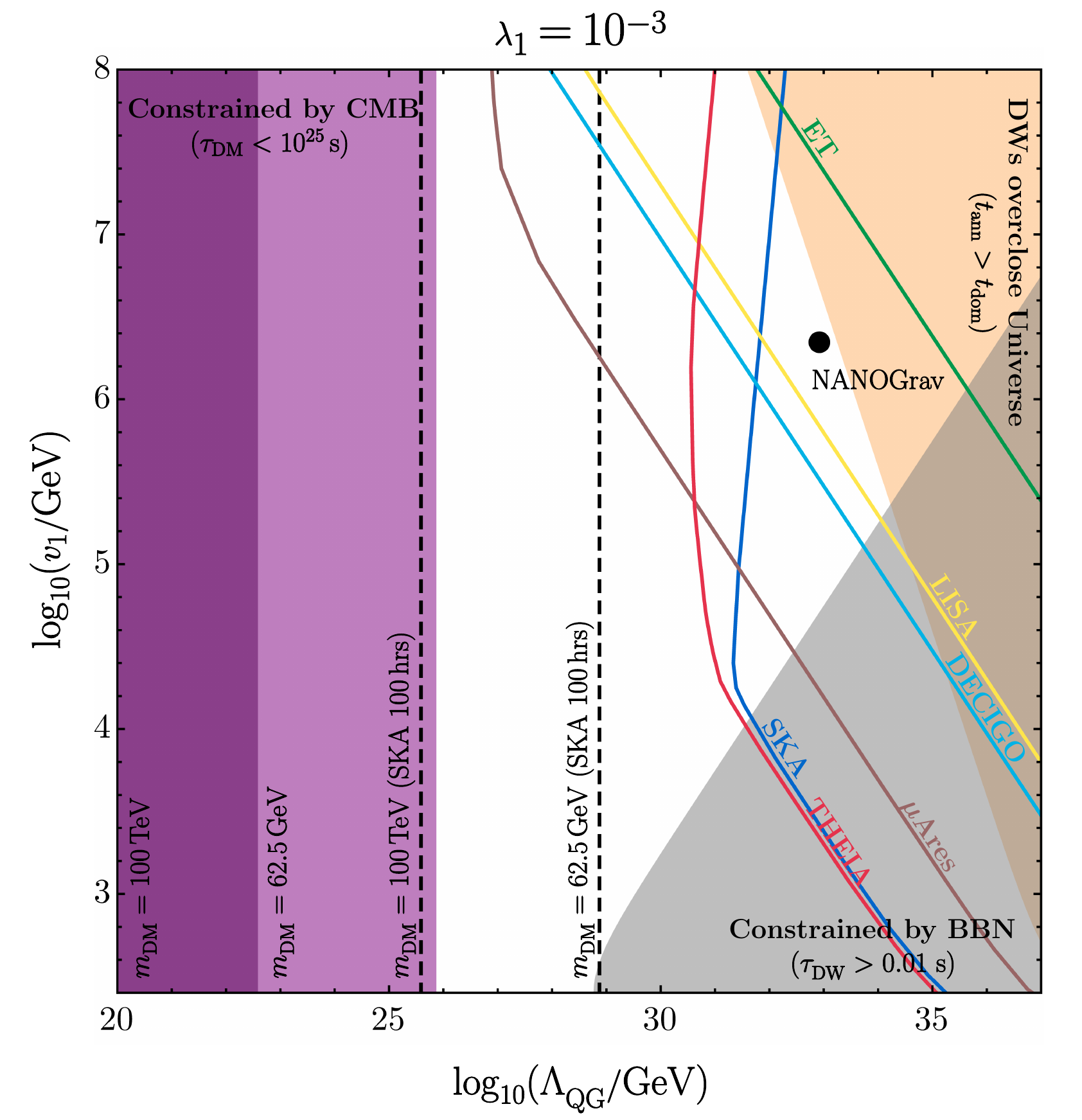}
  \includegraphics[width=0.32\linewidth]{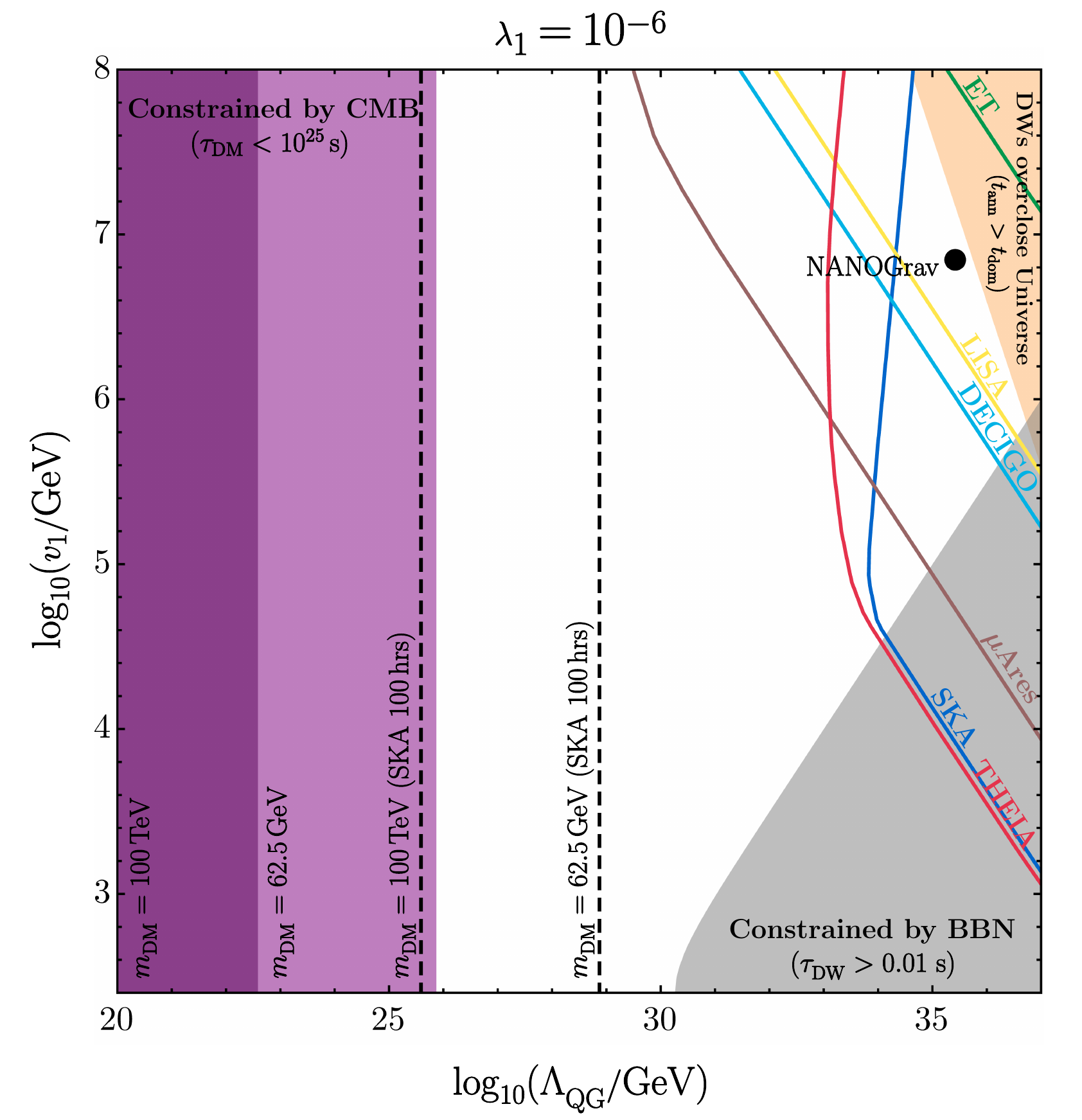}
  \includegraphics[width=0.97\textwidth]{fermisummary_fig21d.png}
  \caption{Combined constraints on $\Lambda_{\rm QG}$ (the scale of QG) and vacuum expectation value of scalar responsible for spontaneous breaking of $Z_2$ symmetry (denoted by $v_1$ in the top panel and $v_s$ in the bottom panel) from indirect DM detection and GWs observations with varying model parameter $\lambda_1$ (self quartic coupling of the scalar responsible for the spontaneous breaking of $Z_2$ symmetry) of \cite{King:2023ayw} in the top panel while varying DM mass of freeze-in DM of \cite{King:2023ztb} in the bottom panel. The top panel is taken from \cite{King:2023ayw} and the bottom panel is taken from \cite{King:2023ztb}. }
  \label{fig:summary}
\end{figure}

Like  CS, the GW resulting from the DW can also play an interesting role in understanding the physics of the early Universe \cite{King:2023ayw,King:2023ztb,Barman:2023fad}. For example, recent works like \cite{King:2023ayw,King:2023ztb}, have explored the phenomenological consequences of breaking discrete global symmetries in quantum gravity (QG) \footnote{On the other hand \cite{Barman:2023fad} discusses the possibility of detecting the scale of Dirac leptogenesis through GW generated from DW annihilation.}. Motivated by Swampland global symmetry conjecture \cite{Banks:1988yz,Banks:2010zn,Harlow:2020bee} which says that there exists no exact global symmetry in effective field theories (EFTs) arising from UV theories of QG, they studied how quantum gravity effects, manifested through the breaking of discrete symmetry responsible for both Dark Matter and Domain Walls, can have observational effects through CMB observations and gravitational waves. They show that the existence of very small bias terms stems from the QG \footnote{Work like \cite{Lu:2023mcz} showed that a small bias can also be achieved in a clockwork axion model where the QCD axion potential induced by the QCD phase transition can serve as an energy bias leading to the annihilation of DW } which
allows DWs to annihilate, thus stopping them from dominating the energy density of the Universe. To illustrate this idea, in \cite{King:2023ayw}, they considered a simple
model with two singlet scalar fields together with two $Z_2$ symmetries, one being responsible for DM stability, and the other spontaneously broken and responsible for DWs.\footnote{For the fermionic DM see \cite{King:2023ztb}. Explicit due to QG causes the DM to be allowed to mix with the SM neutrinos after the electroweak symmetry breaking. This causes the DM to become unstable. Such DM decay puts constraints on the QG operator from the CMB, BBN, galactic and extragalactic diffuse $X/\gamma$-ray background, SKA, neutrinoless double beta decay.} Moreover, both the $Z_2$ symmetries are assumed to be explicitly broken by QG effects by operators at the same mass dimension
and with the same effective Planck scale. They showed that this hypothesis led to observable GW signatures from the annihilation of DWs, which are
correlated with the decaying DM signatures constrained by CMB observations.\footnote{This is also a possible explanation for the results of the NANOGrav~\cite{NANOGrav:2023gor, NANOGrav:2023hvm}, the EPTA~\cite{Antoniadis:2023ott, Antoniadis:2023zhi}, the PPTA~\cite{Reardon:2023gzh} and the CPTA~\cite{Xu:2023wog}} Fig. \ref{fig:summary} shows constraints obtained on the QG scale. 

GW detectors have given us a window to early universe cosmology complementary to any other probes previously developed. A strong and well-understood source of GWs in the early universe could give us unprecedented ability to probe cosmological energy evolution of the early universe far earlier than previously attainable. These works have also demonstrated that topological defects if they exist, would be excellent standard candles to achieve these aims.


\section{Scalar induced gravitational waves}\label{sec:SIGW}


It was first mentioned by Tomita \cite{tomita1967non} then rediscovered by Matarrese, Pantano, and Saez \cite{matarrese1993general,matarrese1994general} that large scalar perturbations can act as a source for tensor perturbations at second order.  For a curvature perturbation power spectrum, ${\cal P}_{\cal R}$, the approximate scaling of scalar induced gravitational waves are\footnote{In this review we focus on gravitational waves induced by curvature perturbations generated by ultra slow roll or by sudden changes in the equation of state. It is also possible to induce gravitational waves from isocurvature perturbations \cite{Domenech:2021and,Bhaumik:2022pil,Lozanov:2023aez}, or through spectator fields during inflation \cite{Biagetti:2013kwa}. For a review see \cite{Domenech:2023jve}}
\begin{equation}
    \Omega _{\rm GW} ^{\rm induced} h^2 \sim \frac{1}{12} \Omega _{r,0} h^2 {\cal P}^2 _{\cal R}\sim 10^{-6}{\cal P}^2 _{\cal R}. 
\end{equation}
In vanilla cosmology, the scalar power spectrum after inflation can be inferred from CMB extrapolation ${\cal P} _{\cal R}\sim 10^{-9}$ \cite{akrami2020planck} which means that absent any enhancement, $\Omega _{\rm GW} ^{\rm induced} h^2 \sim 10^{-24}$, which is far too small for any detector in the foreseeable future. The enhancement can occur during a period of ultra-slow roll inflation, a large change in the field angle in multifield inflation, or due to the evolution of scalar perturbations after inflation \cite{Domenech:2021ztg}. For example, a sudden change in the equation of state can dramatically enhance the gravitational wave spectrum \cite{hooshyar2009gravitational,alabidi2013observable,inomata2019gravitational,inomata2019enhancement}. Scalar-induced gravitational waves have a remarkable range of applications. For instance, if there exists an instability scale as hinted at in the Standard Model \cite{Altarelli:1994rb,Casas:1996aq,Hambye:1996wb,Sher:1988mj,Arnold:1989cb,Isidori:2001bm,Ellis:2009tp,Elias-Miro:2011sqh,Degrassi:2012ry,Bezrukov:2012sa,Bednyakov:2015sca,Buttazzo:2013uya,Espinosa:2016nld}, there should be a scalar induced gravitational wave spectrum \cite{Espinosa:2017sgp}. As the running of the Higgs quartic self-interaction could be affected by any number of hypothetical particles in the mass range between the weak scale and the instability scale, gravitational wave detectors may be the only way to tell if the electroweak vacuum is stable. Further, if the Baryon asymmetry was produced during the Affleck Dine mechanism, there was likely a period of Q-ball domination whose sudden decay results in a gravitational wave spectrum, one of the few ways to test this paradigm \cite{White:2021hwi}. Finally, if one produces primordial Black holes that are too small to survive after Big Bang nucleosynthesis before evaporating via Hawking radiation, these black holes can dominate the Universe and their decay rate is again surface-to-volume suppressed which results in a sudden transition from matter to radiation \cite{Inomata:2020lmk}. We summarize the range of applications in section \ref{sec:applications} in slightly more depth. For now, we emphasize the extraordinary promise of this source of gravitational waves.

In this section, we review the current formalism and our understanding of both causes of scalar-induced gravitational waves and present some recent applications.

\subsection{Enhanced scalar power spectrum during inflation}

 When large scalar perturbations reenter the horizon after inflation they can produce a detectable source of gravitational waves \cite{Garcia-Bellido:2017mdw,Ezquiaga:2017fvi,Ballesteros:2017fsr,Li:2023xtl,Li:2023qua,Wang:2023sij,Zhao:2023joc,Yu:2023jrs} (for a review, see \cite{Domenech:2021ztg}). Consider the slow roll parameters for a scalar field in an expanding background 
 \begin{equation}
\epsilon = - \frac{\dot{H}}{H}  \ ,    \epsilon _2 = \frac{2 \ddot{\phi}}{H \dot{\phi}} + 2 \epsilon  \ .
 \end{equation}
The slow roll conditions are of course that the above parameters are small. However, near an inflection point when $V'\to 0$, the Klein Gordon equation in an expanding background yields $\ddot{\phi} = -3H \dot{\phi}$ and we have
\begin{equation}
\epsilon_2 =-6+2 \epsilon \to |\epsilon _2| >>1 \ .
\end{equation}
So we have the surprising result that on the one hand, the slow roll (SR) approximation assumes that the potential is flat, but if it is too flat the approximation breaks down. In this ``ultra slow roll'' (USR) regime the potential barely affects the evolution and the number of e-folds contrasts with slow roll as follows\footnote{ For simplicity, we consider the ultra slow roll mechanism. In multifield inflation there is an additional mechanism where the inflaton undergoes a large degree of angular motion in field space, see for instance ref. \cite{Palma:2020ejf} and the references therein}
\begin{equation}
   \Delta N =  \left\{ \begin{array}{cc} \frac{1}{m_{\rm pl}} \int _{\phi _1}^{\phi _2}\frac{d \phi}{2 \epsilon} & {\rm SR} \\
         -\int \frac{dH}{\epsilon H}  & {\rm USR}    \end{array} \right.
\end{equation}
and the scalar power spectrum grows exponentially for modes near the Hubble scale during the period of USR, ${\cal P}_{\cal R} \sim e^{6 \Delta N}$. To calculate the scalar power spectrum accurately during ultra-slow roll inflation, one needs to solve the coupled equations of motion for the inflaton and metric perturbation with the metric expanded to second order in perturbations. Transforming to Fourier space yields the Mukhanov-Sasaki equations \cite{Sasaki:1986hm,Mukhanov:1988jd} 
\begin{eqnarray}
    \frac{d^2 u_k}{d \eta ^2} + \left(  k^2 - \frac{1}{z}\frac{d^2z}{d \eta ^2} \right) u_k=0
\end{eqnarray}
where $u=-z {\cal R}$ and $z=\frac{1}{H}\frac{d\phi}{d \eta}$. If the second term in brackets dominates, we have an exponentially growing mode. It is clear that this is achieved when $\dot{\phi}$ is small, that is the ultra-slow roll regime.
 With the possibility of such large fluctuations, it is worth noting that if the fluctuations are large enough, one might not only induce gravitational waves at second order, but primordial black holes \cite{Carr:1975qj,saito2009gravitational,saito2010gravitational,Garcia-Bellido:2016dkw} which themselves can leave a gravitational wave signal \cite{Inomata:2016rbd,Wang:2019kaf,Escriva:2022duf,Ireland:2023avg,Wang:2023ost,Zhao:2022kvz}.\footnote{ This is one of many mechanisms to produce primordial black holes (for recent reviews see \cite{Carr:2016drx,Sasaki:2018dmp,Green:2020jor,Carr:2020xqk,Carr:2020gox,Villanueva-Domingo:2021spv,Escriva:2022duf,Ozsoy:2023ryl})} This means that if one sees PBHs, one can see the corresponding primordial GW background induced by scalar perturbations. In particular, the scalar power spectrum can be enhanced if there is a period where the scalar field is near an inflection point. In this review we focus on cosmological sources of gravitational waves, so will only make brief comments about PBHs through this mechanism. PBHs are formed when $\delta \rho/ \rho \gtrsim \delta _c$. Fluctuations during inflation follow a Gaussian distribution around zero, so the support for a large enough fluctuation is negligible in vanilla inflation unless there is an enhancement in the scalar power spectrum. One typically needs ${\cal P} _{\cal R} \gtrsim 10^{-2}$ to have a substantial abundance of pbhs, implying $\Omega _{\rm GW} h^2 \gtrsim 10^{-10}$ which is expected to be detectable. Conversely, the absence of such an observable gravitational wave spectrum can rule out the production of PBHs through inflation. There is a debate about whether this is achievable during single-field inflation and one might need a multifield inflation model to realize this mechanism \cite{Kawai:2021edk,Kristiano:2022maq,Kristiano:2023scm,Riotto:2023gpm,Choudhury:2023vuj,Choudhury:2023fjs,Choudhury:2023jlt,Choudhury:2023rks,Bhattacharya:2023ysp,Firouzjahi:2023ahg,Choudhury:2023hvf}. 

\subsection{Modelling of scalar induced gravitational waves}
\begin{figure}
     \centering
     \includegraphics[width=0.7\textwidth]{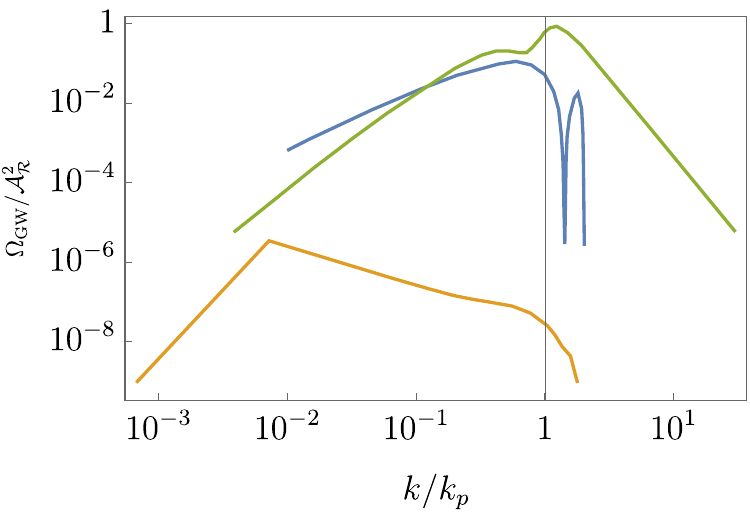}
     \caption{Gravitational wave spectrum from induced gravitational waves from benchmarks involving a Dirac delta function power spectrum and a speed of sound near unity (blue/middle curve) a Gaussian power spectrum (yellow/lower) and a broken power law spectrum (green/upper). Data taken from ref. \cite{Domenech:2021ztg} and the references therein.  Note that the speed of sound and equation of state is different for each of the spectra and the figure is not intended to be indicative of the strength of each signal but rather its shape. Details on each benchmark can again be found in ref. \cite{Domenech:2021ztg}}
     \label{fig:sigw}
 \end{figure}

The precise shape of the scalar power spectrum, and therefore the gravitational wave spectrum, will depend upon the specifics of the potential. So in principle, this leads to a wide range of possible signal shapes to search for. Fortunately, there are several ways to parametrize categories of power spectra. Common categories are delta function spectra, log normal spectral, and a broken power law. Let us begin with the delta function case. A sharply peaked, almost monochromatic spectra appear to be only possible in multifield inflation,\cite{cai2019primordial,kawasaki1998primordial,frampton2010primordial,kawasaki2013primordial,inomata2017inflationary,pi2018scalaron,cai2018primordial,chen2019primordial,chen2020dirac} (to see a discussion of this cannot be achieved during single field inflation, see \cite{byrnes2019steepest,ozsoy2020slope,palma2020seeding}). In such a case we can parametrize the curvature perturbation power spectrum as follows,
\begin{equation}
 {\cal P} _{\cal R} = {\cal A} _{{\cal R}} \delta (\ln (k/k_p))   
\end{equation}
where $k_p$ is the typical scale at which scalar fluctuations are occurring and ${\cal A}_{\cal R}$ is the amplitude of the fluctuations.  
 We assume instantaneous reheating for simplicity as this allows us to just assume that the induced GWs remain constant after their production (aside of course from redshifting). The resulting GW spectrum for instantaneous reheating occurring at time $\tau _{\rm rh}$ and $k_{\rm rh}= {\cal H}_{\rm rh}$ is,
\begin{equation}
    \Omega _{\rm GW , rh} (k) = \frac{2}{3} \left(\frac{k_p}{k_{\rm rh}} \right) ^2 \left( 1-\frac{k}{4 k_p^2}\right)^2 \bar{I_{\rm RD} (k/k_{\rm rh},k/k_p)} \Theta (2 k_p - k) \ ,
\end{equation}
we will expand on the definition of $I_{\rm rad}$ later, but it describes how the gravitational waves are sourced by scalar perturbations during an era of radiation domination.
More generally, one can model a finite width with a log normal peak,
\begin{equation} {\cal P}_{\cal R} = \frac{\cal A_{\cal R}}{\sqrt{2 \pi} \Delta} \exp \left[-\frac{\ln ^2 (k/k_p)}{2 \Delta ^2} \right] \ .
\end{equation}
 For example, refs \cite{cai2019primordial,kawasaki1998primordial,frampton2010primordial,kawasaki2013primordial,inomata2017inflationary,pi2018scalaron,cai2018primordial,chen2019primordial,chen2020dirac,ashoorioon2021eft} considered narrow peaks and for broad peaks see refs. \cite{espinosa2018cosmological,kohri2018semianalytic,ando2018formation,ando2018primordial,frampton2010primordial,garcia1996density,yokoyama1998chaotic,Cang:2022jyc,Cang:2023ysz}. In this case the gravitational wave spectrum has the form
\begin{equation}
    \Omega _{\rm GW , \Delta }(k) = {\rm erf} \left( \frac{1}{\Delta} \sinh ^{-1} \frac{k}{2 k_p} \right) \frac{2}{3} \left(\frac{k_p}{k_{\rm rh}} \right) ^2 \left( 1-\frac{k}{4 k_p^2}\right)^2 \bar{I_{\rm RD} (k/k_{\rm rh},k/k_p)} \Theta (2 k_p - k)
\end{equation}
Finally, in the case of single field inflation \cite{byrnes2019steepest,ozsoy2020slope,palma2020seeding,atal2019role}, we have more predictability and the power spectrum can be described by a broken power law
\begin{equation}
    {\cal P} _{\rm cal R} = {\cal A} _{\cal R} \left\{ \begin{array}{cc}
         \left( \frac{k}{k_p} \right)^{n_{\rm IR}} & k\leq k_p \\
     \left( \frac{k}{k_p} \right)^{-n_{\rm UV}}    & k \geq k_p
    \end{array} \right.
\end{equation}
with a resulting gravitational wave spectrum \cite{liu2020analytical,atal2021probing,xu2020gravitational,shannon2015gravitational,riccardi2021solving},
\begin{eqnarray}
    \Omega _{\rm GW,rh} (k \ll k_p) &\sim&  3 {\cal A}_{\cal R}^2 \left( \frac{1}{2 n_{\rm IR}-3} + \frac{1}{2 n_{\rm UV}+3} \right)\left( \frac{k}{k_p} \right)^2 \ln ^2 \left( \frac{k}{k_{p}} \right)\\
     \Omega _{\rm GW,rh} (k \gg k_p, n_{\rm UV}<4) &\sim&  \frac{1}{48} {\cal A}_{\cal R}^2 \left( 41+\frac{16 n_{\rm UV}}{\sqrt{16-n_{\rm UV}^2}} \right) \left( \frac{k}{k_p} \right)^{-2 n_{\rm UV}}
\end{eqnarray}
Clearly, in all cases, we require the amplitude to be much larger than the CMB extrapolation, $A_s >> 10^{-9}$ as occurs in vanilla inflation where the slow roll conditions never become invalid due to a period of ultra slow roll. The gravitational wave spectra of all three types of sources we show in Fig. \ref{fig:sigw} with the benchmarks taken from ref \cite{Domenech:2021ztg}.

\subsection{Sudden changes in the equation of state}

Even if the curvature power spectrum is no larger than the CMB extrapolation, if there is an early period of matter domination that suddenly decays faster than the Hubble time one can induce a sizeable gravitational wave spectrum.  During matter domination, the scalar perturbations grow and if the transition to radiation is faster than a Hubble time, there is a resonant enhancement of the induced gravitational waves which can result in a striking sharp gravitational wave spectrum \cite{inomata2019enhancement}. 
Let us consider the scenario where there was an early period of matter domination which suddenly ends when $\eta = \eta_R$.\footnote{For the gravitational wave signal involving a transition from/to kination see ref. \cite{Harigaya:2023mhl}. Further, how the gravitational wave spectrum is affected by a general constant equation of state was considered in refs. \cite{Domenech:2019quo,Domenech:2020kqm}}

\begin{figure}
     \centering
     \includegraphics[width=0.7\textwidth]{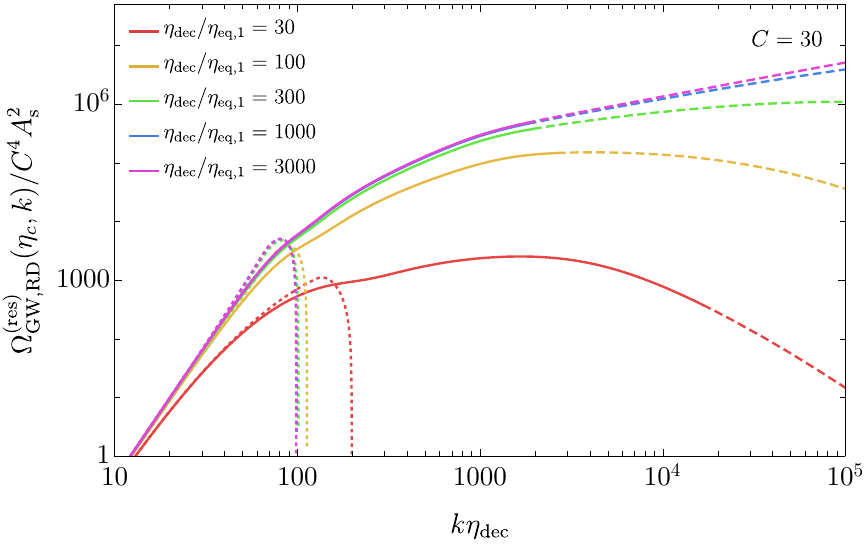}
     \caption{Shape of the gravitational wave spectrum for periods of matter domination that last so long as to render linear perturbations inadequate. The dotted line is the component of the spectrum arising from the linear regime, the solid from the non-linear regime, and the dashed line refers to a period where the calculations in the figure are inaccurate. Figure taken from ref. \cite{Kawasaki:2023rfx} with parameter $C$ explained within}
     \label{fig:SIGWNL}
 \end{figure}
\begin{figure}
    \centering
    \includegraphics[width=0.7\textwidth]{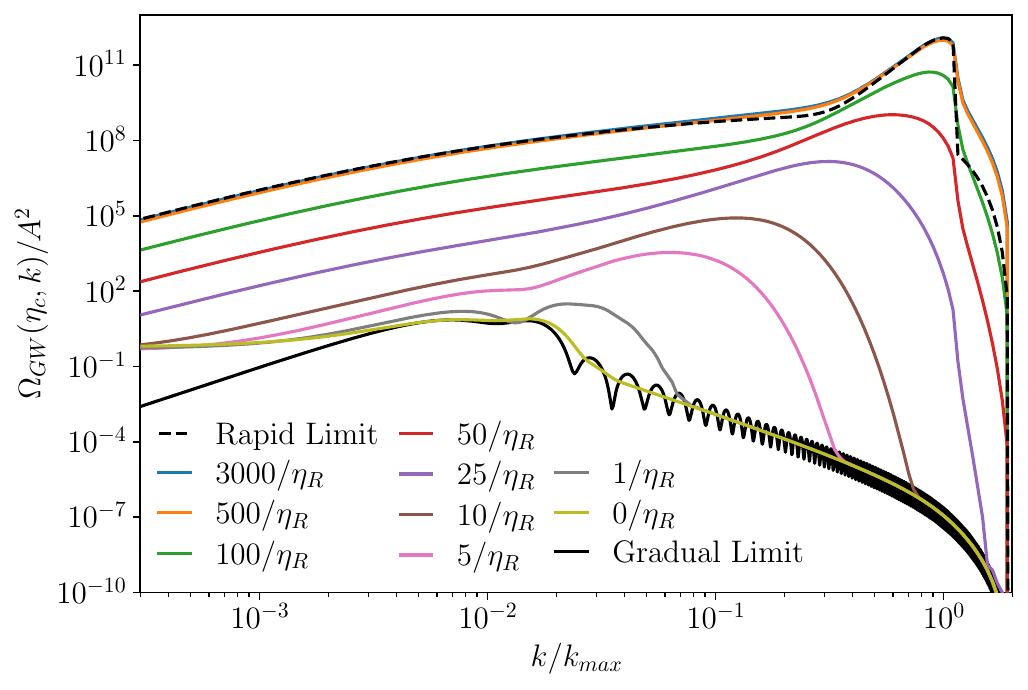}
    \caption{Shape of scalar induced gravitational wave spectra as a function of how rapidly the matter disappears. Higher curves are progressively shorter transitions. Figure taken from ref \cite{Pearce:2023kxp} }
    \label{fig:SIGWtransition}
\end{figure}

The power spectrum of tensor modes sourced by curvature perturbations is given by,
\begin{widetext}
\begin{equation}
    \overline{\mathcal{P}(\eta,k)}=4\int_{0}^{\infty}dv\int_{|1-v|}^{1+v}du\left(\frac{4v^2-(1+v^2-u^2)^2}{4vu}\right)^2\overline{I^2(u,v,k,\eta,\eta_R)}\mathcal{P}_{\zeta}(uk)\mathcal{P}_{\zeta}(vk)
\end{equation}
\end{widetext}
where the power spectrum of curvature perturbations is given by
\begin{equation}
{\mathcal{P}} _\zeta (k) = \Theta (k_{\rm max} - k) A_s \left( \frac{k}{k_\ast} \right) ^{n_s-1}    
\end{equation}
with $A_s=2.1\times 10^{-9}$ the amplitude of the pivot scale, $n_s=0.97$ is the spectral tilt, $k_\ast=0.05 {\rm Mpc}^{-1}$ the pivot scale where all of these values we take from ref. \cite{Planck:2018vyg}. As the scalar perturbations grow during matter domination, we eventually enter a regime where it is no longer valid to consider linear perturbations. This occurs when $k_{\rm max}\leq 470/\eta_r$ and there is no accessible way to calculate the spectrum in this case. A recent simulation seemed to show that the gravitational wave spectrum continues to grow as on enters the non-linear regime \cite{Kawasaki:2023rfx}\footnote{In \cite{Fernandez:2023ddy}, the authors did N-body simulations, incorporating the gravitational effects of (particle-like) matter decays and radiation, without relying on fitting functions found elsewhere. They found a substantial enhancement of the resulting induced GW spectrum on scales that went nonlinear and parameterize its properties.}. 
The $I$ function in the above describes the time dependence of the gravitational waves and is the convolution of the appropriate greens function and a source function, $f$,
\begin{equation}
\label{eq:I}
    I(u,v,k,\eta,\eta_R)=\int_0^x d\bar{x}\frac{a(\bar{\eta})}{a(\eta)}kG_k(\eta,\bar{\eta})f(u,v,\bar{x},x_R)
\end{equation}
with $(x,x_R)=(k \eta ,k \eta _R)$ and the source function is given by the expression 
    \begin{eqnarray}
&&f(u,v,\bar{x},x_R)= \nonumber \\ && \frac{3(2(5+3w)\Phi(u\bar{x})\Phi(v\bar{x})+4\mathcal{H}^{-1}(\Phi'(u\bar{x})\Phi(v\bar{x})+\Phi(u\bar{x})\Phi'(v\bar{x}))+4\mathcal{H}^{-2}\Phi'(u\bar{x})\Phi'(v\bar{x}))}{25(1+w)} \nonumber \\ 
    \end{eqnarray}
Here,  $\Phi$ is the transfer function of the gravitational potential. The Green functions for radiation and matter domination have different functional forms, so it is convenient to split Eq. \ref{eq:I},
\begin{eqnarray}
    && I(u,v,k,\eta,\eta _R)\nonumber = \\ && \int _0 ^{x_r} d \bar{x} \frac{1}{2(x/x_r)-1}  \left( \frac{\bar{x}}{x_r} \right)^2 k G_k ^{ {\rm MD} } (\eta , \bar{\eta}) f(u, v, \bar{x}, x_r) \nonumber \\ && + \int _{x_r} ^x \bar{x} \left( \frac{2(\bar{x}/x_r)-1}{2(x/x_r)-1} \right) k G^{\rm RD} _k (\eta , \bar{\eta}) f(u,v,\bar{x},x_R) \\ 
    &=& I_{{\rm MD}} (u,v,x,x_R) + I_{\rm RD}(u,v,x,x_R) \ .
\end{eqnarray}
As the power spectrum involves the square of the above, the scalar-induced gravitational waves will involve a radiation term, a matter term, and a cross term,
\begin{equation}
    \Omega _{\rm GW} = \Omega _{\rm RD} + \Omega _{\rm MD} + \Omega _{\rm cross} \ .
\end{equation}
The cross term is negative definite and vanishes when the transition is instantaneous. So the faster the transition, the stronger the gravitational waves. For a sudden transition, the gravitational wave spectrum we expect has the approximate form,
\begin{equation}
    \frac{\Omega _{\rm GW}(\eta _c , k)}{A_s^2} \sim \left\{ \begin{array}{cc}
        0.8 & (x_{\rm R}\lesssim 150 x_{\rm max,R}^{-5/3} \\
       3\times 10^{-7} x_{\rm R}^3 x_{\rm max , R}^5  & (150 x_{\rm max, R}^{-5/3} \lesssim x_{\rm R} \ll  1 \\ 
       1\times 10^{-6} x_{\rm R} x_{\rm max,R}^5 & (1 \ll x_R  \lesssim x_{\rm max,R} ^{5/6} \\ 
       7 \times 10^{-7 } x_{\rm R}^7 & (x _{\rm max,R} \lesssim x_{\rm R} \lesssim x_{\rm max, R} \\ 
       {\rm sharp drop} & (x_{\rm max, R} \lesssim X_{\rm R} \leq 2 x_{\rm max, R})
    \end{array} \right.
\end{equation}
This approach of course assumes an instantaneous transition from matter to radiation, and that the period of matter domination did not last so long that it is no longer valid to make use of linear perturbations. One might naturally ask what happens to the signal if these two assumptions break down. This question is only just being probed by the community, nonetheless, let us briefly make some qualitative remarks based on some recent work. First, if one enters the non-linear regime, the peak of the gravitational wave spectrum does not grow as quickly with respect to the duration of the matter domination. Further, the resonance flattens out considerably, see Fig. \ref{fig:SIGWNL}. When the timescale of transition is in between the limits of instantaneous and Hubble time, the signal flattens out and the peak becomes smaller as seen in \ref{fig:SIGWtransition}. Such a smooth interpolation between the signals predicted in the instantaneous and gradual limits lends hope to the possibility that one can distinguish between differing scenarios that result in an early period of matter domination.
 



\subsection{Gauge invariance}


Early work found that the predictions of second-order gravitational waves can vary by orders of magnitude due to an unphysical gauge dependence \cite{Hwang:2017oxa}. The issue is that tensor modes are gauge-independent, so if we define
\begin{eqnarray}
    \rho _{\rm GW}\sim \langle h^{\prime ij}h^\prime _{ij} \rangle 
\end{eqnarray}
and perform a gauge transformation
\begin{equation}
    x^\mu \to x^\mu - \xi ^\mu , \quad  {\rm where }\ \xi ^\mu = (T,\partial ^i L)
\end{equation}
one gets a gauge dependence in our prediction of the gravitational wave spectrum,
\begin{equation}
    \rho _{\rm GW} \to \rho _{\rm GW} + \langle \partial _i T \partial _j T \partial ^i T \partial ^j T \rangle +\cdots \ .
\end{equation}
There is no obvious way to remove the spurious gauge dependence by choosing a gauge invariant variable, as it is not clear which gauge invariant variables should go in the above equation to correspond to the real gravitational wave a detector would observe. As yet the issue is unsolved, yet a partial solution discussed in \cite{DeLuca:2019ufz,Inomata:2019yww,Yuan:2019fwv,Lu:2020diy,Chang:2020tji}. One finds in a Universe with a fixed equation of state and a non-zero speed of sound, the Newtonian gauge and synchronous gauge yield the same predictions for gravitational waves. In general, there are a class of gauges that yield the same results under a set of conditions \cite{Domenech:2020xin}
\begin{itemize}
    \item[1] We consider only sub-horizon scales
    \item[2] tensor modes can be approximated as freely propagating
    \item[3] The gauge we use is suitable for small distance calculations.
    \end{itemize}
The second condition is essentially insisting that we calculate the gravitational wave energy density after the source has become negligible.
 If such conditions are satisfied then the correction to the tensor mode from a gauge transformation is suppressed by a factor $H^2/k^2$ and the gravitational wave spectrum is roughly gauge invariant. As yet, a more satisfying and complete answer remains elusive and is one of the challenges facing the field.

    \subsection{Applications of scalar induced gravitational waves}\label{sec:applications}

Let us conclude this section by sampling the range of applications scalar-induced gravitational waves have. We will begin with one of the most striking applications of scalar-induced gravitational waves - that it may be possible to discern whether the Higgs vacuum is absolutely stable \cite{Espinosa:2018eve}. During inflation, so long as the Higgs mass is less than Hubble, its vacuum expectation value will randomly walk with kicks of $\pm O(H/2\pi)$ until the Higgs field ends up in the catastrophic vacuum. The classical motion of the Higgs begins to dominate and the Higgs rapidly descends into the abyss. At the end of inflation, the Higgs potential obtains thermal corrections to its potential which can shift the catastrophic vacuum to begin at a larger field value \cite{DelleRose:2015bpo} and the Higgs may harmlessly oscillate around its minimum with an amplitude that decays as the Universe expands. The situation is slightly different from the ultra-slow roll scenario as the Higgs is a spectator field during inflation. Nonetheless, the motion of the Higgs field, $h_c$, at the end of inflation causes curvature perturbations 
\begin{equation}
    {\cal P}_{\zeta} (t_{\rm dec}) = \frac{\rho_h^2(t_{\rm dec)}}{\rho_{\rm tot}^24} \left( \frac{H}{2 \pi } \right)^2 \left( \frac{\dot{h}_c(t_e)}{h_c(t_e)\dot{h}_c (t_k)} \right)^2
\end{equation}
where $t_{\rm dec}$ is the time at the end of inflation. The shape of the resulting gravitational wave spectrum is highly sensitive to the precise instability scale.

Another application is the so-called poltergeist mechanism \cite{Inomata:2020lmk} (see also \cite{Domenech:2020ssp,Bhaumik:2020dor,Dalianis:2020gup,Kozaczuk:2021wcl,Haque:2021dha,Domenech:2021wkk,Bhaumik:2022pil,Bhaumik:2022zdd,Papanikolaou:2020qtd,Papanikolaou:2022chm,Basilakos:2023xof,Basilakos:2023jvp}). Black holes below a threshold mass of $10^9$g evaporate before the onset of Big Bang nucleosynthesis. However, they can dominate the energy budget of the Universe as their density dilutes like matter rather than radiation. When they evaporate, it is only through Hawking radiation at the event Horizon. This means the evaporation rate is surface-to-volume suppressed and accelerates once the rate of mass loss is comparable to the expansion of the Universe. This is precisely the conditions needed to create scalar-induced gravitational waves. Such light, evaporating black holes could partly explain the Hubble tension \cite{Hooper:2019gtx,Nesseris:2019fwr,Lunardini:2019zob}, be crucial in the explanation for why there is more matter than anti-matter \cite{Toussaint:1978br,Turner:1979zj,Turner:1979bt,Barrow:1990he,Majumdar:1995yr,Upadhyay:1999vk,Dolgov:2000ht,Bugaev:2001xr,Bugaev:2001xr,Baumann:2007yr,Fujita:2014hha,Hamada:2016jnq,Barman:2021ost,Barman:2022gjo,Barman:2022pdo} or generate dark matter particles \cite{Fujita:2014hha,Lennon:2017tqq,Morrison:2018xla,Gehrman:2023esa,Domenech:2023mqk}. The poltergeist mechanism can also be used to probe high-scale supersymmetry, as the flat directions in SUSY tend to fragment into Q-balls which, for much of the parameter space relevant to high-scale SUSY, merge and collapse into primordial black holes that are light enough to create a signal \cite{Flores:2023dgp}. The range of masses of primordial black holes that can be detected by proposed gravitational wave experiments are $10^2$g-$10^9$g \cite{Inomata:2020lmk}

Q-balls are another source of early matter domination \cite{Kasuya:2022cko}. One of the core paradigms for explaining why there is more matter than anti-matter relies on the existence of flat directions during Supersymmetry having a non-zero lepton or baryon number. Such directions fragment into Q-balls with a finite baryon or lepton charge. Much like primordial black holes, the Q-balls can only decay at the surface of the soliton leading to a surface-to-volume suppression. The rate of decay is slightly different from primordial black holes which means that there is hope in distinguishing the signals. Recent work found that the mechanism generally resulted in a visible scalar-induced gravitational wave spectrum \cite{White:2021hwi}.

\section{Other significant sources}\label{sec:HFsources }
In this review, we have focused on three types of gravitational wave sources that we see as the most promising as well as the most relevant to next-generation gravitational wave detectors, predicting signals in the nHz to kHz frequency band. There are many other sources, some of which are too significant to omit, though we neglect a full section on them because the likely signal is high frequency or there has been little recent progress. For the sake of brevity, we regret that we do not cover some of the more exotic sources in the literature including
oscillons \cite{gleiser1994pseudostable,amin2010flat,amin2010inflaton,amin2012oscillons,zhou2013gravitational,amin2013k,lozanov2014end,antusch2016parametric,antusch2016impact,antusch2017gravitational,antusch2018oscillons,antusch2018can,lozanov2018self,lozanov2018self,antusch2019properties,sang2019stochastic,lozanov2019gravitational,hiramatsu2021gravitational}, brane worlds \cite{rubakov1983we,randall1999out,seahra2005detecting,clarkson2007gravitational}, audible axions \cite{Machado:2018nqk,Machado:2019xuc,Ratzinger:2020oct} and pre-big bang cosmology \cite{gasperini2003pre,gasperini2007string,brustein1995gravitational}. We refer the reader to these references for more information.  However, in this section, we cover the case of blue-tilted gravitational wave spectra from inflation, gravitational waves as a Big Bang thermometer - perhaps the only measure of the reheating temperature, and gravitational waves from (p)reheating. 
\subsection{Inflation}

It is well known that gravitational waves arising from inflation result in B modes in the CMB \cite{Grishchuk:1974ny,starobinsky1979relict,rubakov1982graviton,Fabbri:1983us,Abbott:1984fp,Drewes:2019rxn,Drewes:2023bbs}. For single-field inflation, the actual signal from the CMB peaks at an ultralow frequency, many orders of magnitude below what even pulsar timing arrays are sensitive to. Worse, the spectrum of tensor modes predicted by single field inflation is red-tilted, meaning that excluding probes of the CMB, the prediction of the amplitude of a SGWB is formidably small in the frequency range relevant to next-generation experiments. However, in multifield inflation, it is possible to have a blue-tilted spectrum. The usual notation is to denote the tensor spectral index as $n_t$ and the scalar-tensor ratio as $r$, in which case we can write the gravitational wave spectrum in the case of a constant spectral tilt as a simple power law,
\begin{equation}
    \Omega _{\rm GW} (f) = 2.1 \times 10^{-6} g_\ast (f) \left( \frac{g^0_{\ast , s}}{ g_{\ast , s}(f) } \right)^{4/3} r A_s \left(\frac{f}{f_{\rm cmb}} \right)^{n_t}  {\cal T}(f) \ .
\end{equation}
where $g_{\ast ,s}^0=3.93$ is the number of effective relativistic degrees of freedom today, $g_\ast  (f),g_{\ast,s}(f)$ is the number of relativistic degrees of freedom when mode $k=2\pi f$
 re-entered the horizon, $A_s = 2.1 \times 10^{-9}$ \cite{Planck:2018vyg} is the amplitude of the scalar power spectrum and ${\cal T}$ is a transfer function connecting reheating to radiation domination \cite{kuroyanagi2015blue,Kuroyanagi:2020sfw}
 \begin{equation}
     {\cal T} (f) \sim \frac{\Theta (f_{\rm end} - f)}{1- 0.22 (f/f_{\rm rh})^{3/2} + 0.64 (f/f_{\rm rh})^2} \ .
 \end{equation}
The high-frequency cutoff of the gravitational wave spectrum, which marks the end of inflation is set by the reheating temperature and the Hubble rate at the end of inflation,
\begin{equation}
    f_{\rm end} = \frac{1}{2 \pi} \left( \frac{g_{\ast ,s}^0}{g^{\rm rh}_{\ast , s}} \right) ^{1/3} \left( \frac{\pi^2 g^{\rm rh }_\ast}{90} \right) ^{1/3} \frac{T_{\rm rh}^{1/3} H_{\rm end}^{1/3}T_0}{M_p ^{2/3}}
\end{equation}
with $T_0=2.73K$ today's CMB temperature. Naively, it is not easy to generate a positive value of $n_t$, as in single field inflation at lowest order in slow-roll parameters, one finds a negative tilt $n_T  = -2\epsilon$ \cite{liddle2000cosmological}. However, in axion vector inflation \cite{Cook:2011hg,Anber:2012du,Adshead:2013nka,Adshead:2013qp,Namba:2015gja,Dimastrogiovanni:2016fuu,Adshead:2016omu,Caldwell:2017chz,Adshead:2017hnc,Niu:2023bsr}, massive gravity \cite{Fujita:2018ehq} and other non-minimal scenarios \cite{Piao:2004tq,Satoh:2008ck,Kobayashi:2010cm,Endlich:2012pz,Kawai:2017kqt,Kawai:2017kqt,Fujita:2018ehq,Niu:2022quw} it is possible to have a blue tilted spectrum.

\subsection{(p)reheating}
Khlebnikov and Tkachev first pointed out the production of GW during the era of reheating~\cite{Khlebnikov:1996mc}. Since reheating is an outcome of almost all the inflationary scenarios,\footnote{For a recent exception see ref \cite{Eggemeier:2022gyo,Fernandez:2023ddy}} a GW generated during this period carries all the information of the inflationary period together with the reheating era. This is because they remain decoupled after they are produced and hence can act as a probe of the interaction strength between the inflation and the other fields. See refs. \cite{Easther:2006gt,Garcia-Bellido:2007nns,Dufaux:2007pt}  for the details and ref for a summary~\cite{Hyde:2015gwa}.

Once the inflation has ended, the inflaton field oscillates around its minimum to produce the elementary particles that interact among themselves to reach thermal equilibrium.  In scenarios like chaotic inflation \cite{Linde:1981mu,Albrecht:1982wi,Linde:1983gd}, if these oscillations are huge and coherent, the particles can be produced rapidly through resonant enhancement and the mechanism is popularly known as the parametric resonance \cite{Traschen:1990sw,Kofman:1994rk,Kofman:1997yn}. During this phase, the
amplitudes of the inflaton field grow exponentially and the
growth rate is dominated by the Mathieu characteristic
exponent.  This stage of rapid particle production is known as \emph{preheating}. Here the particles produced are not in equilibrium and another phase is needed to thermalize the radiation. On the other hand, in the case of hybrid inflation \cite{Linde:1993cn} one can have a different kind of preheating known as \emph{tachyonic preheating} \cite{Linde:1993cn,Felder:2000hj}. Here, the secondary field (waterfall field) descends from the maxima of its potential and oscillates around its minimum. In principle, around the maxima of the potential, there might exist a region where the quadratic mass of the field turns negative and the field fluctuations grow exponentially. Chaotic inflation and hybrid inflation are the most studied inflationary scenarios in which GW production during the following preheating phase has been investigated. The process of production of GWs in both scenarios is almost the same. Here, the GW production is sourced from inhomogeneities that cannot be neglected as they act as a source term in the GW equation of motion. These inhomogeneities in the energy density of the Universe are produced as a result of the highly pumped modes (large oscillations of the fields) during this phase. For some models of preheating, the gravitational wave spectra can be so large that bounds on $\Delta N_{\rm eff}$ already constrain the preheating model \cite{Adshead:2018doq,Adshead:2019igv,Adshead:2019lbr}.

 \begin{figure}
     \centering
     \includegraphics[scale=1.1]{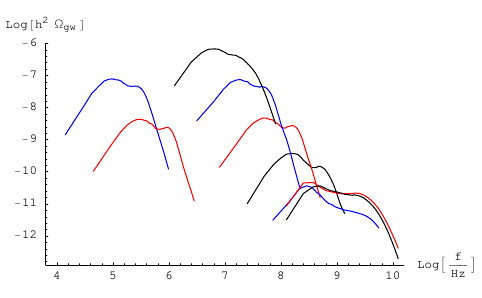}
     \caption{GW spectra from numerical simulations generated after hybrid inflation. The figure is taken from \cite{Guzzetti:2016mkm}. }
     \label{fig:GW_preheating}
 \end{figure}

Including the rate of expansion of the Universe between the time of emission and the present time, the GW that we see today is rescaled as  \cite{Dufaux:2007pt},
\begin{equation}
	\Omega_{\rm GW}h^{2}=\frac{S_{k}\left(\tau_{f}\right)}{a_{\rm j}^{4}\rho_{\rm j}}\left(\frac{a_{\rm j}}{a_{\ast}}\right)^{1-3w}\left(\frac{g_{\ast}}{g_{0}}\right)^{-1/3}\Omega_{\rm rad}h^{2}\,,
\end{equation}
where $\Omega_{\rm rad}h^{2}=4.3\times 10^{-5}$, $\rho_{\rm j}$ is the total energy-density at $t=t_{\rm j}$ a time after the jump of the equation of state and $t_{\ast}$ denotes the time when the thermal equilibrium was established. While $\tau_{\rm f}$ denotes the time when the GW source becomes negligible,  $S_{k}$ carries the information about the amount of GW produced at the time the source was present.
The frequency evaluated today is,
\begin{equation}	
f\equiv\frac{k}{a_{\rm j}\rho_{\rm j}^{1/4}}\left(\frac{a_{\rm j}}{a_{\ast}}\right)^{1-\frac{3}{4}\left(1+w\right)}4\times 10^{10}\,\mathrm{Hz}\,,
\end{equation}
where $k$ is the comoving wave number of a given fluctuation.

For preheating following chaotic inflation or hybrid inflation, the mean value of $w$ in the intermediate stage reaches $w=1/3$ soon after the end of inflation \cite{Dufaux:2007pt,Dufaux:2008dn}, so that the factor $\left(a_{\rm j}/a_{\ast}\right)^{1-3w}$
can be neglected. In this case the previous relations read
\begin{equation}\label{omegabu}
	f=\frac{k}{a_{\rm j}\rho_{\rm j}^{1/4}}\,4\times 10^{10}\,\mathrm{Hz} \qquad\mbox{and}\qquad
	\Omega_{\rm GW}h^{2}=\frac{S_{k}\left(\tau_{\rm f}\right)}{a_{\rm j}^{4}\rho_{\rm j}}\,9.3\times 10^{-6}\,.
\end{equation}

\noindent Fig. \ref{fig:GW_preheating} shows the variation of GW spectrum with frequency for the case of hybrid inflation. As can be seen from Fig. \ref{fig:GW_preheating}, the GW spectra generated during the preheating are mostly blue-tilted and lies beyond the accessible range of the present and future detectors. At this stage, we would like to point out that in recent times it was shown that one can get GW signals during the preheating era that lies within the sensitivity of some present and future GW experiments. We will not delve into the details of these scenarios but we refer readers to \cite{Figueroa:2022iho,Ghoshal:2022jdt} and the references therein for details.
\subsection{Graviton production as a big bang thermometer}
Production of SGWB by thermal plasma is guaranteed in the early Universe \cite{Ghiglieri:2015nfa,Ghiglieri:2020mhm}\footnote{In \cite{Drewes:2023oxg}, authors have shown that if one goes beyond the SM, there exists an upper bound on the thermal GW background. }. The energy density of this SGWB scales with the maximum temperature $T_{\rm{max}}$ of the plasma attained at the initial time of the Big Bang era, then peaks in the microwave range. The standard hot Big Bang cosmology gives an excellent description of the Universe when the SM bath was radiation-dominated and the Universe's temperature was around a few MeV~\cite{Kawasaki:1999na,Kawasaki:2000en,Giudice:2000ex,Hasegawa:2019jsa}. Unfortunately, it fails to predict the temperature of the bath before the radiation-dominated era and hence the temperature before this era could be arbitrarily large. However, one can provide an upper bound on this maximum temperature from the Planck scale $i.e.~T_{\rm{max}}\lesssim M_P$~\cite{Sakharov:1966fva}. For a temperature higher than this bound the quantum gravity effect cannot be ignored. In the regime where $T_{\rm{max}}> M_P$, the gravitons reach thermal equilibrium and attain a black body spectrum that would decouple around $T_{\text{dec}}\simeq M_P$~\cite{Kolb:1990vq}\footnote{If there is a SM instability scale, the $T_{\rm max}$ is around GUT scale \cite{DelleRose:2015bpo}}  and thereafter, the black body spectrum redshifts with the expansion of the Universe. The prediction for the relative abundance
of gravitational waves is that of a blackbody spectrum with an effective temperature obtained by
redshifting the decoupling temperature $M_P$ by the expansion of the universe between the decoupling time and the present~\cite{Kolb:1990vq}\footnote{Physics beyond the Standard Model may modify the cosmic GW background spectrum, making it a potential testing ground for BSM physics. We refer readers to \cite{Muia:2023wru} for details. On the other hand, a recent work \cite{Vagnozzi:2022qmc} discusses how a GW signal generated from thermalized graviton, if ever detected,  could also be used to rule out inflation.}. This follows simply from noting that after decoupling, gravitons would stop interacting and start to propagate freely, with their momenta redshifting due to the expansion and hence \cite{Ghiglieri:2015nfa,Ghiglieri:2020mhm,Ringwald:2020ist,Ghiglieri:2022rfp}
\begin{equation}
  \Omega_{\rm Eq.}(f)=\,\frac{16\pi^2}{3M_P^2 H_0^2}\frac{ f^4}{e^{2\pi f/ T_{\rm grav}}-1}, 
T_{\rm grav}=\, \frac{a(T=M_P)}{a(T=T_0)}M_P= \left(\frac{g_{*s}({\rm fin})}{g_{*s}(M_P)}\right)^{1/3} T_0.
\label{eq:OmegaEqCGMB}
 \end{equation}

\noindent where $g_{*s}({\rm fin}) = 3.931\pm 0.004$~\cite{Saikawa:2018rcs} is the number of effective degrees of freedom of the entropy density after neutrino decoupling. Here, $T_0$ is the current CMB temperature, and for the equilibrated spectrum with the peak frequency,
\begin{eqnarray}
\label{req:peak_freq_cgmb_1}
f_{\rm peak}^{\Omega} 
\approx  \frac{3.92}{2\pi}     \left[\frac{g_{*s}({\rm fin})}{g_{*s}(M_{\rm P})}\right]^{1/3}   T_0 
\simeq   74\,{\rm GHz} 
\,  
\left[\frac{g_{*s}(M_{\rm P})}{106.75}\right]^{-1/3}
\,.
\end{eqnarray}


On the other hand, for $T_{\rm{max}}< M_P$, when the Planck suppressed gravitational interaction rates are slower than the expansion rate of the Universe the gravitons are not expected to thermalize. Even in this scenario, the out-of-equilibrium gravitational excitations can still be generated from the thermal plasma, and $T_{\rm{max}}$ can be probed by the GW with the redshifted  GW spectrum given as~\cite{Ghiglieri:2015nfa,Ghiglieri:2020mhm,Ringwald:2020ist,Ghiglieri:2022rfp},

\begin{eqnarray}
\label{eq:Omega_CGMB_analytic_expression}
{{h^2}}\,\Omega  (f)
&\approx  &
\frac{1440\sqrt{10}}{2\pi^2} \,  
\,{{h^2}}\,\Omega_\gamma\,
\frac{\left[g_{*s}({\rm fin})\right]^{1/3}}
{\left[g_{*s}(T_{\rm max})\right]^{5/6}}\,
\frac{ f^3}{T_0^3}\, 
\frac{T_{\rm max}}{M_P} \,
\hat\eta\left(
T_{\rm max},2\pi \,
\left[\frac{g_{*s}(T_{\rm max})}{g_{*s}({\rm fin})}\right]^{1/3}\,
\frac{f}{T_0}
\right)
\\
&= & 
4.03\times 10^{-12} 
\, 
\left[\frac{T_{\rm max}}{M_P}\right]
\left[\frac{g_{*s}(T_{\rm max})}{106.75}\right]^{-5/6}
\left[ \frac{f}{\rm GHz}\right]^3
\hat\eta\left(
T_{\rm max},2\pi \,
\left[\frac{g_{*s}(T_{\rm max})}{g_{*s}({\rm fin})}\right]^{1/3}\,
\frac{f}{T_0}
\right)
 \;.
\nonumber
\end{eqnarray}
with peak frequency,
\begin{eqnarray}
\label{req:peak_freq_cgmb}
f_{\rm peak}^{\Omega} 
\approx  \frac{3.92}{2\pi}     \left[\frac{g_{*s}({\rm fin})}{g_{*s}(T_{\rm max})}\right]^{1/3}   T_0 
\simeq   74\,{\rm GHz} 
\,  
\left[\frac{g_{*s}(T_{\rm max})}{106.75}\right]^{-1/3}
\,,
\end{eqnarray}

\begin{figure}
    \centering
    \includegraphics[width=0.7\textwidth]{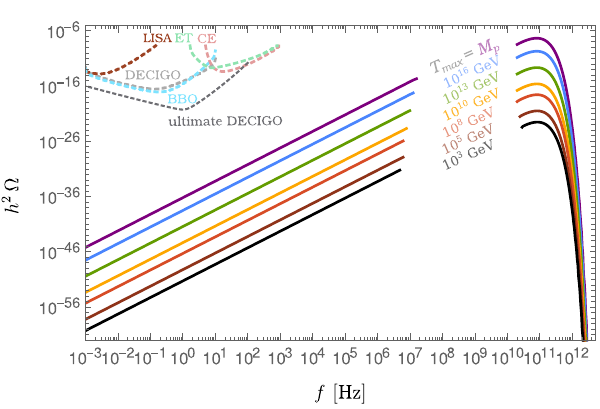}
    \caption{Shape of gravitational wave spectra generated from the primordial thermal plasma for different values of maximum temperature $T_{\rm{max}}$. Figure is taken from ref \cite{Ringwald:2020ist} }
    \label{GW_themal_plasma}
\end{figure}

$\Omega_\gamma \equiv 2.4728(21)\times 10^{-5}/h^2$ the present fractional energy density of the CMB photons, with temperature $T_0 = 2.72548(57)$\,K and $f$ being the present day GW frequency. Finally, $\hat{\eta}$ is a dimensionless source term the details of which can be found in \cite{Ghiglieri:2015nfa,Ghiglieri:2020mhm,Ringwald:2020ist,Ghiglieri:2022rfp}. As seen from Eq. \ref{eq:Omega_CGMB_analytic_expression}, the amplitude scales approximately linearly with $T_{\rm max}$ and hence can play the role of a hot Big Bang thermometer. Considering the broken power law and using Eq. \ref{eq:Omega_CGMB_analytic_expression} and Eq. \ref{req:peak_freq_cgmb}, one can plot the variation of the GW spectra with the frequency for different values $T_{\rm max}$ as shown in Fig. \ref{GW_themal_plasma} and as expected a larger $T_{\rm max}$ corresponds to a larger amplitude of the GW spectrum. It is interesting to point out that, up to the factor $\left[\frac{g_{*s}({\rm fin})}{g_{*s}(T_{\rm max})}\right]^{1/3}$, the peak frequency coincides with the CMB peak frequency, 
\begin{equation}
f_{\rm peak}^{\Omega_{\rm CMB}} 
\simeq  \frac{3.92}{2\pi} \    T_0 \simeq 223\, {\rm GHz} 
\,,
\end{equation}
of the present energy fraction of the CMB per logarithmic frequency, 
\begin{equation}
\label{eq:CMB}
\Omega_{\rm CMB} (f)
=
\frac{16\,\pi^2 }{3 H_0^2 M_P^2} \frac{f^4}{{\rm e}^{2\pi f/T_0}-1}
\,.
\end{equation}

\noindent As can be seen, these high-frequency GW spectra remain out of reach of the current and future GW experiments.
The authors of \cite{Drewes:2023oxg} showed that refs. \cite{Ghiglieri:2015nfa,Ringwald:2020ist} have missed the suppression of the GW production rate on superhorizon scales. Including this suppression, one obtains the following Fig. \ref{GW_themal_plasma_2} ,
\begin{figure}
    \centering
    \includegraphics[width=0.7\textwidth]{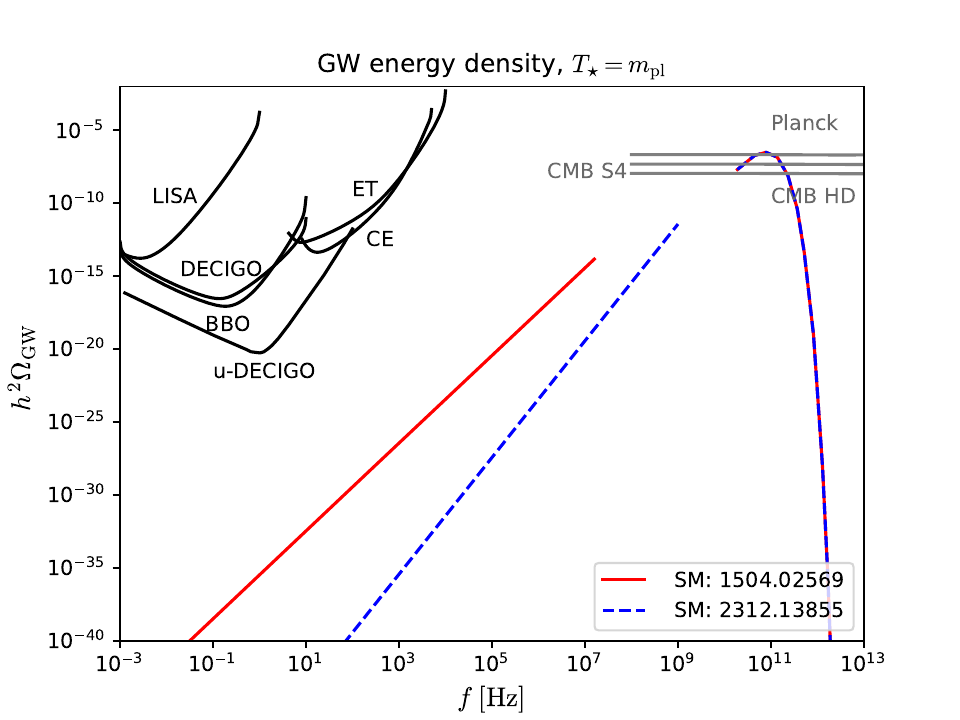}
    \caption{Comparision of the shape of gravitational wave spectra generated from the primordial thermal plasma with (blue/dashed ) and without (red/solid) taking into account the suppression of the GW production rate on superhorizon scales. Here the maximum temperature is denoted by $T_{\rm{\star}}$. We thank Marco Drewes for preparing and sharing this plot. }
    \label{GW_themal_plasma_2}
\end{figure}
\section{Conclusions}\label{sec:conclusions}
Gravitational wave cosmology is one of the most promising avenues for discovering physics beyond the Standard Model. Next-generation detectors promise to have cosmologically significant strain sensitivities over 12 decades of frequencies - from the nanoHz to the kHz. In section \ref{sec:detector} we outlined three major strategies for gravitational wave detection - pulsar timing arrays, astrometry, and of course interferometry both terrestrial and in space.

In sections \ref{sec:phase}-\ref{sec:SIGW} we covered three very active subfields in gravitational cosmology which cover, in this review's opinion, the three major sources that next-generation detectors will be sensitive to. First-order cosmic phase transitions can produce a source from the collision of bubbles, the explosion of sound, and the aftermath of turbulence. We surveyed the many possible scenarios that can produce a cosmological phase transition, though the electroweak phase transition continues to inspire the most interest. This source is arguably the most difficult to understand as it requires taming of finite temperature perturbation theory and understanding how the behaviour of bubbles and sound shells translates to precise predictions of spectra. So far, the community requires simulations to shed light on both issues and the theory is still maturing. 

We covered an array of topological defects including hybrid defects in section \ref{sec:top}. The string signals are expected to grow with the scale of spontaneous symmetry breaking, as is the cause of textures. Domain walls by contrast have a peak frequency and strength that is determined not just by the surface tension, but the lifetimes of the walls. In all three cases, there can be a detectable signal arising from physics at scales many orders of magnitude higher than what can be reached at the LHC. Successive symmetry-breaking steps can result in hybrid defects, with each possible combination having its own unique gravitational wave signature, indicating the possibility of learning about a potentially rich cosmic history. The most severe source of theoretical uncertainty is undoubtedly the mismatch between field-theoretic and string simulations of cosmic strings. This conflict appears to still be in flux and needs to be resolved. We conclude this section by discussing some of the applications of GW from topological defects.

Finally, in section \ref{sec:SIGW} we considered scalar-induced gravitational waves. Despite being lumped under the same category, two radically different types of scenarios can result in scalar-induced gravitational waves - a period of ultra-slow roll inflation and a sudden change in the equation of state. Both sources have a theoretical uncertainty due to gauge invariance and it is hard to rigorously pin down the severity of this issue. There are, however, plausible arguments that we at least qualitatively understand the gravitational wave signal. 

In the case of all three signals, there are no known mechanisms under which the Standard Model could provide the conditions needed to generate such a primordial source. So any detection of such a background would be a smoking gun of new physics. Moreover, all three sources potentially can involve physics at a higher scale than we can reach at the LHC. In the case of phase transitions, the frequency is loosely proportional to the scale of the transition, meaning that it is not a probe of very high-scale physics. However, the other two scenarios are less constrained. In the case of strings, the signal is expected to grow with the scale of symmetry breaking, whereas the peak frequency of signals from domain walls and resonantly enhanced scalar-induced gravitational waves corresponds to the lifetime of a macroscopic object. The range of applications to test high-scale physics is already quite broad and we expect it to grow over the next few decades.

There are other sources of gravitational waves, some of which we grouped together in section \ref{sec:HFsources }, which we only briefly summarized due to the fact that their detection is loosely connected to next-generation detectors such as inflation,\footnote{We say that is loosely connected rather than unconnected due to the possibility of a blue tilted spectrum} or because the source is high frequency. 

The discovery of gravitational waves has provided a method for observing the first moment of creating, where deficiencies in our understanding of what lies beyond the Standard Model matter the most. Gravitational wave cosmology is a new and rapidly growing field that has recently arisen in response to this challenge and opportunity and it feels as though we are just beginning to understand the potential of this field.



{\bf Acknowledgements.} We are grateful to David Dunsky, Jessica Turner, Peter Athron, Carlos Tamarit, Yann Gouttenoire, Wenyuan Ai, Guillem Domenech and Qaisir Shafi for some useful discussions. GW acknowledge the STFC Consolidated Grant ST/L000296/1 and RR acknowledges
financial support from the STFC Consolidated Grant ST/T000775/1.



\bibliography{refs}
\bibliographystyle{apsrmp4-2}.

\end{document}